\documentclass[reprint,
 amsmath,amssymb,
 aps,
amsbsy,
bm,
amsfonts
]{revtex4-2}
\usepackage{booktabs}
\usepackage{amsmath,bm}
\usepackage[thinc]{esdiff}
\usepackage{amsbsy}
\usepackage{xcolor}
\usepackage{placeins}
\usepackage{mathtools}
\usepackage{bbold}
\usepackage{wrapfig}
\usepackage[para,online,flushleft]{threeparttable}
\usepackage{floatrow}
\usepackage{siunitx}
\usepackage{graphicx} 
\usepackage{amsmath}
\usepackage{amssymb}
\usepackage{graphicx}
\usepackage{amsfonts}
\newcommand{\vvec}[1]{\mathbf{#1}}
\newcommand{\eqnref}[1]{Eq.\ \ref{#1}}

\newcommand{\figref}[1]{Fig.\ \ref{#1}}
\newcommand{\secref}[1]{Sec.\ \ref{#1}}
\newcommand{\tabref}[1]{Tab.\ \ref{#1}}
\newcommand{\Fbcal}{\pmb{\mathcal{F}}}

\newcommand{\reals}{\mathbb{R}}

\newcommand{\norm}[1]{\left\lVert#1\right\rVert}
\newcommand{\mysum}{\bigoplus_{} }

\usepackage{graphicx}
\usepackage{dcolumn}
\usepackage{bm}
\usepackage{hyperref}

\begin{document}
\title{Stretched and measured neural predictions of complex network dynamics}

\author{Vaiva Vasiliauskaite}
\affiliation{Computational Social Science, ETH Z\"urich, 8092 Z\"urich, Switzerland}
\author{Nino Antulov-Fantulin}\email{anino@ethz.ch}
\affiliation{%
Computational Social Science, ETH Z\"urich \& Aisot Technologies AG, Z\"urich, Switzerland}

\date{\today}
\begin{abstract}
Differential equations are a ubiquitous tool to study dynamics, ranging from physical systems to complex systems, where a large number of agents interact through a graph with non-trivial topological features. 
Data-driven approximations of differential equations present a promising alternative to traditional methods for uncovering a model of dynamical systems, especially in complex systems that lack explicit first principles.
A recently employed machine learning tool for studying dynamics is neural networks, which can be used for data-driven solution finding or discovery of differential equations. 
Specifically for the latter task, however, deploying deep learning models in unfamiliar settings—such as predicting dynamics in unobserved state space regions or on novel graphs—can lead to spurious results. Focusing on complex systems whose dynamics are described with a system of first-order differential equations coupled through a graph, we show that extending the model's generalizability beyond traditional statistical learning theory limits is feasible. However, achieving this advanced level of generalization requires neural network models to conform to fundamental assumptions about the dynamical model. Additionally, we propose a statistical significance test to assess prediction quality during inference, enabling the identification of a neural network's confidence level in its predictions.
\end{abstract}

\maketitle

\section{Introduction}

Coupled differential equations serve as a fundamental modeling tool for dynamical systems, enabling classical analyses such as stability and control. In its simplest form, a dynamical system is defined as a set of coupled ordinary differential equations $\dot{\vvec{x}}(t) = \Fbcal(\vvec{x},t)$ that describe the rate of change of a dependent variable $\vvec{x}$ at time $t$. Discovering a dynamical model entails the task of finding a suitable vector field $\Fbcal$, and requires a deep understanding of first principles from, e.g.\ fluid or solid mechanics, as well as insights derived from experiments and, above all, creativity.

In today's data-rich world, there is an allure to leverage this abundant resource for synthesizing $\Fbcal$. One popular approach involves utilizing symbolic regression analysis to determine the elementary functions that constitute $\Fbcal$~\cite{ Brunton2016DiscoveringSystems, Gao2022,aiFeynman,schmidt2009distilling}. However, these techniques assume that parsimonious enumeration of the dynamics in terms of known basis functions exists, is finite, and is possible to determine. This may not always be the case~\cite{gilpin2020learning}, especially if no prior knowledge about a system is available~\cite{hillar2012comment}. To guarantee that any vector field can be expressed with symbolic regression, one would need to resort to a dictionary that spans infinite dimensional functional spaces (as is the case for a Fourier basis). 
Generally, the system identification task solved by symbolic regression using experimental data is NP-hard for both classical and quantum systems~\cite{learningPhyDataNPhard,virgolin2022symbolic}.

\begin{figure}[!ht]
    \centering
    \includegraphics[width=\linewidth]{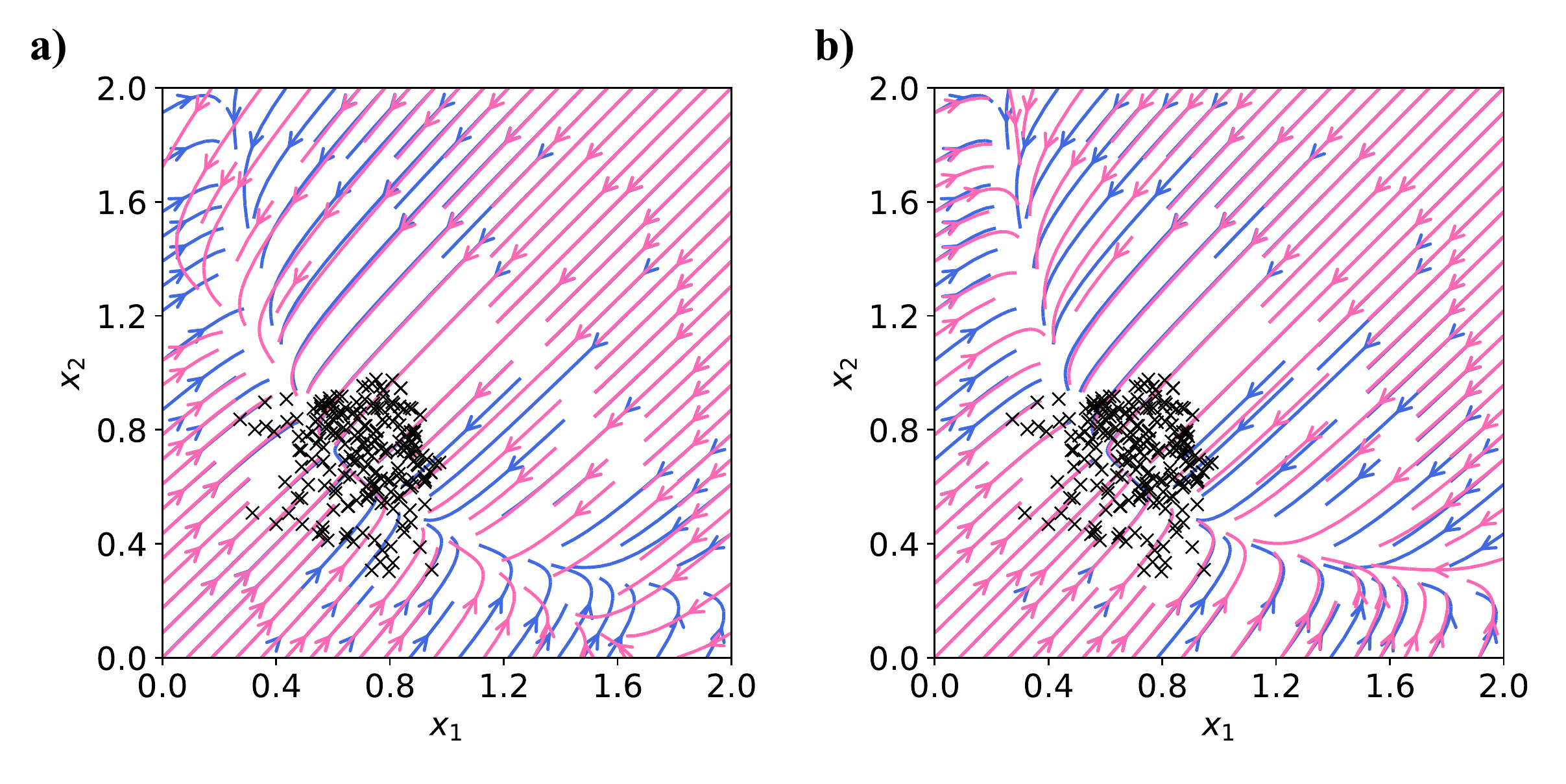}
    \caption{Reconstruction of a two-dimensional vector field representing mass-action kinetics (MAK) i.e., protein-protein interaction dynamics~\cite{Barzel2013UniversalityDynamics,voit2000computational} (see \tabref{tab1} for definition) \textbf{a)} utilizing a feed-forward neural network, and \textbf{b)} utilizing a dedicated graph neural network that conforms to inductive biases, appropriate for modeling dynamics described as a system of coupled ordinary differential equations. The application of these biases extends the domain of vector-field approximation for MAK dynamics well beyond the range of our training data, indicated as black crosses. }
    \label{fig:neural-approx-motivation}
\end{figure}

Instead of identifying the elementary functions that constitute $\Fbcal$, an alternative and appealing method approximates the vector field using a feed-forward neural network~\cite{Zang2020NeuralNetworks,Murphy2021DeepNetworks}, which is a universal function approximator for continuous functions between two Euclidean spaces~\cite{hornik1989multilayer, cybenko1992approximation}. Within the bounded support of training data, the accuracy of approximation of a deep learning model is guaranteed by the Universal Approximation Theorems (UAT) for various deep neural networks including those with arbitrary depth~\cite{yarotsky2017error, kidger2020universal}, bounded depth and width~\cite{maiorov1999lower}, and permutation invariant neural networks operating on sets or graphs~\cite{wagstaff2022universal, zaheer2017deep, xu2018powerful}. 

While UAT does not give a recipe for how to find the weights of neural networks, potentially non-globally optimal solutions found through gradient descent in practice give good approximations~\cite{wang2023implicit,zhao2022combining,arora2022understanding,du2019gradient,locatello2019challenging} for various tasks related to dynamical systems, such as control~\cite{bottcher2022ai,asikis2022neural,jin2020pontryagin} and forecasting~\cite{Srinivasan2022ParallelNetworks,pathak2018model}.

However, if a neural network is used to understand dynamical systems by, e.g.\ extrapolating predictions in regions of state space that were not observed during training, it needs to have an ability to generalize. Such generalization capacity is not guaranteed by UAT and has traditionally been studied through the lens of the Statistical Learning Theory (SLT)~\cite{vapnik1991necessary, bartlett2002rademacher}, which provides different generalization bounds~\cite{tucker1959generalization} for trained models that rely on the assumption of the independent and identically distributed (i.i.d.)\ sampling of training and test data. Importantly, learning to generalize is impossible without appropriate inductive biases~\cite{gerbelot2023applying,zhang2019identity,Goyal2022InductiveCognition}. 
An architecture that disrespects the physical realism of the true dynamical model (adheres to wrong inductive biases) may lead to an abrupt cut-off from the domain within which the model performs well, i.e.\ the training setting. 

\figref{fig:neural-approx-motivation} illustrates the importance of inductive biases for generalization of the model of dynamics. The dynamics depicted here is described by a system of two first order ordinary differential equations $\dot{x}_1=F-Bx_1^b+ Rx_2$, and $\dot{x}_2=F-Bx_2^b+ Rx_1$. Note that 
there exists a mirror symmetry between the variables $x_1$ and $x_2$. The neural network model in \figref{fig:neural-approx-motivation}\textbf{a)} does not account for this symmetry and does not generalize beyond the training data's support. The prediction of a dedicated graph neural network model, as shown in \figref{fig:neural-approx-motivation}\textbf{(b)}, which is proposed in this paper, utilizes an inductive bias that the vector field is invariant under the transformation $x_1\leftrightarrow x_2$. This allows the model to form predictions about the vector field beyond the training domain.
This example not only illustrates one possible violation of UAT assumptions, but also a possibility to stretch the model's generalization capacity with appropriate inductive biases. The second kind of violation often comes in forecasting problems, which is even more challenging, since it involves making consecutive predictions in time, where numerical errors accumulate~\cite{goyal2022inductive,fotiadis2023disentangled}. 

In this paper, we show that graph neural networks' capacity to generalize can be \emph{stretched} to form predictions in scenarios that go beyond the classical boundaries of UAT and SLT. However, to achieve this level of generalization, the neural network models have to adhere to basic assumptions about the vector field that describes the dynamics. To \emph{measure} the limitations of neural network generalization capacity during inference, we also define a statistical significance test, driven by the allowed fluctuations of the model variance.

We concentrate on a sub-domain of dynamical systems known as complex systems. Complex systems are often modeled as networks (graphs) composed of a large number of interdependent, internally equivalent elements called agents~\cite{boccaletti2006complex,vasiliauskaite2020understanding}. The emergent behavior and globally observed features of these systems arise from the local interactions among agents. Examples of complex systems include ecosystems, economies, the brain, as well as social networks. We consider a general class of dynamical models on networks~\cite{Barzel2013UniversalityDynamics} where the change in the state of each agent $i$, denoted as $\dot{\vvec{x}}_i(t)\in\reals^{ k}$, depends not only on its own state $\vvec{x}_i(t)$ but also on the sum of the states of its neighbors:
\begin{eqnarray}\label{eq:dynamical_system_barzel}
     \dot{\vvec{x}}_i(t) &=& \textbf{L}(\vvec{x}_i(t)) + \bigoplus_j A_{ij}(t) \textbf{Q}(\vvec{x}_i(t), \vvec{x}_j(t))\\
     &=&\Fbcal(\vvec{x}_i(t),\vvec{x}(t),\vvec{A}(t)),\nonumber
\end{eqnarray}
where $\vvec{x}(t)\in\reals^{n\times k}$ is a tensor that collects states of all $n$ nodes at time $t$.
This system can be described by a system of first order ordinary differential equations (ODEs), where $\vvec{A}(t)\in\mathbb{R}^{n\times n}$ represents a potentially time-varying network adjacency matrix of a graph $\mathcal{G}$, $\textbf{L}$ is a function that describes self-interactions, $\textbf{Q}$ is a function that models pairwise interactions between neighbors, and $\mysum$ denotes an aggregation function. 
Dynamical systems of arbitrary size and connectivity structure can be described using the same functions $\textbf{L},\textbf{Q}, \mysum$ thereby entailing the same type of dynamics, only on a different network. 

By making appropriate choices for the functions $\textbf{L}$, $\textbf{Q}$, and $\mysum$, \eqnref{eq:dynamical_system_barzel} can represent a wide variety of models for complex network dynamics~\cite{Barzel2013UniversalityDynamics}, including biochemical dynamics, birth-death processes, spreading processes, gene regulatory dynamics~\cite{Barzel2013UniversalityDynamics}, as well as chaotic~\cite{Sprot2008ChaoticNetworks}, diffusive~\cite{delvenne2015diffusion}, oscillatory~\cite{rodrigues2016kuramoto}, neuronal~\cite{rabinovich2006dynamical} dynamics.
It is worth noting that for many complex systems, the dynamical model, i.e., the functional forms of $\textbf{L}$, $\textbf{Q}$, and $\mysum$ remains unknown, as there are no first principles~\footnote{Such as the principle of relativity that leads to, e.g.\ the mass-energy equivalence.} from which such models can be derived \emph{ab initio}, making data-driven approaches a particularly alluring option. Note that although the \eqnref{eq:dynamical_system_barzel} is general, it does not encompass all categories of dynamical systems, including non-local~\cite{riascos2021random} and stochastic~\cite{gleeson2013binary} dynamics, which are avenues for future research. Furthermore, in this study we will concentrate on a simple case of local interactions-driven deterministic, autonomous, time-invariant dynamics that occurs on undirected, unweighted, static graphs.

The paper is organized as follows. In \secref{sec:model}, we introduce a prototype neural network model, denoted as $\pmb{\Psi}$, designed to approximate the dynamical system $\Fbcal$ described by \eqnref{eq:dynamical_system_barzel}, as well as a general learning setting for the task of approximating $\Fbcal$. In \secref{sec:hierarchy}, we define the generalization capacity of a neural model for dynamics through a series of test settings that extent beyond the conventional boundaries of SLT. In \secref{stat_sig}, we outline the statistical significance test that identifies the limit of a neural network's generalization capacity. Lastly, in \secref{sec:results} we assess the generalization capacity of the proposed model as well as our ability to quantify prediction accuracy. 

\section{A Graph Neural Network Model for the Vector Field of Complex Network Dynamics}\label{sec:model}

To approximate a dynamical system with a neural network, we need to learn the vector field $\Fbcal:\reals^{n\times k} \rightarrow \reals^{n\times k}$ which describes the change in the state of system's variables $\vvec{x}(t)$. While we are free to choose any neural network for the task, some models allow for more efficient training and better generalization than others. A natural candidate neural network is that which mimics the structure of the function that is approximated. Therefore, for dynamics on a complex network, a neural network model should resemble \eqnref{eq:dynamical_system_barzel}. Since the functions $\mathbf{L},\mathbf{Q}, \mysum$ are independent of the graph structure and the system size, a neural network model may also posit such independences. A graph neural network (GNN)~\cite{Scarselli2009TheModel,Xu2018HowNetworks} $\pmb{\Psi}$ is naturally suited for this task, as it is an universal function approximator and exhibits permutation invariance with resepct to the ordering of nodes, ${\textbf{x}}_i$. 
Furthermore, the number of parameters in a GNN scale w.r.t.\ $ k$, and not with $n$, thus significantly reducing the dimensionality of the task.

Since \eqnref{eq:dynamical_system_barzel} consists of separate and potentially different functions that describe the self- and the neighbor-interactions, namely $\mathbf{L}$ and $\mathbf{Q}$, we separate $\pmb{\Psi}$ into two separate functions:
\begin{eqnarray}\label{eq:dynamic_systems_approx}
\dot{\vvec{x}}
&=& \pmb{\psi}^{\text{self}}(\vvec{x})+ \pmb{\psi}^{\text{nbr}}(\vvec{x}), 
\end{eqnarray}
Here $\pmb{\psi}^{\text{self}}(\cdot)$ is a simple feed forward neural network with one or more hidden layers. Each hidden layer applies a linear transformation followed by a non-linearity $\sigma$ to an output of the previous layer. Starting from an input $\vvec{h}_{0}$ which is a tensorized $\vvec{x}$, the transformation at each layer $\alpha\in\{1,...,N\}$ is defined as
\begin{equation*}
\vvec{h}_{\alpha} =  \sigma{(\vvec{h}_{\alpha-1}\vvec{W}_{\alpha}^{\top} +\vvec{b}_{\alpha})}.
\end{equation*}
Here $\vvec{W}_{\alpha}\in\mathbb{R}^{d_{\alpha}\times d_{\alpha-1}}$ is a weight matrix, and $\textbf{b}\in\mathbb{R}^{d_{\alpha}}$ is a bias term. The output of the $N^{\text{th}}$ --- last --- layer is the output of the neural network.

The neighbor-interaction term $\pmb{\psi}^{\text{nbr}}$ must include quadratic terms $\vvec{x}_i\vvec{x}_j$ that are a likely functional form of $\vvec{Q}$. A prototypical single-layer graph neural network, such as a convolutional graph neural network does not simply satisfy such a condition. However, including multiple graph neural layers may cause other problems. Note that \eqnref{eq:dynamical_system_barzel} considers local interactions: per infinitesimal unit of time, a signal propagates from a node to its immediate neighbors. A multi-layer graph neural network would include terms $\vvec{A}^{\kappa}$ that allow for $\kappa$-hop interactions via length $\kappa$ walks in a network at a timescale smaller than the infinitesimal $\mathrm{d}t$ thereby subdividing $ \mathrm{d}t$ to $\kappa$ intervals and breaking an assumption of temporal locality. 
 
To ensure that $\pmb{\psi}^{\text{nbr}}$ includes quadratic terms, but no powers of adjacency matrix, we define it as a pair of feed forward neural networks, integrated through a graph adjacency matrix~\footnote{In place of $\vvec{A}$, one may consider, e.g., a single-layer graph convolution~\cite{kipf2016semi}, degree-scaled adjacency matrix, or a laplacian matrix.  
 } and an aggregation function:
\begin{eqnarray}\label{eq:GNN}
\pmb{\psi}^{\text{nbr}}(\vvec{x}) 
&=& \pmb{\psi}^{\mysum}\left[\vvec{A}\odot \left( \pmb{\psi}^{q_1}(\vvec{x})^{\top_{1}} \times_{b}\pmb{\psi}^{q_2}(\vvec{x})^{\top_{2}} \right) \right].
\end{eqnarray}
Here $\pmb{\psi}^{\mysum{}}(\cdot),\pmb{\psi}^{q_1}(\cdot), \pmb{\psi}^{q_2}(\cdot)$ are feed-forward neural networks (single-layer or multi-layer), an operator $\odot$ denotes a standard ``broadcasted" element-wise multiplication, while $\times_{b}$ indicates a ``batched" matrix-matrix product; $\top_{1}, \top_{2}$ denote specific transpose operations. $\pmb{\psi}^{\mysum{}}$ is an invariant pooling layer, which is an operation that maintains the invariance with respect to the order of inputs~\cite{wagstaff2022universal, zaheer2017deep, xu2018powerful} when aggregating neighbor interaction terms. Other forms of a GNN are possible, as long as they adhere to relevant inductive biases for dynamical systems), see Sec.\ I in the Supplemental Information (SI) for further discussion, along with the details for all neural network mappings for the GNN described here.

The importance of inductive biases is revealed through comparison of the model outlined here to other graph neural network models~\cite{GraphConv,ResGatedGraphConv,SAGEConv,ChebConv,GATConv}, that take the graph structure into the account but may not conform to other biases regarding complex network dynamics. We found that a graph neural network that: \textbf{(i)} separates self- and neighbor-interaction terms; \textbf{(ii)} includes only $\kappa=1$ power of adjacency matrix; and \textbf{(iii)} includes quadratic terms $\vvec{x}_i\vvec{x}_j$,
significantly outperforms in its generalization ability other state-of-the-art deep graph neural networks~\cite{pyG}: SAGEConv~\cite{SAGEConv}, GraphConv~\cite{GraphConv}, ResGatedGraphConv~\cite{ResGatedGraphConv}, GATConv~\cite{GATConv}, ChebConv~\cite{ChebConv}.
To further test our hypothesis on appropriate inductive biases, we have reduced the tested deep GNNs to shallow (single graph convolutional layer) models. We found that the shallow models perform better than their multi-layer counterparts.
Notably, the \textsc{ResGatedGraphConv} model~\cite{bresson2017residual}, which closely aligns with the model discussed in this section, demonstrates good generalization ability. The detailed results of this analysis are presented in Sec.\ IV of the SI. 

\subsection{Learning setting}
The training and test data are defined as $\mathcal{D} = \{(\vvec{x}, \vvec{y})\}, \text{ s.t.\ }\vvec{x} \in \reals^{n\times  k}, \vvec{y} \in \reals^{n\times k}$. The best neural approximation $\pmb{\Psi}^{*}$ is obtained by minimizing the loss $\mathcal{L}$ between the true labels $\vvec{y}$ and the predicted labels $\hat{\vvec{y}}$ in the training data:
\begin{eqnarray*}
    \pmb{\Psi}^{*} &=& \arg \min_{\pmb{\Psi}:\reals^{n\times k} \rightarrow \reals^{n \times k}} \mathop{\mathbb{E}}_{\mathcal{P}(\vvec{x}, \vvec{y})} \mathcal{L}(\vvec{\hat{y}}, \vvec{y}),\\ 
 && \text{ where \ }\vvec{\hat{y}} = f(\pmb{\Psi}, \vvec{x}, \pmb{\theta}),
\end{eqnarray*}
and $\pmb{\theta}$ are non-trainable parameters necessary to estimate $\hat{\vvec{y}}$, e.g.\ adjacency matrix $\vvec{A}$. In the subsequent sections, we will omit the superscript ``$*$'' and refer to the best model obtained after training as $\pmb{\Psi}$. The results presented in the subsequent sections were obtained using the normalized $\ell^1$ norm:
\begin{eqnarray}\label{eq:loss}
\mathcal{L} &=& \frac{1}{Z}\Big[\mathbf{E}_{  (\vvec{x},\vvec{y})\in\mathcal{D}} \big[ \norm{\vvec{y}- f(\pmb{\Psi}, \vvec{x}, \pmb{\theta})}_1 \\
&+&  \lambda \mathbf{Var}_{i}\norm{\vvec{y}_i- f_i(\pmb{\Psi}, \vvec{x}, \pmb{\theta})}_1 \big]\Big],\nonumber
\end{eqnarray}
where $\norm{\cdot}_1$ denotes the $\ell^1$ norm and the normalization $Z=nk$. $\ell^1$ is a preferred choice due to its robustness in high-dimensional spaces which are often susceptible to the curse of dimensionality when using other metrics like $\ell^2$~\cite{aggarwal2001surprising}. Additionally, its inherent interpretability is particularly suitable for our generalization analysis.

The loss per each sample, or node, can be further weighed to reduce bias towards optimizing for common nodes at a cost of poor predictions at rare nodes~\cite{Murphy2021DeepNetworks}. Instead of such weighting, we adopt a regularizer that penalizes solutions with high variance in the loss across nodes, without a priori defining which nodes are important. 

In \secref{sec:prediction} and \secref{sec:non_iid_forecast}, we will consider cases where labels originate from a vector field $\vvec{y}=\Fbcal(\vvec{x})$ and $\vvec{\hat{y}}=\pmb{\Psi}(\vvec{x})$. In \secref{sec:prediction} and \secref{sec:non_iid_forecast}\textit{a}, we will consider analytically computed $\Fbcal(\vvec{x})$, whereas in \secref{sec:non_iid_forecast}\textit{b} it will be computed numerically, following a Newton's quotient rule. Lastly, in \secref{sec:realdata}, we will discuss training with labels derived from noisy trajectories: $\vvec{{y}}=I[\Fbcal,\vvec{x},\vvec{A},t_1,t_2] + \pmb{\varepsilon}$ and predicted labels defined as $\vvec{\hat{y}}=I[\pmb{\Psi},\vvec{x},\vvec{A},t_1,t_2]$. 
Here $\pmb{\varepsilon}$ denotes observational noise, and numerical integration is denoted as 
\begin{equation}\label{eq:integral}
I[g,\vvec{x},\vvec{A},t_1,t_2]=\vvec{x}+ \int^{t_2}_{t_1} g(\vvec{x}',\vvec{A})\text{d}\tau.
\end{equation}
Note that the time step $\text{d}\tau$ in the numerical integration may be different for the true and the predicted labels.

Let us further denote probability density functions associated with a training setting by $\phi$, and with a test setting by $\omega$. For training, we consider two types of datasets: $\vvec{x} \sim \phi_{x}(\vvec{x})$ are i.i.d.\ samples from a pre-defined distribution, or non-i.i.d.\ samples $\{\vvec{x}(t)\}$ obtained by numerical integration $\vvec{x}(t) = I[\pmb{\Fbcal},\vvec{x}_0,t_0,t]$ with a random initial condition $\vvec{x}_0\sim \phi_{x_0}(\vvec{x})$. 

\section{Generalizability of a neural network model of dynamics}\label{sec:hierarchy}

Neural networks belong to the class of over-parameterized models, whose training through gradient descent and generalization are currently not fully understood within classical SLT~\cite{belkin2019reconciling, zhang2021understanding,jakubovitz2019generalization}\footnote{Novel SLT frameworks, including algorithm stability~\cite{hardt2016train, bousquet2002stability}, algorithm robustness~\cite{xu2012robustness}, PAC-Bayes theory~\cite{bartlett2017spectrally,mcallester1999pac}, compression and sampling~\cite{arora2018stronger,giryes2022function} are active fields of research and could possibly shed light on generalization in this class of models.}. Furthermore, since real training data would be in the form of non-i.i.d.\ time series trajectories that likely do not comprehensively cover the state space, it remains unclear whether such data is adequate to reconstruct $\Fbcal$. Therefore, in this section, we introduce various test settings for neural models of complex network dynamics and define a model's generalization capacity as its ability to make accurate predictions across increasingly challenging test settings, as well as test settings that increasingly diverge from the training setting.

A neural network may be used to perform three distinct tasks, which, in the context of this paper, we define as \emph{approximation, prediction, and forecasting}. \textbf{Approximation} of dynamics pertains to the ability of accurately learning the vector field. In other words, a good approximation is achieved if the trained model possesses a low 'in-sample' loss, i.e., the loss on the training data. Approximation capacity of neural networks is generally guaranteed by UAT. 

The next two tasks, prediction and forecasting, relate to model's increasing generalizability capacity. 
\textbf{Prediction} of dynamics considers how well a model performs ``out-of-sample'', i.e.\ when it is tested on a data which is not part of the training set. For example, in standard SLT, one considers model's ability to extrapolate to unseen input datapoints which are sampled from the same probability distribution function as the training data
. A more generalizable model is that which can achieve a loss as low as during training in more general settings, for example, when the equivalence between probability distributions is relaxed, or when the training and test data have different support. We note that SLT offers no performance guarantees when the statistical properties of the test data deviate from those of the training data. Furthermore, when the support of the test data is not equivalent to the support of the training data, UAT is no longer valid, since it is built upon an assumption that an approximated function has a compact support~\footnote{Some UAT results cover density in non-compact domains, e.g.~\cite{kidger2020universal}. Nonetheless, the authors proceeded with the assumption that a target function maps to zero outside of a given support.}. It is worth noting that the last condition is typically absent in traditional machine learning approaches, as an input standardization step ensures that the model never receives inputs outside the range of values it was trained on~\cite{anysz2016influence}. However, in the context of dynamical systems, standardization would modify the dynamics' outcomes and break connections to physical reality, as distinct inputs would be non-injectively mapped to the same standardized values. 

Specifically in the context of complex network dynamics, prediction may also entail the substitution of the graph utilized during training with an alternative graph. To study network effects, a neural model needs to form reliable predictions when the interaction rules are preserved, but the dynamical system is different in terms of who interacts with whom, or the network size. For example, one can consider a test graph from the canonical ensemble~\cite{bianconi2009entropy} of the training graph~
\footnote{Alternatively, 
one may consider a micro-canonical ensemble by employing, e.g.\ a configuration model~\cite{park2004statistical}, thereby imposing harder constraints of fixed number of edges.}, or a graph with different size.

Lastly, a special type of prediction is \textbf{forecasting}, whereby a model can be used to form $m$-step predictions to the future. Accurate long-term predictions of dynamics are difficult for both numerical calculators and neural networks alike, because error accumulation can cause divergence in finite time~\cite{fotiadis2023disentangled}. 

All in all, we underscore that a model can strive for various degrees of generalization, from approximation to predictions and, finally, forecasting. It is also important to consider the data which was used during training: a model trained using non-i.i.d.\ time series data may adopt biases that would lead to poorer prediction capacity outside the training setting. Therefore, in \secref{sec:results} we will assess the performance of the GNN presented in \secref{sec:model} in performing all of these tasks, using both data that adheres to SLT assumptions as well as time series data.

\section{Neural Network Null Model}\label{stat_sig}
As a means to gauge the representativeness of the prediction during inference, we suggest using a statistical hypothesis test, based on a dedicated null model.

\begin{figure}[!h]
    \centering
    \includegraphics[width=1\linewidth]{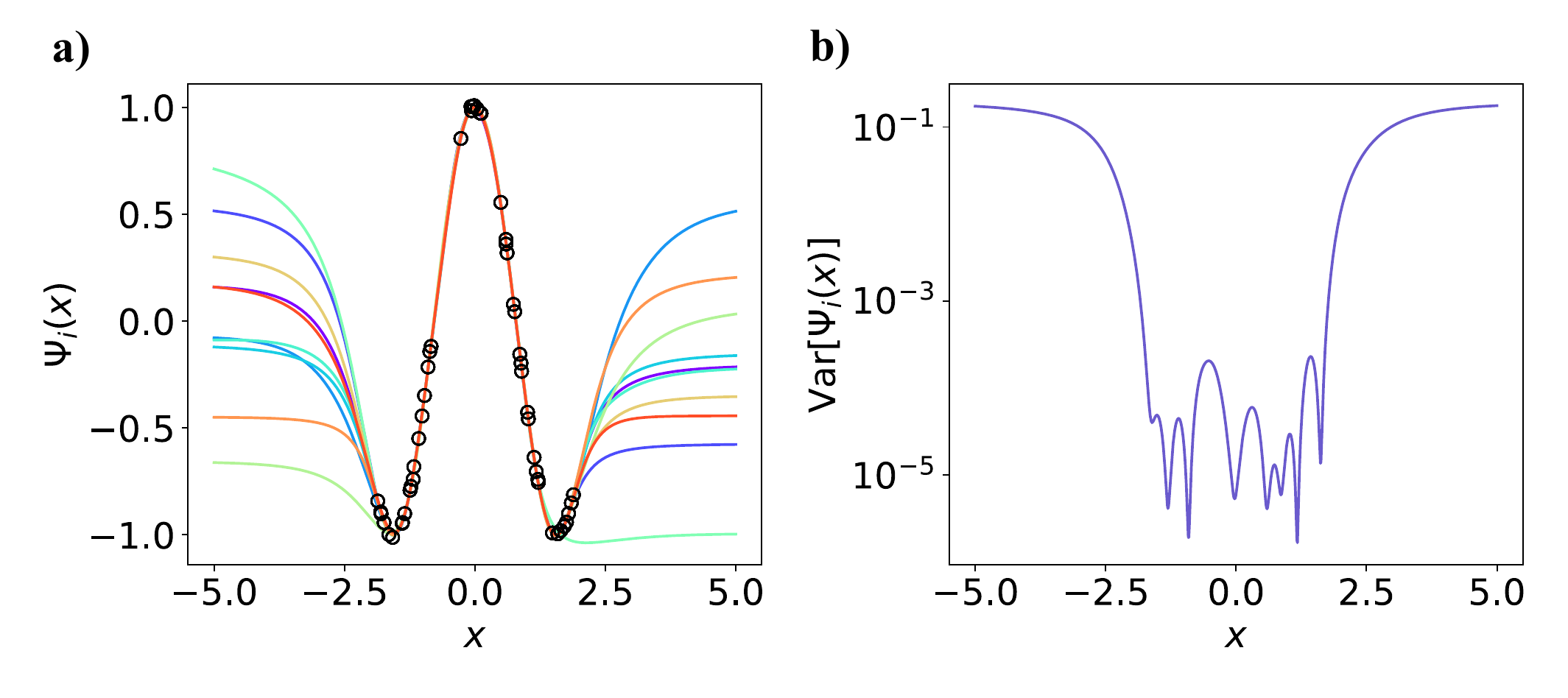}
    \caption{An ensemble of 10 overparameterized feed forward neural networks $\{ \Psi_m(x) \}$ trained independently to approximate $\mathcal{F}(x)=\cos{2x}$ within the range $[-2,2]$ (50 training samples are indicated with black circles). \textbf{a)} Predictions of the models outside the training range, $x\in[-5,5]$. \textbf{b)} Sample variance across ensemble of neural networks $\{ \Psi_m(x) \}$. }
    \label{fig_var_range}
\end{figure}

\paragraph*{Null model} Let us consider an ensemble $\{ \pmb{\Psi}_m \}$ of overparameterized neural networks that are trained on bootstrapped versions of training data $\mathcal{D}$.
Each $\pmb{\Psi}_m$ is a realization of a random variable, where the sources of randomness are the stochastic nature of the optimization algorithm and the initialization of weights.
Our ansatz is that an ensemble of neural networks disagrees more on the estimate of the vector field $\Fbcal$ as the test samples diverge from the training setting.
\figref{fig_var_range} shows an example of such behaviour, where the variance of overparameterized neural networks $\{ \pmb{\Psi}_m \}$ indeed increased once we departed from the training range.

\paragraph*{$d$-statistic} One possible statistic to quantify acceptable amount of model variance is the variance across neural networks in the prediction of $i^{\text{th}}$ node's derivative i.e.\
$d(\vvec{x}_i) = \text{Var}(\pmb{\Psi}_m(\vvec{x}_i))$  --- the variance term in the bias-variance decomposition~\cite{hastie2009elements}. In the case that the node state variable $\vvec{x}_i$ is multi-dimensional ($k>1$, but not too large), one can generate the null distribution $f_d(\xi)$ and perform a statistical test for each dimension separately. 

To estimate the $d$-statistic, we first train a total of $M$
neural networks using, for each, a different bootstrapped sample of the same dataset $\mathcal{D}$.
The distribution of $d$-statistic, $f_d(\xi)$ is  then obtained by repeatedly taking a size $m$ sub-sample of neural networks and computing a $d$-statistic of some input $\vvec{x}_i$.   
By analyzing variance in such a bootstrapped dataset, we estimate the effect of changes in the training data on the estimated models, thereby performing a form of stability analysis.

\paragraph*{Significance test} Since we expect the variance across models to increase, i.e.\ to fall to the right of the null distribution $f_d(\xi)$, an appropriate significance test is right-tailed. A null hypothesis $\text{H}_0$ tests whether, for a given test data point $\vvec{x}^{*}_i$, a corresponding value of the $d$-statistic, $d^*$ comes from a null distribution $f_d(\xi)$ that is generated by the null model.
We reject this hypothesis if the $p$-value, defined as
\begin{equation}
   p:= 1 - \int_0^{d^*}f_d(\xi)\text{d} \xi \equiv \int_{d^*}^{\infty}f_d(\xi)\text{d} \xi  \leq \alpha,
\end{equation}
where we set $\alpha=0.05$ in the discussed analysis.

The significance test cannot tell whether the estimated derivative is close to the ground truth derivative, as the test does not estimate the bias term in bias-variance decomposition. 
Nevertheless, as is evident in \figref{fig_var_range} as well as \figref{fig_significance} discussed in the next section, the $d$-statistic is correlated with the average loss on the test dataset. As the total error can be decomposed into the variance and bias terms, high variance necessarily indicates high error. The opposite, however, is not necessarily true, since an ensemble of models could be very certain about a very wrong prediction.

\begin{table*}[!ht]
\centering
  \begin{threeparttable}
  \caption{The percent relative error (all numbers are in $\%$ unit), \eqnref{eq_l1_relative} of neural network approximation and predictions shows robust genralization ability. The reported results are sample mean and standard deviation of $\eta$. The first results column reports approximation accuracy or the error on the training data $\mathcal{D}_{\text{train}}$, where the input graph $\mathcal{G}=\mathcal{G}_{\text{train}}$. The remaining columns report model prediction accuracy or the error on the test data $\mathcal{D}_{\text{train}}$. From the left, we sample test data from the same distribution as training data, namely $\mathcal{D}:\{\vvec{x}\sim \mathcal{U}(0,1),\Fbcal(\vvec{x})\}$; we then use a different distribution: $\mathcal{D}:\{\vvec{x}\sim \mathcal{B}(5,2),\Fbcal(\vvec{x})\}$. In both cases, we form predictions on the graph used during training. Lastly, we use test data from the same distribution as training data but consider novel input graphs: first, we sample from the training graph's network ensemble ($\mathcal{P}_{\text{train}}(\mathfrak{G})$: ER, $n=100,p=0.1$), and, lastly, we sample from a different network ensemble ($\mathcal{P}_{\text{test}}(\mathfrak{G})$: ER, $n=100,p=0.6$).
    \label{tab1}}
    \begin{tabular}{c c c c c c c c c }
      \toprule
                Dynamics    & $L$ & $Q$ 
                & \begin{tabular}{@{}c@{}}Train data \\ $\vvec{x}\sim\mathcal{U}(0,1)$ \\ 
                $\mathcal{G}\equiv \mathcal{G}_{\text{train}}$\end{tabular} 
                & \begin{tabular}{@{}c@{}} Test data \\
                $\vvec{x}\sim\mathcal{U}(0,1)$ \\ 
                $\mathcal{G}\equiv \mathcal{G}_{\text{train}}$\end{tabular} 
                & \begin{tabular}{@{}c@{}}Test data \\ $\vvec{x}\sim\mathcal{B}(5,2)$ \\ $\mathcal{G}\equiv \mathcal{G}_{\text{train}}$\end{tabular}
                & \begin{tabular}{@{}c@{}}Test data \\ $\vvec{x}\sim\mathcal{U}(0,1)$ \\ $\mathcal{G}\sim \mathcal{P}_{\text{train}}(\mathfrak{G})$\end{tabular}
                & \begin{tabular}{@{}c@{}}Test data \\ $\vvec{x}\sim\mathcal{U}(0,1)$ \\ $\mathcal{G}\sim \mathcal{P}_{\text{test}}(\mathfrak{G})$ 
                \end{tabular} 
                \\
      \midrule
Heat& -- & $B(x_j-x_i)$  & $ 0.8 \pm 0.08 $&$ 0.81 \pm 0.08 $&$ 1.06 \pm 0.1 $&$ 0.8 \pm 0.8 $ & $ 0.72 \pm 0.72 $ \\
MAK & $F-Bx_i^b$ & $Rx_j$ & $ 0.55 \pm 0.07 $&$ 0.55 \pm 0.07 $&$ 0.53 \pm 0.08 $&$ 0.56 \pm 0.56 $ & $ 0.48 \pm 0.48 $ \\
MM &  $-Bx_i$& $R\frac{x_j^h}{1+x_{j}^h}$ & $ 1.79 \pm 0.26 $&$ 1.81 \pm 0.29 $&$ 1.48 \pm 0.29 $&$ 1.8 \pm 1.8 $ & $ 2.66 \pm 2.66 $ \\
PD& $-Bx_i^b$ & $Rx_j^a$ & $ 2.6 \pm 0.3 $&$ 2.59 \pm 0.29 $&$ 2.23 \pm 0.32 $&$ 2.59 \pm 2.59 $ & $ 1.27 \pm 1.27 $ \\
SIS &$-Bx_i$ &$R(1-x_i)x_j$   & $ 0.29 \pm 0.03 $&$ 0.29 \pm 0.03 $&$ 0.31 \pm 0.02 $&$ 0.29 \pm 0.29 $ & $ 0.87 \pm 0.87 $  \\
    \bottomrule
    \end{tabular}
  \end{threeparttable}
\end{table*}

\begin{figure*}[!hbt]
    \centering
    \includegraphics[width =\linewidth]{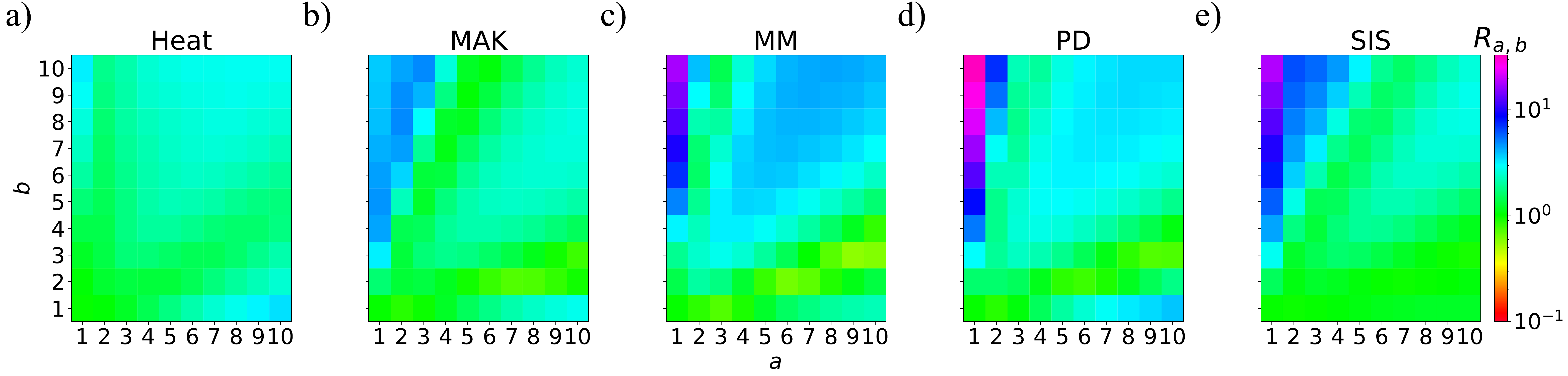}
    \caption{The ratio $R_{a,b}$ between prediction error and the approximation error shows robust generalisation for small changes in distribution parameters. The prediction error is a sample mean obtained when test data $\mathcal{D}:\{\vvec{x}\sim \mathcal{B}(a,b),\Fbcal(\vvec{x})\}$, the approximation error is a sample mean obtained using $\mathcal{D}_{\text{train}}$.}
    \label{fig_app_loss_beta}
\end{figure*}

\section{Approximation, prediction and long-range forecasting of dynamics}\label{sec:results}

In this section, we detail the primary findings of this paper, focusing on neural approximation, prediction and forecasting of complex network dynamics. We commence by examining scenarios in \secref{sec:prediction} under the premise that the function $\Fbcal$ is accessible for analytical estimation. In this section, we also employ training data drawn from a uniform distribution $\mathcal{U}(0,1)$, ensuring uniform coverage of the state space. This approach minimizes model bias towards any specific regions within the state space. In \secref{sec:non_iid_forecast}, we relax the assumption of i.i.d.\ training data to explore the models' ability to predict and to forecast dynamics given more realistic training conditions. Additionally, in this section we employ the statistical significance testing to quantify the models' confidence in the prediction. We end this section with a discussion on considerations regarding training with noisy and irregularly sampled time series data. The training details are listed in Sec.\ III of SI.

\subsection{Approximation and prediction of complex network dynamics}\label{sec:prediction}

Here we consider approximation and prediction of five models of complex network dynamics with $k=1$ variable per node: mass-action kinetics (MAK), population dynamics (PD), Michaelis–Menten (MM) equation, susceptible–infectious–susceptible (SIS) model discussed in~\cite{Barzel2013UniversalityDynamics}, as well as heat diffusion (Heat)~\cite{newman2018networks}. The true functional forms of these models are listed in \tabref{tab1}. The training graph is sampled from an Erd\"os-R\'enyi (ER) network ensemble~\cite{erdHos1960evolution} with $n=100$ nodes and edge probability $p=0.1$. To quantify the performance of a model, trained using a graph $\mathcal{G}_{\text{train}}$ and data samples $\mathcal{D}_{\text{train}}$, we consider the percent relative error computed using $\ell^1$ norm:
\begin{equation}\label{eq_l1_relative}
    \eta (\vvec{x}\in \mathcal{D},\mathcal{G}) = \frac{\norm{\pmb{\Psi}(\vvec{x},\mathcal{G}) - \Fbcal(\vvec{x},\mathcal{G})}_1}{\norm{\Fbcal(\vvec{x},\mathcal{G})}_1} \cdot 100,
\end{equation}
where $\vvec{x}$ is a sample from a dataset $\mathcal{D}$, and $\mathcal{G}$ is an input graph with adjacency matrix $\vvec{A}$. 

In \tabref{tab1}, we compare the sample mean $\langle \eta (\vvec{x},\mathcal{G} ) \rangle_{\vvec{x}\in \mathcal{D}}$ in approximation to various prediction scenarios. When $\vvec{x}\in \mathcal{D}_{\text{train}}, \mathcal{G}=\mathcal{G}_{\text{train}}$, $\langle \eta\rangle$ is an in-sample error, or the approximation accuracy. Across different models, we observe that the in-sample error is less than $3\%$ for all dynamics. Furthermore, when test data is sampled from the training data's probability distribution $\mathcal{U}(0,1)$, the prediction accuracy is equal to the approximation accuracy, indicating a lack of overfitting. However using test data sampled from a different --- beta distribution ($\vvec{x} \sim \mathcal{B}(5,2)$) --- reveals small deviations from the in-sample error: the prediction accuracy may be better or worse than the approximation accuracy, depending on the dynamics. To study this effect in greater detail, we considered the ratio $R_{a,b}$ between prediction and approximation accuracy where the former is computed using samples from a beta distribution, characterized by parameters $a,b$. As shown in \figref{fig_app_loss_beta}, $R_{a,b}$ varies minimally depending on the distribution of the test data, indicating that \emph{it is possible to form accurate predictions on test data that has different statistical properties from training data}. Note that for most dynamics, the biggest discrepancy between prediction and approximation accuracy is in the region $a=1,b>5$, where the beta distribution becomes increasingly right-skewed, meaning that a large portion of the distribution's mass is concentrated near $0$, with a long tail extending towards $1$. As the entries in $\vvec{x}$ tend towards zero, $\pmb{\Psi}(\vvec{x})$ converges to a constant value that remains unaffected by the inputs. Under these conditions, the difference between $\Fbcal(\vvec{x})$ and $\pmb{\Psi}(\vvec{x})$ is solely attributed to the bias terms $\textbf{b}$ within the linear layers. These bias terms may not necessarily provide an accurate approximation of the constant value to which $\pmb{\Psi}(\vvec{x})$ is converging.

\begin{figure}[!h]
    \centering
    \includegraphics[width = \linewidth]{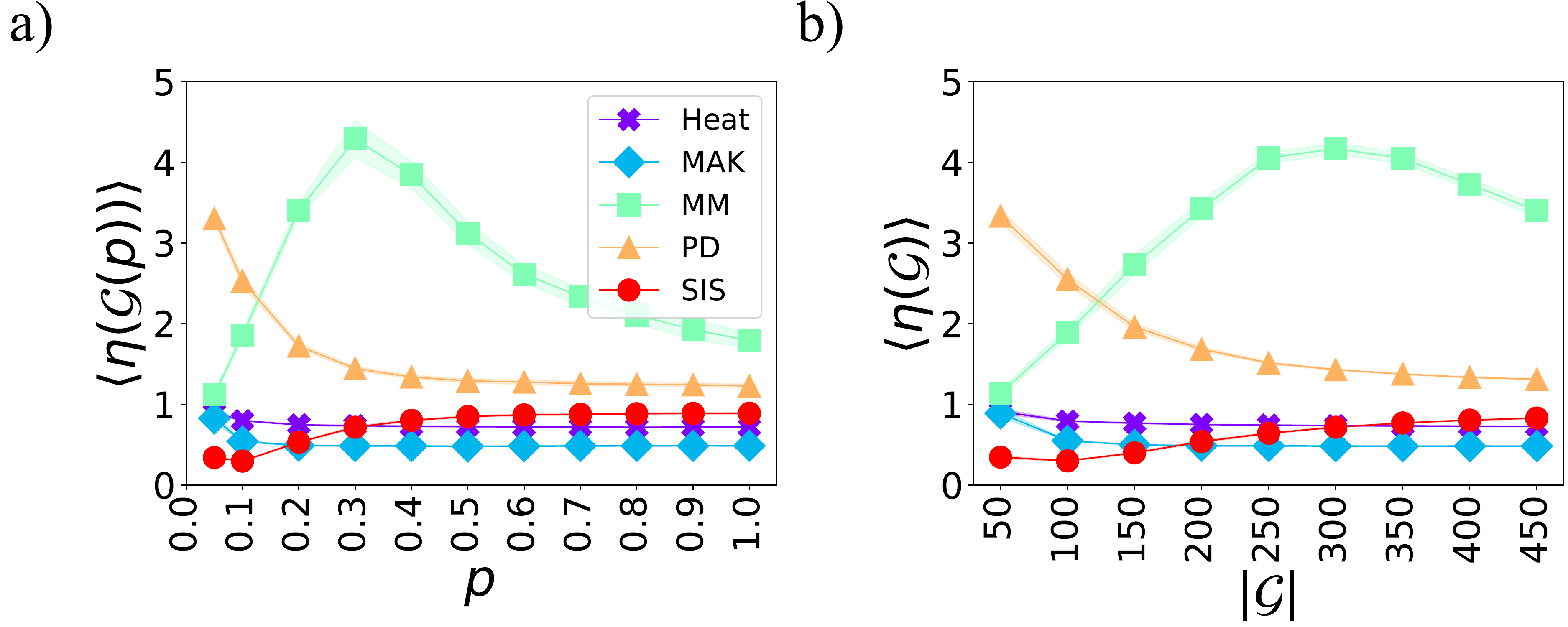}
    \caption{Prediction error when the train graph is replaced by a novel graph, as a function of $\textbf{a)}$ the density of the novel graph $\mathcal{G}$, $\textbf{p}$, and $\textbf{b)}$ the size of the novel graph, $|\mathcal{G}|$. The neural networks were tested using samples from uniform distribution and $10$ different ER graphs with parameters $p,n$. The lack of correlation between the error and the size or a density of a graph suggests that neural network models of dynamics can be repurposed to form predictions on novel graphs not observed during training. }
    \label{loss_vs_delta}
\end{figure}

Next, we study the prediction quality of models trained on one graph and applied on novel graphs. In the last two columns of \tabref{tab1}, we consider two new graphs, one sampled from the ensemble of the training graph $\mathcal{G}_{\text{train}}$ ($\mathcal{G}\sim \mathcal{P}_{\text{train}}(\mathfrak{G})$), another graph sampled from a denser Erd\"os-R\'enyi ensemble ($\mathcal{G}\sim \mathcal{P}_{\text{test}}(\mathfrak{G})$), with $n=100$ and $p=0.6$). The prediction accuracy matches the approximation accuracy for predictions on a graph from $\mathcal{P}_{\text{train}}$; however, when applied to graphs from an ensemble with statistical properties differing from those of the training graph, the accuracy of these predictions diverges from the approximation accuracy. The larger errors observed in MM and PD dynamics could be attributed to the inherent non-linearity of their equations. Overall, however, as we shown in \figref{loss_vs_delta}, a neural network that approximated dynamics in a system of size $n$ is able to make predictions on different graphs of different size and connectivity, while maintaining a small relative error. It implies that \emph{the learned representation encapsulates a form of conditional independence between the dynamics it models and the particular training graph, and a neural network can be successfully reused to model dynamics on another graph}.

\begin{figure*}[!hbt]
    \centering
    \includegraphics[width=1\linewidth]{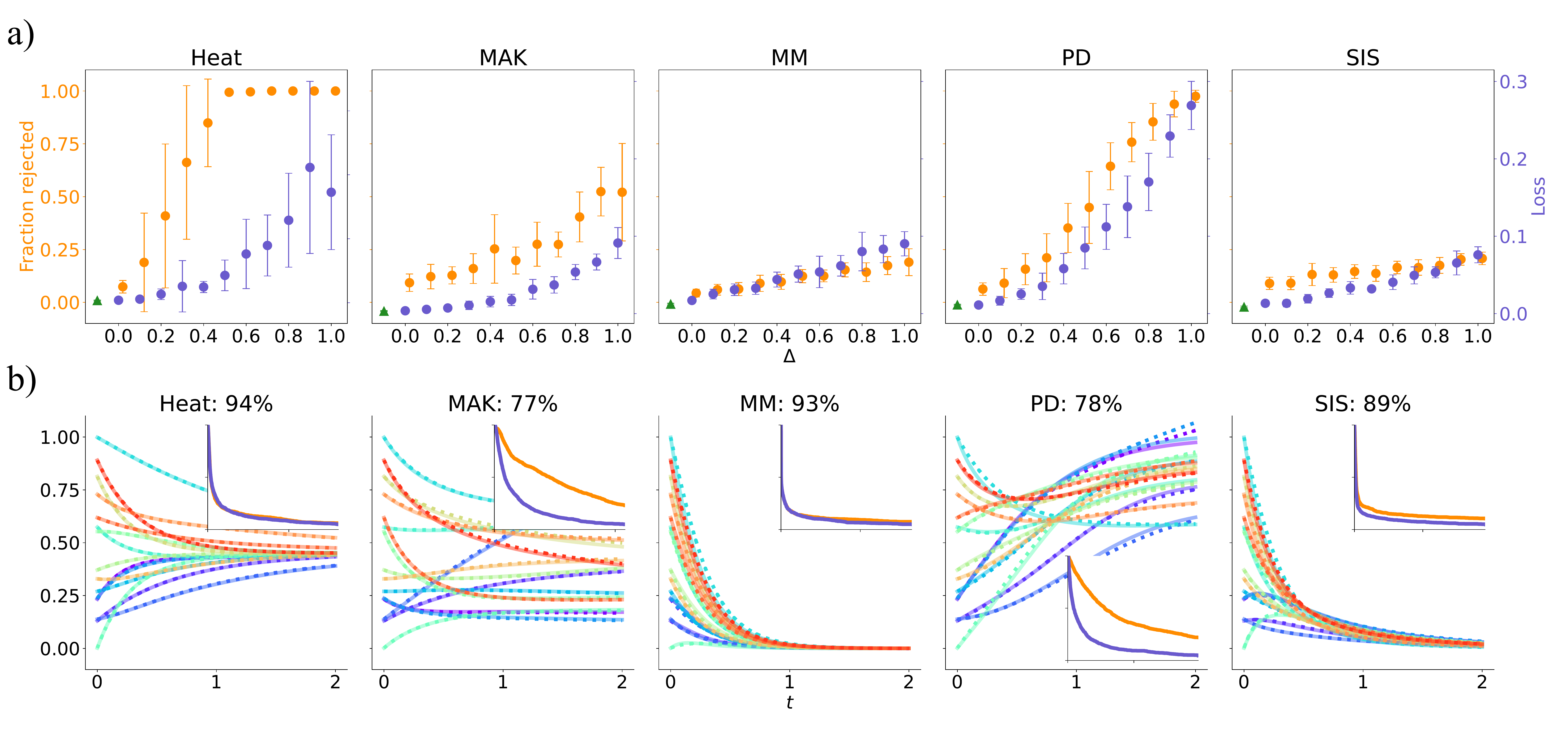}
    \caption{Forecasts of five complex network dynamics are accurate both on a graph used during training, as well as on novel input graphs. The $d$-statistic successfully identifies the limit of model's generalization. 
     \textbf{a)} compares the training loss (green triangle) with the test loss (purple circles), as a function of $\Delta$, where the initial value $\vvec{x}(t_0)\sim\mathcal{U}(0,1)+\Delta$. The orange circles indicate the fraction of rejected data points based on the $d$-statistic for a given $\Delta$. Here the test graph is isomorphic to the training graph.
     In $\textbf{b)}$, we changed the test graph to a larger one, $n=15,p=0.3$ and contrast the true dynamics (solid lines) and the neural network forecasts (dotted lines) for a new set of initial values that are sampled from the training distribution. The insets contrast the cumulative distribution of the $d$-statistic in training data (purple) and in test data (orange) and show the distributions in the range up to a critical value for the significance level of $5\%$. The title reports the percentage of accepted data points, i.e.\ the percentage of datapoints for which the corresponding $d$-statistic fell within the 95\% of the null (purple) distribution. }
    \label{fig_significance}
\end{figure*}

\subsection{Long-range forecasting of dynamics from time series data}\label{sec:non_iid_forecast}

Now we turn our attention to forecasting and training using more realistic data, derived from time series. In this section, we also study the $d$-statistic, and its potential to quantify the reliability of the forecasts.

\paragraph{Forecasting $k=1$ dynamics with a neural network trained using analytical derivatives} As a first test, we again consider $k=1$ dynamics defined in the previous section. To construct the $d$-statistic, a set of $50$ neural networks were trained on data $\mathcal{D}=\{\vvec{x},\Fbcal(\vvec{x})\}$ where $\vvec{x}$ was taken from five time series trajectories, and the dynamics were simulated on an ER random graph $\mathcal{G}_{\text{train}}$ with $n=10,p=0.5$. The trajectories were generated by solving the initial value problem, with initial conditions sampled from $\mathcal{U}(0,1)$. 

Given that $\mathcal{F}(\vvec{x})$ approaches zero as the dynamics approaches a steady state, we opt not to consider $\eta$ for evaluating the accuracy for individual data samples $\vvec{x}$. Instead, as detailed in \tabref{tab2}, we present the percent relative error for the full forecasted trajectory:
\begin{equation}\label{eq:eta_traj}\small
    \eta_{\text{traj}}(\vvec{x}_0,\mathcal{G}) = \frac{\sum_{t_r} \norm{I[\pmb{\Psi},\vvec{x}_0,\vvec{A},t_0,t]-I[\Fbcal,\vvec{x},\vvec{A},t_0,t] }_1}{\sum_{t_r} \norm{I[\Fbcal,\vvec{x},\vvec{A},t_0,t]}_1}\cdot 100
\end{equation}
where the integral $I$ is defined in \eqnref{eq:integral},  $\vvec{A}$ is an adjacency matrix of $\mathcal{G}$, the initial conditions $\vvec{x}_0$ are novel for the neural networks, $t_0=0$, $t_r\in \{t_0,t_1,...,t_R\}$, $\delta_r=t_{r+1}-t_r = 0.0101$ $\forall r$. The results suggest that forecasts on the training graph $\mathcal{G}_{\text{train}}$ consistently achieve accuracy $\eta_{\text{traj}}(\vvec{x}_0,\mathcal{G}_{\text{train}})$ significantly below $5\%$. Moreover, neural networks trained on a graph with $n=10$ nodes can effectively make long-term forecasts not only on the graphs they were initially trained on but also on new, unseen graphs, such as an Erdős-Rényi (ER) graph with $n=15$ nodes and $p=0.3$. \figref{fig_significance}b showcases the dynamics on $\mathcal{G}_{\text{test}}$ as predicted by one of the trained neural networks, accompanied by a measured $d$-statistic that evaluates the neural network's confidence in its predictions.

\begin{table}[ht]
\centering
\begin{tabular}{lll}
\hline
Dynamics & $\mathcal{G}\equiv \mathcal{G}_{\text{train}}$& $\mathcal{G}\equiv \mathcal{G}_{\text{test}}$ \\
\hline
Heat & $ 0.14 \pm 0.05 $ & $ 0.19 \pm 0.07 $ \\
MAK & $ 0.39 \pm 0.11 $ & $ 0.56 \pm 0.15 $ \\
MM & $ 4.02 \pm 0.53 $ & $ 5.07 \pm 0.76 $ \\
PD & $ 0.54 \pm 0.07 $ & $ 1.6 \pm 0.13 $ \\
SIS & $ 1.75 \pm 0.28 $ & $ 2.62 \pm 0.51 $ \\
\hline
\end{tabular}
\caption{Forecasting accuracy in terms of the percent relative error for the full forecasted trajectory, $\eta_{\text{traj}}(\vvec{x}_0,\mathcal{G})$, defined in \eqnref{eq:eta_traj}. All values are in \% units. The left column reports the accuracy of forecasting dynamics on $\mathcal{G}_{\text{train}}$, whereas the second column reports the accuracy in the forecast on a novel graph, here, ER with $n=15,p=0.3$. In both cases, the initial value $\vvec{x}_0\sim \mathcal{U}(0,1)$. }
\label{tab2}
\end{table}

While the forecasts in \figref{fig_significance}b) looks accurate across different dynamics, the $d$-statistic indicated low confidence in the MAK and PD forecasts despite the seemingly precise predictions, suggesting a ``false negative'', where the prediction is accurate, but the model ensemble is not confident in it. However, generally, we observe a consistent correlation between the $d$-statistic and out-of-sample loss. This is illustrated in \figref{fig_significance}a), where a monotonic increase in error is observed as the initial condition increasingly diverges from the domain of the initial condition that was used to generate training trajectories. This increase is accompanied by a rise in data points rejected by the statistical significance test, highlighting the $d$-statistic's potential as a marker for the model's operational domain. 

\begin{figure*}[!htb]
    \centering
    \includegraphics[width=1\linewidth]{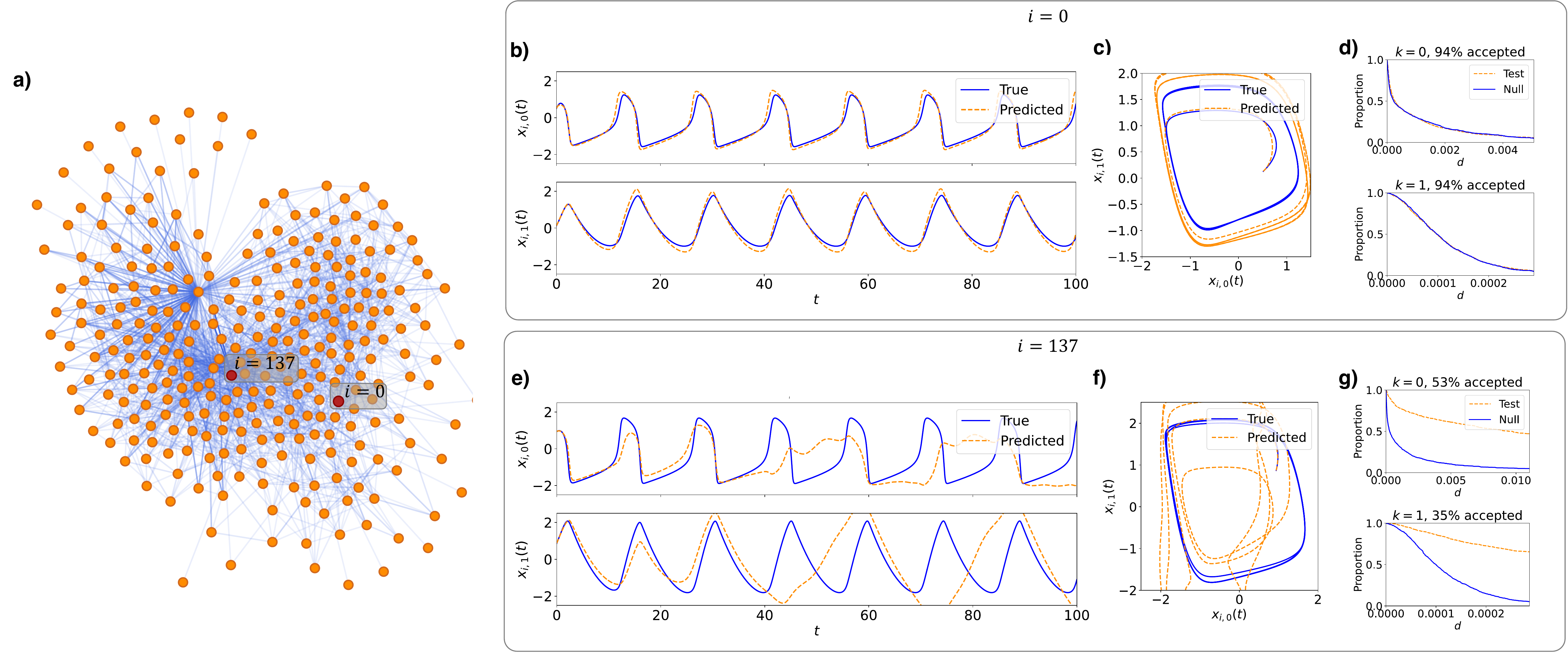}
    \caption{Forecasting FitzHugh-Nagumo Neuronal Dynamics ($k=2$) on a C.\ elegans connectome graph. The forecasts is produced solving an initial value problem for a novel initial condition sampled from a uniform distribution. Panel \textbf{a)} illustrates the connectome, with edge transparency reflecting the synaptic weight magnitude. Panels \textbf{b)} and \textbf{e)} compare true (blue dashed line) versus model-predicted (orange solid line) neuronal activity time series for two nodes highlighted in \textbf{a)}. Panels \textbf{c)} and \textbf{f)} show the phase space diagrams corresponding to these nodes. Panels \textbf{d)} and \textbf{g)} present the cumulative distribution of the $d$-statistic using training data (null distribution), as well as the predicted time series (test distribution). This comparison allows us to deduce the fraction of accepted datapoints in each time series, and conclude that the forecast for a node $i=0$ is sufficient, whereas the forecast for a node $i=137$ is unreliable. }
    \label{fig_celegans}
\end{figure*}

\paragraph{Forecasting $k=2$ dynamics with a neural network trained using numerical derivatives} So far, we have operated with an assumption that $\Fbcal$ is available. However, in practical situations, the training data would be derived from time series $\{\vvec{x}(t)\}$. Importantly, $\dot{\vvec{x}}=\Fbcal(\vvec{x})$ would not be an observable. To address this, if we collect $R$ signal samples $\vvec{x}(t_r)$ at times $t_r\in \{t_0,t_1,...,t_R\}$, we can approximate the labels for our training data numerically, e.g.\ $\vvec{y}=\vvec{\dot{x}}_r \approx \frac{\vvec{x}(t_{r+1}) - \vvec{x}(t_{r})}{\delta_r}$, where $\delta_r=t_{r+1}-t_r$.

 To test the ability of neural networks to forecast the dynamics in this case, we study Fitzhugh-Nagumo ($k=2$) neuronal dynamics (FNH)~\cite{rabinovich2006dynamical} on an empirical network of Caenorhabditis elegans (C.\ elegans) connectome~\cite{white1986structure}\footnote{Retrieved from \url{https://networks.skewed.de/net/celegansneural}}, with $n=279$ neurons. The dynamics for each node is defined as~\cite{Gao2022AutonomousData} 
\begin{eqnarray*}
    \frac{\text{d}x_{i,1}}{\text{d}t} &=& x_{i,1}-x_{i,1}^3 - x_{i,2}-\frac{\rho}{k_i^{\text{in}}} \sum_{j=1}^n A_{ij}(x_{j,1}-x_{i,1})\\ 
    \frac{\text{d}x_{i,2}}{\text{d}t}&=& a + b x_{i,1} + cx_{i,2},\nonumber
\end{eqnarray*}
where $\rho,a,b,c$ are constants detailed in the SI. 
This dynamics describes neural spikes, with nonlinear dynamics of membrane voltage $x_{i,1}$ of sodium channel reactivation, and  $x_{i,2}$ potassium channel deactivation after external stimulus. Note that this type of non-linear dynamics exhibits rich bifurcation properties~\cite{rocsoreanu2012fitzhugh}. 
We constructed training data from one time series trajectory and trained an ensemble of $M=20$ neural networks to evaluate the quality of predictions of a novel trajectory using the $d$-statistic. 
As \figref{fig_celegans} shows, overall, the graph neural network model is capable of approximating neuronal dynamics in C. elegans and forming long-term forecasts using novel initial values. However, the prediction quality may vary across nodes: the forecast for a node $0$ is accurate for the full forecasting window (\textbf{b,c}), whereas for a node $137$ it is reasonably accurate for the first cycle up to $t=10$, which is equivalent to $\approx 1000$ forecasted segments, after which the forecast diverges from the true trajectory (\textbf{e,f}). By considering the variance in model prediction, we observe that the prediction for a node $0$ is accompanied by a high acceptance rate of the $d$-statistic, whereas the prediction for a node $137$ is deemed largely insignificant.

The exact reasons of the correlation between out-of-sample loss and the variance in model prediction, while not fully explored within the scope of this paper, could be attributed to inherent properties of the dynamical systems under study, such as Lipschitz continuity, and the neural network's intrinsic spectral learning bias towards low-frequency components~\cite{rahaman2019spectral,ronen2019convergence,xu2019training}. Therefore, some part of the generalization out-of-range may be due to the dominance of low frequencies in the decomposition of functions $\Fbcal$ we considered.

\subsection{Learning \& predictions in presence of noise and irregular sampling}\label{sec:realdata}

\begin{figure}[!ht]
    \centering
    \includegraphics[width=1\linewidth]{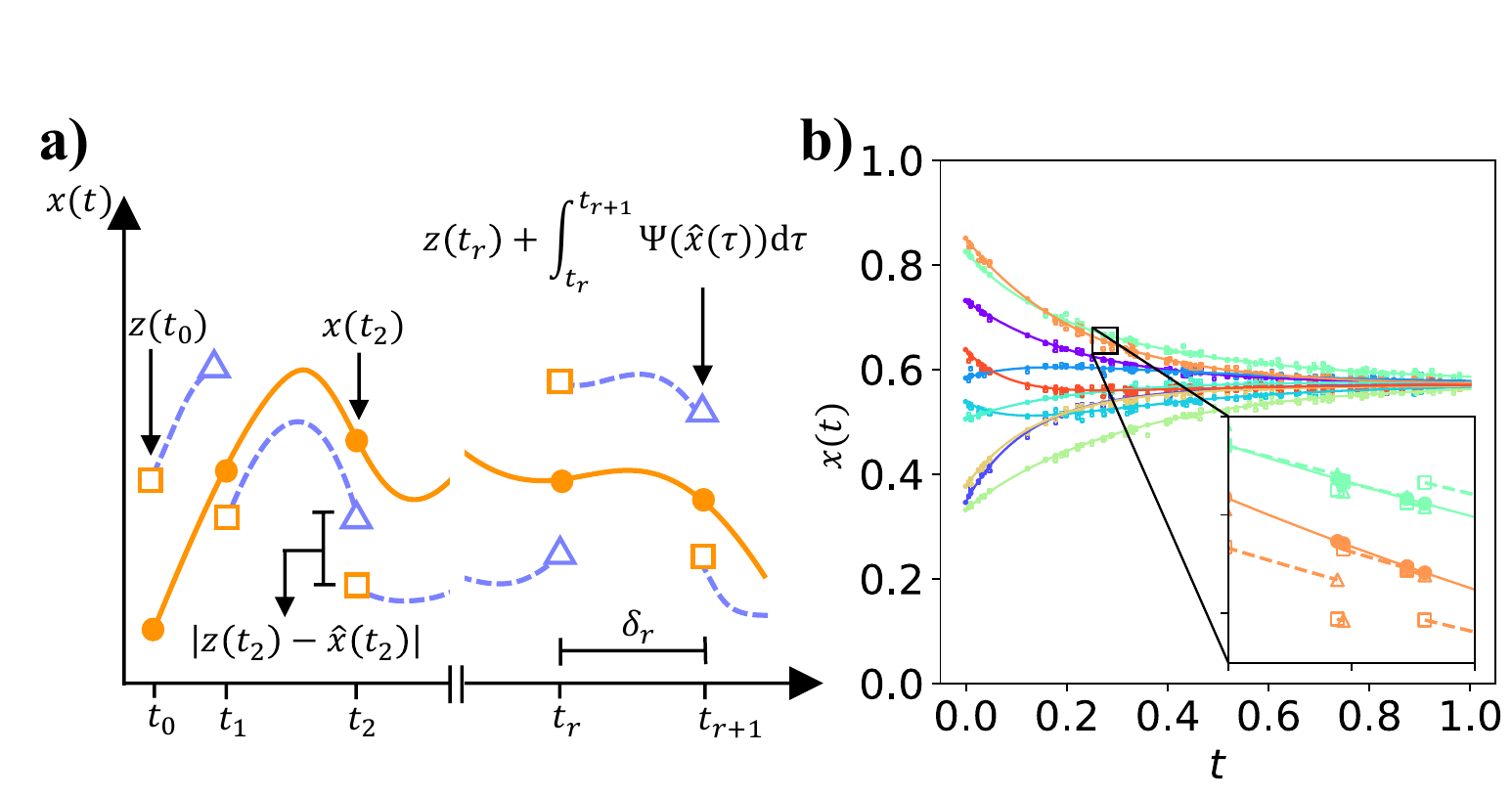}
    \caption{\textbf{a)} Illustration of training a neural network from irregularly sampled one dimensional signal $x(t_r)$, $t_r \in [0,R]$ (orange solid line). True labels are the observed signal ${z}(t_r)$, compared during training to predicted labels $\hat{{x}}(t_{r})={z}(t_{r-1})+ \int_{t_{r-1}}^{t_r}{\Psi}({\hat{x}}(\tau))\text{d}\tau$. Here the integral is evaluated numerically using infinitesimal time $\Delta t$. 
    \textbf{b)} shows neural network approximations of heat diffusion dynamics when a noisy, irregularly sampled signal used in training. In the inset, the square scatter points indicate the temporary initial values which are inputs to the neural network, whereas the triangles represent the predicted labels. }
    \label{fig_real}
\end{figure}

The dataset derived from time series could furthermore be sampled at potentially irregular intervals and be subject to observational noise: assuming additive noise, the observed signal is $\vvec{z}[\vvec{x}(t)]=\vvec{x}(t) + \pmb{\varepsilon}(t)$. If $\delta_r$ is time-varying, numerically approximating the derivative $\dot{\vvec{x}}$ may introduce large numerical errors, especially if the signal is noisy. As an alternative, one could learn $\pmb{\Psi}$ at a much higher frequency $\Delta t$ to allow estimates $\vvec{x}(t_r)$, i.e.\ $\hat{\vvec{x}}(t_r)$ from $\vvec{x}(t_{r-1})$ using numerical integration. We set $\delta_r\gg \Delta t$ so that $\frac{\Delta t}{\delta_r}\approx 0$ $\forall r$. This procedure is illustrated in \figref{fig_real}a) 
and uses the following loss function:
\begin{equation}\small
    \mathcal{L}=\sum_{r=0}^{R} \norm{ \vvec{z}[\vvec{x}(t_r)] - \vvec{z}[\vvec{x}(t_{r-1})]-\int_{t_{r-1}}^{t_{r}} \pmb{\Psi}(\vvec{\hat{x}}(\tau)) d\tau  }_1 .
    \label{eq:ode_loss}
\end{equation}
Here the gradient descent is computed through a forward computational graph, employing an ODE solver~\footnote{For efficiency, an adjoint sensitivity method could also be used~\cite{chen2018neural}}. 
In Sec.\ V of SI, we derive the gradient update rule for loss in
\eqnref{eq:ode_loss}. 

\figref{fig_real}b) shows an example of training the neural network to approximate heat diffusion from irregularly sampled, noisy time series data. Approximating the rate of change in the system state is possible even in a realistic setting where the state variables are observed with noise and at irregular sampling intervals. 

To test the effect of noise, we trained another neural network using the same data with observational noise removed. We then considered a test error $\textbf{E}_{\mathcal{D}:{\vvec{x}\sim \phi_{x_0}(\vvec{x})}}\left[\norm{\hat{\vvec{x}}(\delta)-\vvec{x}(\delta)}_1\right]$, where $\hat{\vvec{x}}(\delta)$ is a prediction, obtained by numerically integrating $\pmb{\Psi}$ starting with $\vvec{x}$ over the interval $t=[0,\delta]$, with a step size $\Delta t$, whereas $\vvec{x}(\delta)$ is obtained by integrating $\Fbcal$. The errors of the two models are not significantly different (respectively, $0.09\pm 0.06$ for the model with noise and $0.07\pm 0.04$ for the model without the noise). In contrast, a random untrained neural network yields a significantly larger loss of $0.24  \pm  0.03$. 

Overall, the neural networks can approximate various dynamical models and extrapolate predictions even when statistical properties of the input data, or the graph structure change, transcending the formal boundaries of SLT and UAT. We also showed that training and generalization are also possible using time series data. As expected, there exists a limit for how much generalization can be achieved, however, we note the presence of regularity that is observed in the gradual increase in test loss. Using the proposed $d$-statistic we can therefore evaluate the confidence in the inferred prediction, even in cases when we depart from standard SLT, UAT assumptions of i.i.d.\ sampling and a compact support.

\section*{Conclusion and discussion}
We presented a comprehensive framework to approximate, predict and forecast dynamics, defined as a system of ordinary differential equations coupled via a complex network (graph), a ubiquitous model of dynamics in complex systems. We introduced a toolbox to analyze the quality of neural approximations with the standard ruler of statistical learning theory, as well as in more diverse settings: (i) when statistical properties for train and test data differ, (ii) train samples are non-i.i.d., (iii) the ground truth functions have a non-compact support (iv) training and test graphs differ, and (vi) when trained models are used to form long-range forecasts. We showed that if a model adheres to the basic assumptions about the vector field that describes dynamics, the trained model is not only expressive within the conditions that it was trained on, but also has potential to generalize beyond boundaries set out by UAT and by SLT. The confidence in the inferred predictions within these diverse test settings can also be reliably estimated with a dedicated null model and a statistical test.

The set of tools we have presented could be extended to understand both the limits and benefits of deep learning models for complex dynamical systems. While we observed some generalization capacity, we note that if the functions that constitute $\Fbcal$, namely, $\textbf{L}$ and $\textbf{Q}$ are not learnt accurately, the model's generalization is impeded. More exotic training settings hold potential to help resolve these issues. We also suggest studying neural trainability through the property of dynamic isometry and the mean-field theory~\cite{chen2018dynamical,pennington2017resurrecting}, enforcing physical constraints e.g.\ Lipschitz continuity of dynamics~\cite{Gouk2021RegularisationContinuity}, and expanding the range of tools from SLT~\cite{belkin2019reconciling, zhang2021understanding,jakubovitz2019generalization}.  
We also limited our analysis to the simplest case where the dynamics is deterministic and autonomous, and the graphs are static, undirected, connected and fully known. It will be important to study the effects of increased complexity of the structure on the approximations. 

More broadly, we emphasize that applications of deep learning models of dynamical system go beyond mere predictions within statistical confines of training data. Therefore it is imperative that these models be constructed to maintain accuracy with conditions, that were not encountered during training, e.g.\ when test data has a different support to training data. Expanding our proposed framework holds a potential of enabling forecasting, modeling and, ultimately, understanding a wide spectrum of complex dynamical systems.
\vspace{1cm}
\section*{Acknowledgements}{This work is supported by the European Union – Horizon 2020 Program under the scheme “INFRAIA-01-2018-2019 – Integrating Activities for Advanced Communities”, Grant Agreement n.871042, “SoBigData++: European Integrated Infrastructure for Social Mining and Big Data Analytics” (http://www.sobigdata.eu).}

\FloatBarrier

\begin{thebibliography}{90}%
\makeatletter
\providecommand \@ifxundefined [1]{%
 \@ifx{#1\undefined}
}%
\providecommand \@ifnum [1]{%
 \ifnum #1\expandafter \@firstoftwo
 \else \expandafter \@secondoftwo
 \fi
}%
\providecommand \@ifx [1]{%
 \ifx #1\expandafter \@firstoftwo
 \else \expandafter \@secondoftwo
 \fi
}%
\providecommand \natexlab [1]{#1}%
\providecommand \enquote  [1]{``#1''}%
\providecommand \bibnamefont  [1]{#1}%
\providecommand \bibfnamefont [1]{#1}%
\providecommand \citenamefont [1]{#1}%
\providecommand \href@noop [0]{\@secondoftwo}%
\providecommand \href [0]{\begingroup \@sanitize@url \@href}%
\providecommand \@href[1]{\@@startlink{#1}\@@href}%
\providecommand \@@href[1]{\endgroup#1\@@endlink}%
\providecommand \@sanitize@url [0]{\catcode `\\12\catcode `\$12\catcode
  `\&12\catcode `\#12\catcode `\^12\catcode `\_12\catcode `\%12\relax}%
\providecommand \@@startlink[1]{}%
\providecommand \@@endlink[0]{}%
\providecommand \url  [0]{\begingroup\@sanitize@url \@url }%
\providecommand \@url [1]{\endgroup\@href {#1}{\urlprefix }}%
\providecommand \urlprefix  [0]{URL }%
\providecommand \Eprint [0]{\href }%
\providecommand \doibase [0]{https://doi.org/}%
\providecommand \selectlanguage [0]{\@gobble}%
\providecommand \bibinfo  [0]{\@secondoftwo}%
\providecommand \bibfield  [0]{\@secondoftwo}%
\providecommand \translation [1]{[#1]}%
\providecommand \BibitemOpen [0]{}%
\providecommand \bibitemStop [0]{}%
\providecommand \bibitemNoStop [0]{.\EOS\space}%
\providecommand \EOS [0]{\spacefactor3000\relax}%
\providecommand \BibitemShut  [1]{\csname bibitem#1\endcsname}%
\let\auto@bib@innerbib\@empty
\bibitem [{\citenamefont {Brunton}\ \emph {et~al.}(2016)\citenamefont
  {Brunton}, \citenamefont {Proctor}, \citenamefont {Kutz},\ and\ \citenamefont
  {Bialek}}]{Brunton2016DiscoveringSystems}%
  \BibitemOpen
  \bibfield  {author} {\bibinfo {author} {\bibfnamefont {S.~L.}\ \bibnamefont
  {Brunton}}, \bibinfo {author} {\bibfnamefont {J.~L.}\ \bibnamefont
  {Proctor}}, \bibinfo {author} {\bibfnamefont {J.~N.}\ \bibnamefont {Kutz}},\
  and\ \bibinfo {author} {\bibfnamefont {W.}~\bibnamefont {Bialek}},\
  }\bibfield  {title} {\bibinfo {title} {{Discovering governing equations from
  data by sparse identification of nonlinear dynamical systems}},\ }\href
  {https://doi.org/10.1073/PNAS.1517384113/-/DCSUPPLEMENTAL/PNAS.1517384113.SAPP.PDF}
  {\bibfield  {journal} {\bibinfo  {journal} {Proceedings of the National
  Academy of Sciences of the United States of America}\ }\textbf {\bibinfo
  {volume} {113}},\ \bibinfo {pages} {3932} (\bibinfo {year}
  {2016})}\BibitemShut {NoStop}%
\bibitem [{\citenamefont {Gao}\ and\ \citenamefont {Yan}()}]{Gao2022}%
  \BibitemOpen
  \bibfield  {author} {\bibinfo {author} {\bibfnamefont {T.-T.}\ \bibnamefont
  {Gao}}\ and\ \bibinfo {author} {\bibfnamefont {G.}~\bibnamefont {Yan}},\
  }\bibfield  {title} {\bibinfo {title} {{Autonomous inference of complex
  network dynamics from incomplete and noisy data}}\ }\href
  {https://doi.org/10.1038/s43588-022-00217-0}
  {10.1038/s43588-022-00217-0}\BibitemShut {NoStop}%
\bibitem [{\citenamefont {Udrescu}\ and\ \citenamefont
  {Tegmark}(2020)}]{aiFeynman}%
  \BibitemOpen
  \bibfield  {author} {\bibinfo {author} {\bibfnamefont {S.-M.}\ \bibnamefont
  {Udrescu}}\ and\ \bibinfo {author} {\bibfnamefont {M.}~\bibnamefont
  {Tegmark}},\ }\bibfield  {title} {\bibinfo {title} {Ai feynman: A
  physics-inspired method for symbolic regression},\ }\href@noop {} {\bibfield
  {journal} {\bibinfo  {journal} {Science Advances}\ }\textbf {\bibinfo
  {volume} {6}},\ \bibinfo {pages} {eaay2631} (\bibinfo {year}
  {2020})}\BibitemShut {NoStop}%
\bibitem [{\citenamefont {Schmidt}\ and\ \citenamefont
  {Lipson}(2009)}]{schmidt2009distilling}%
  \BibitemOpen
  \bibfield  {author} {\bibinfo {author} {\bibfnamefont {M.}~\bibnamefont
  {Schmidt}}\ and\ \bibinfo {author} {\bibfnamefont {H.}~\bibnamefont
  {Lipson}},\ }\bibfield  {title} {\bibinfo {title} {Distilling free-form
  natural laws from experimental data},\ }\href@noop {} {\bibfield  {journal}
  {\bibinfo  {journal} {science}\ }\textbf {\bibinfo {volume} {324}},\ \bibinfo
  {pages} {81} (\bibinfo {year} {2009})}\BibitemShut {NoStop}%
\bibitem [{\citenamefont {Gilpin}\ \emph {et~al.}(2020)\citenamefont {Gilpin},
  \citenamefont {Huang},\ and\ \citenamefont {Forger}}]{gilpin2020learning}%
  \BibitemOpen
  \bibfield  {author} {\bibinfo {author} {\bibfnamefont {W.}~\bibnamefont
  {Gilpin}}, \bibinfo {author} {\bibfnamefont {Y.}~\bibnamefont {Huang}},\ and\
  \bibinfo {author} {\bibfnamefont {D.~B.}\ \bibnamefont {Forger}},\ }\bibfield
   {title} {\bibinfo {title} {Learning dynamics from large biological data
  sets: machine learning meets systems biology},\ }\href@noop {} {\bibfield
  {journal} {\bibinfo  {journal} {Current Opinion in Systems Biology}\ }\textbf
  {\bibinfo {volume} {22}},\ \bibinfo {pages} {1} (\bibinfo {year}
  {2020})}\BibitemShut {NoStop}%
\bibitem [{\citenamefont {Hillar}\ and\ \citenamefont
  {Sommer}(2012)}]{hillar2012comment}%
  \BibitemOpen
  \bibfield  {author} {\bibinfo {author} {\bibfnamefont {C.}~\bibnamefont
  {Hillar}}\ and\ \bibinfo {author} {\bibfnamefont {F.}~\bibnamefont
  {Sommer}},\ }\bibfield  {title} {\bibinfo {title} {Comment on the article"
  distilling free-form natural laws from experimental data"},\ }\href@noop {}
  {\bibfield  {journal} {\bibinfo  {journal} {arXiv preprint arXiv:1210.7273}\
  } (\bibinfo {year} {2012})}\BibitemShut {NoStop}%
\bibitem [{\citenamefont {Cubitt}\ \emph {et~al.}(2012)\citenamefont {Cubitt},
  \citenamefont {Eisert},\ and\ \citenamefont {Wolf}}]{learningPhyDataNPhard}%
  \BibitemOpen
  \bibfield  {author} {\bibinfo {author} {\bibfnamefont {T.~S.}\ \bibnamefont
  {Cubitt}}, \bibinfo {author} {\bibfnamefont {J.}~\bibnamefont {Eisert}},\
  and\ \bibinfo {author} {\bibfnamefont {M.~M.}\ \bibnamefont {Wolf}},\
  }\bibfield  {title} {\bibinfo {title} {Extracting dynamical equations from
  experimental data is np hard},\ }\href
  {https://doi.org/10.1103/PhysRevLett.108.120503} {\bibfield  {journal}
  {\bibinfo  {journal} {Phys. Rev. Lett.}\ }\textbf {\bibinfo {volume} {108}},\
  \bibinfo {pages} {120503} (\bibinfo {year} {2012})}\BibitemShut {NoStop}%
\bibitem [{\citenamefont {Virgolin}\ and\ \citenamefont
  {Pissis}(2022)}]{virgolin2022symbolic}%
  \BibitemOpen
  \bibfield  {author} {\bibinfo {author} {\bibfnamefont {M.}~\bibnamefont
  {Virgolin}}\ and\ \bibinfo {author} {\bibfnamefont {S.~P.}\ \bibnamefont
  {Pissis}},\ }\bibfield  {title} {\bibinfo {title} {Symbolic regression is
  np-hard},\ }\href@noop {} {\bibfield  {journal} {\bibinfo  {journal} {arXiv
  preprint arXiv:2207.01018}\ } (\bibinfo {year} {2022})}\BibitemShut {NoStop}%
\bibitem [{\citenamefont {Barzel}\ and\ \citenamefont
  {Barab{\'{a}}si}(2013)}]{Barzel2013UniversalityDynamics}%
  \BibitemOpen
  \bibfield  {author} {\bibinfo {author} {\bibfnamefont {B.}~\bibnamefont
  {Barzel}}\ and\ \bibinfo {author} {\bibfnamefont {A.~L.}\ \bibnamefont
  {Barab{\'{a}}si}},\ }\bibfield  {title} {\bibinfo {title} {{Universality in
  network dynamics}},\ }\href {https://doi.org/10.1038/nphys2741} {\bibfield
  {journal} {\bibinfo  {journal} {Nature Physics 2013 9:10}\ }\textbf {\bibinfo
  {volume} {9}},\ \bibinfo {pages} {673} (\bibinfo {year} {2013})}\BibitemShut
  {NoStop}%
\bibitem [{\citenamefont {Voit}(2000)}]{voit2000computational}%
  \BibitemOpen
  \bibfield  {author} {\bibinfo {author} {\bibfnamefont {E.~O.}\ \bibnamefont
  {Voit}},\ }\href@noop {} {\emph {\bibinfo {title} {Computational analysis of
  biochemical systems: a practical guide for biochemists and molecular
  biologists}}}\ (\bibinfo  {publisher} {Cambridge University Press},\ \bibinfo
  {year} {2000})\BibitemShut {NoStop}%
\bibitem [{\citenamefont {Zang}\ and\ \citenamefont
  {Wang}(2020)}]{Zang2020NeuralNetworks}%
  \BibitemOpen
  \bibfield  {author} {\bibinfo {author} {\bibfnamefont {C.}~\bibnamefont
  {Zang}}\ and\ \bibinfo {author} {\bibfnamefont {F.}~\bibnamefont {Wang}},\
  }\bibfield  {title} {\bibinfo {title} {{Neural Dynamics on Complex
  Networks}},\ }in\ \href {https://doi.org/10.1145/3394486.3403132} {\emph
  {\bibinfo {booktitle} {Proceedings of the ACM SIGKDD International Conference
  on Knowledge Discovery and Data Mining}}}\ (\bibinfo  {publisher}
  {Association for Computing Machinery},\ \bibinfo {year} {2020})\ pp.\
  \bibinfo {pages} {892--902}\BibitemShut {NoStop}%
\bibitem [{\citenamefont {Murphy}\ \emph {et~al.}(2021)\citenamefont {Murphy},
  \citenamefont {Laurence},\ and\ \citenamefont
  {Allard}}]{Murphy2021DeepNetworks}%
  \BibitemOpen
  \bibfield  {author} {\bibinfo {author} {\bibfnamefont {C.}~\bibnamefont
  {Murphy}}, \bibinfo {author} {\bibfnamefont {E.}~\bibnamefont {Laurence}},\
  and\ \bibinfo {author} {\bibfnamefont {A.}~\bibnamefont {Allard}},\
  }\bibfield  {title} {\bibinfo {title} {{Deep learning of contagion dynamics
  on complex networks}},\ }\bibfield  {journal} {\bibinfo  {journal} {Nature
  Communications}\ }\textbf {\bibinfo {volume} {12}},\ \href
  {https://doi.org/10.1038/s41467-021-24732-2} {10.1038/s41467-021-24732-2}
  (\bibinfo {year} {2021})\BibitemShut {NoStop}%
\bibitem [{\citenamefont {Hornik}\ \emph {et~al.}(1989)\citenamefont {Hornik},
  \citenamefont {Stinchcombe},\ and\ \citenamefont
  {White}}]{hornik1989multilayer}%
  \BibitemOpen
  \bibfield  {author} {\bibinfo {author} {\bibfnamefont {K.}~\bibnamefont
  {Hornik}}, \bibinfo {author} {\bibfnamefont {M.}~\bibnamefont
  {Stinchcombe}},\ and\ \bibinfo {author} {\bibfnamefont {H.}~\bibnamefont
  {White}},\ }\bibfield  {title} {\bibinfo {title} {Multilayer feedforward
  networks are universal approximators},\ }\href@noop {} {\bibfield  {journal}
  {\bibinfo  {journal} {Neural networks}\ }\textbf {\bibinfo {volume} {2}},\
  \bibinfo {pages} {359} (\bibinfo {year} {1989})}\BibitemShut {NoStop}%
\bibitem [{\citenamefont {Cybenko}(1992)}]{cybenko1992approximation}%
  \BibitemOpen
  \bibfield  {author} {\bibinfo {author} {\bibfnamefont {G.}~\bibnamefont
  {Cybenko}},\ }\bibfield  {title} {\bibinfo {title} {Approximation by
  superpositions of a sigmoidal function},\ }\href@noop {} {\bibfield
  {journal} {\bibinfo  {journal} {Mathematics of Control, Signals and Systems}\
  }\textbf {\bibinfo {volume} {5}},\ \bibinfo {pages} {455} (\bibinfo {year}
  {1992})}\BibitemShut {NoStop}%
\bibitem [{\citenamefont {Yarotsky}(2017)}]{yarotsky2017error}%
  \BibitemOpen
  \bibfield  {author} {\bibinfo {author} {\bibfnamefont {D.}~\bibnamefont
  {Yarotsky}},\ }\bibfield  {title} {\bibinfo {title} {Error bounds for
  approximations with deep relu networks},\ }\href@noop {} {\bibfield
  {journal} {\bibinfo  {journal} {Neural Networks}\ }\textbf {\bibinfo {volume}
  {94}},\ \bibinfo {pages} {103} (\bibinfo {year} {2017})}\BibitemShut
  {NoStop}%
\bibitem [{\citenamefont {Kidger}\ and\ \citenamefont
  {Lyons}(2020)}]{kidger2020universal}%
  \BibitemOpen
  \bibfield  {author} {\bibinfo {author} {\bibfnamefont {P.}~\bibnamefont
  {Kidger}}\ and\ \bibinfo {author} {\bibfnamefont {T.}~\bibnamefont {Lyons}},\
  }\bibfield  {title} {\bibinfo {title} {Universal approximation with deep
  narrow networks},\ }in\ \href@noop {} {\emph {\bibinfo {booktitle}
  {Conference on learning theory}}}\ (\bibinfo {organization} {PMLR},\ \bibinfo
  {year} {2020})\ pp.\ \bibinfo {pages} {2306--2327}\BibitemShut {NoStop}%
\bibitem [{\citenamefont {Maiorov}\ and\ \citenamefont
  {Pinkus}(1999)}]{maiorov1999lower}%
  \BibitemOpen
  \bibfield  {author} {\bibinfo {author} {\bibfnamefont {V.}~\bibnamefont
  {Maiorov}}\ and\ \bibinfo {author} {\bibfnamefont {A.}~\bibnamefont
  {Pinkus}},\ }\bibfield  {title} {\bibinfo {title} {Lower bounds for
  approximation by mlp neural networks},\ }\href@noop {} {\bibfield  {journal}
  {\bibinfo  {journal} {Neurocomputing}\ }\textbf {\bibinfo {volume} {25}},\
  \bibinfo {pages} {81} (\bibinfo {year} {1999})}\BibitemShut {NoStop}%
\bibitem [{\citenamefont {Wagstaff}\ \emph {et~al.}(2022)\citenamefont
  {Wagstaff}, \citenamefont {Fuchs}, \citenamefont {Engelcke}, \citenamefont
  {Osborne},\ and\ \citenamefont {Posner}}]{wagstaff2022universal}%
  \BibitemOpen
  \bibfield  {author} {\bibinfo {author} {\bibfnamefont {E.}~\bibnamefont
  {Wagstaff}}, \bibinfo {author} {\bibfnamefont {F.~B.}\ \bibnamefont {Fuchs}},
  \bibinfo {author} {\bibfnamefont {M.}~\bibnamefont {Engelcke}}, \bibinfo
  {author} {\bibfnamefont {M.~A.}\ \bibnamefont {Osborne}},\ and\ \bibinfo
  {author} {\bibfnamefont {I.}~\bibnamefont {Posner}},\ }\bibfield  {title}
  {\bibinfo {title} {Universal approximation of functions on sets},\
  }\href@noop {} {\bibfield  {journal} {\bibinfo  {journal} {The Journal of
  Machine Learning Research}\ }\textbf {\bibinfo {volume} {23}},\ \bibinfo
  {pages} {6762} (\bibinfo {year} {2022})}\BibitemShut {NoStop}%
\bibitem [{\citenamefont {Zaheer}\ \emph {et~al.}(2017)\citenamefont {Zaheer},
  \citenamefont {Kottur}, \citenamefont {Ravanbakhsh}, \citenamefont {Poczos},
  \citenamefont {Salakhutdinov},\ and\ \citenamefont {Smola}}]{zaheer2017deep}%
  \BibitemOpen
  \bibfield  {author} {\bibinfo {author} {\bibfnamefont {M.}~\bibnamefont
  {Zaheer}}, \bibinfo {author} {\bibfnamefont {S.}~\bibnamefont {Kottur}},
  \bibinfo {author} {\bibfnamefont {S.}~\bibnamefont {Ravanbakhsh}}, \bibinfo
  {author} {\bibfnamefont {B.}~\bibnamefont {Poczos}}, \bibinfo {author}
  {\bibfnamefont {R.~R.}\ \bibnamefont {Salakhutdinov}},\ and\ \bibinfo
  {author} {\bibfnamefont {A.~J.}\ \bibnamefont {Smola}},\ }\bibfield  {title}
  {\bibinfo {title} {Deep sets},\ }\href@noop {} {\bibfield  {journal}
  {\bibinfo  {journal} {Advances in neural information processing systems}\
  }\textbf {\bibinfo {volume} {30}} (\bibinfo {year} {2017})}\BibitemShut
  {NoStop}%
\bibitem [{\citenamefont {Xu}\ \emph {et~al.}(2018{\natexlab{a}})\citenamefont
  {Xu}, \citenamefont {Hu}, \citenamefont {Leskovec},\ and\ \citenamefont
  {Jegelka}}]{xu2018powerful}%
  \BibitemOpen
  \bibfield  {author} {\bibinfo {author} {\bibfnamefont {K.}~\bibnamefont
  {Xu}}, \bibinfo {author} {\bibfnamefont {W.}~\bibnamefont {Hu}}, \bibinfo
  {author} {\bibfnamefont {J.}~\bibnamefont {Leskovec}},\ and\ \bibinfo
  {author} {\bibfnamefont {S.}~\bibnamefont {Jegelka}},\ }\bibfield  {title}
  {\bibinfo {title} {How powerful are graph neural networks?},\ }\href@noop {}
  {\bibfield  {journal} {\bibinfo  {journal} {arXiv preprint arXiv:1810.00826}\
  } (\bibinfo {year} {2018}{\natexlab{a}})}\BibitemShut {NoStop}%
\bibitem [{\citenamefont {Wang}\ \emph {et~al.}(2023)\citenamefont {Wang},
  \citenamefont {Fu}, \citenamefont {Zhou},\ and\ \citenamefont
  {Yan}}]{wang2023implicit}%
  \BibitemOpen
  \bibfield  {author} {\bibinfo {author} {\bibfnamefont {L.}~\bibnamefont
  {Wang}}, \bibinfo {author} {\bibfnamefont {Z.}~\bibnamefont {Fu}}, \bibinfo
  {author} {\bibfnamefont {Y.}~\bibnamefont {Zhou}},\ and\ \bibinfo {author}
  {\bibfnamefont {Z.}~\bibnamefont {Yan}},\ }\bibfield  {title} {\bibinfo
  {title} {The implicit regularization of momentum gradient descent in
  overparametrized models},\ }in\ \href@noop {} {\emph {\bibinfo {booktitle}
  {Proceedings of the AAAI Conference on Artificial Intelligence}}},\
  Vol.~\bibinfo {volume} {37}\ (\bibinfo {year} {2023})\ pp.\ \bibinfo {pages}
  {10149--10156}\BibitemShut {NoStop}%
\bibitem [{\citenamefont {Zhao}(2022)}]{zhao2022combining}%
  \BibitemOpen
  \bibfield  {author} {\bibinfo {author} {\bibfnamefont {D.}~\bibnamefont
  {Zhao}},\ }\bibfield  {title} {\bibinfo {title} {Combining explicit and
  implicit regularization for efficient learning in deep networks},\
  }\href@noop {} {\bibfield  {journal} {\bibinfo  {journal} {Advances in Neural
  Information Processing Systems}\ }\textbf {\bibinfo {volume} {35}},\ \bibinfo
  {pages} {3024} (\bibinfo {year} {2022})}\BibitemShut {NoStop}%
\bibitem [{\citenamefont {Arora}\ \emph {et~al.}(2022)\citenamefont {Arora},
  \citenamefont {Li},\ and\ \citenamefont
  {Panigrahi}}]{arora2022understanding}%
  \BibitemOpen
  \bibfield  {author} {\bibinfo {author} {\bibfnamefont {S.}~\bibnamefont
  {Arora}}, \bibinfo {author} {\bibfnamefont {Z.}~\bibnamefont {Li}},\ and\
  \bibinfo {author} {\bibfnamefont {A.}~\bibnamefont {Panigrahi}},\ }\bibfield
  {title} {\bibinfo {title} {Understanding gradient descent on the edge of
  stability in deep learning},\ }in\ \href@noop {} {\emph {\bibinfo {booktitle}
  {International Conference on Machine Learning}}}\ (\bibinfo {organization}
  {PMLR},\ \bibinfo {year} {2022})\ pp.\ \bibinfo {pages}
  {948--1024}\BibitemShut {NoStop}%
\bibitem [{\citenamefont {Du}\ \emph {et~al.}(2019)\citenamefont {Du},
  \citenamefont {Lee}, \citenamefont {Li}, \citenamefont {Wang},\ and\
  \citenamefont {Zhai}}]{du2019gradient}%
  \BibitemOpen
  \bibfield  {author} {\bibinfo {author} {\bibfnamefont {S.}~\bibnamefont
  {Du}}, \bibinfo {author} {\bibfnamefont {J.}~\bibnamefont {Lee}}, \bibinfo
  {author} {\bibfnamefont {H.}~\bibnamefont {Li}}, \bibinfo {author}
  {\bibfnamefont {L.}~\bibnamefont {Wang}},\ and\ \bibinfo {author}
  {\bibfnamefont {X.}~\bibnamefont {Zhai}},\ }\bibfield  {title} {\bibinfo
  {title} {Gradient descent finds global minima of deep neural networks},\ }in\
  \href@noop {} {\emph {\bibinfo {booktitle} {International conference on
  machine learning}}}\ (\bibinfo {organization} {PMLR},\ \bibinfo {year}
  {2019})\ pp.\ \bibinfo {pages} {1675--1685}\BibitemShut {NoStop}%
\bibitem [{\citenamefont {Locatello}\ \emph {et~al.}(2019)\citenamefont
  {Locatello}, \citenamefont {Bauer}, \citenamefont {Lucic}, \citenamefont
  {Raetsch}, \citenamefont {Gelly}, \citenamefont {Sch{\"o}lkopf},\ and\
  \citenamefont {Bachem}}]{locatello2019challenging}%
  \BibitemOpen
  \bibfield  {author} {\bibinfo {author} {\bibfnamefont {F.}~\bibnamefont
  {Locatello}}, \bibinfo {author} {\bibfnamefont {S.}~\bibnamefont {Bauer}},
  \bibinfo {author} {\bibfnamefont {M.}~\bibnamefont {Lucic}}, \bibinfo
  {author} {\bibfnamefont {G.}~\bibnamefont {Raetsch}}, \bibinfo {author}
  {\bibfnamefont {S.}~\bibnamefont {Gelly}}, \bibinfo {author} {\bibfnamefont
  {B.}~\bibnamefont {Sch{\"o}lkopf}},\ and\ \bibinfo {author} {\bibfnamefont
  {O.}~\bibnamefont {Bachem}},\ }\bibfield  {title} {\bibinfo {title}
  {Challenging common assumptions in the unsupervised learning of disentangled
  representations},\ }in\ \href@noop {} {\emph {\bibinfo {booktitle}
  {international conference on machine learning}}}\ (\bibinfo {organization}
  {PMLR},\ \bibinfo {year} {2019})\ pp.\ \bibinfo {pages}
  {4114--4124}\BibitemShut {NoStop}%
\bibitem [{\citenamefont {B{\"o}ttcher}\ \emph {et~al.}(2022)\citenamefont
  {B{\"o}ttcher}, \citenamefont {Antulov-Fantulin},\ and\ \citenamefont
  {Asikis}}]{bottcher2022ai}%
  \BibitemOpen
  \bibfield  {author} {\bibinfo {author} {\bibfnamefont {L.}~\bibnamefont
  {B{\"o}ttcher}}, \bibinfo {author} {\bibfnamefont {N.}~\bibnamefont
  {Antulov-Fantulin}},\ and\ \bibinfo {author} {\bibfnamefont {T.}~\bibnamefont
  {Asikis}},\ }\bibfield  {title} {\bibinfo {title} {Ai pontryagin or how
  artificial neural networks learn to control dynamical systems},\ }\href@noop
  {} {\bibfield  {journal} {\bibinfo  {journal} {Nature communications}\
  }\textbf {\bibinfo {volume} {13}},\ \bibinfo {pages} {1} (\bibinfo {year}
  {2022})}\BibitemShut {NoStop}%
\bibitem [{\citenamefont {Asikis}\ \emph {et~al.}(2022)\citenamefont {Asikis},
  \citenamefont {B{\"o}ttcher},\ and\ \citenamefont
  {Antulov-Fantulin}}]{asikis2022neural}%
  \BibitemOpen
  \bibfield  {author} {\bibinfo {author} {\bibfnamefont {T.}~\bibnamefont
  {Asikis}}, \bibinfo {author} {\bibfnamefont {L.}~\bibnamefont
  {B{\"o}ttcher}},\ and\ \bibinfo {author} {\bibfnamefont {N.}~\bibnamefont
  {Antulov-Fantulin}},\ }\bibfield  {title} {\bibinfo {title} {Neural ordinary
  differential equation control of dynamics on graphs},\ }\href@noop {}
  {\bibfield  {journal} {\bibinfo  {journal} {Physical Review Research}\
  }\textbf {\bibinfo {volume} {4}},\ \bibinfo {pages} {013221} (\bibinfo {year}
  {2022})}\BibitemShut {NoStop}%
\bibitem [{\citenamefont {Jin}\ \emph {et~al.}(2020)\citenamefont {Jin},
  \citenamefont {Wang}, \citenamefont {Yang},\ and\ \citenamefont
  {Mou}}]{jin2020pontryagin}%
  \BibitemOpen
  \bibfield  {author} {\bibinfo {author} {\bibfnamefont {W.}~\bibnamefont
  {Jin}}, \bibinfo {author} {\bibfnamefont {Z.}~\bibnamefont {Wang}}, \bibinfo
  {author} {\bibfnamefont {Z.}~\bibnamefont {Yang}},\ and\ \bibinfo {author}
  {\bibfnamefont {S.}~\bibnamefont {Mou}},\ }\bibfield  {title} {\bibinfo
  {title} {Pontryagin differentiable programming: An end-to-end learning and
  control framework},\ }\href@noop {} {\bibfield  {journal} {\bibinfo
  {journal} {Advances in Neural Information Processing Systems}\ }\textbf
  {\bibinfo {volume} {33}},\ \bibinfo {pages} {7979} (\bibinfo {year}
  {2020})}\BibitemShut {NoStop}%
\bibitem [{\citenamefont {Srinivasan}\ \emph {et~al.}(2022)\citenamefont
  {Srinivasan}, \citenamefont {Coble}, \citenamefont {Hamlin}, \citenamefont
  {Antonsen}, \citenamefont {Ott},\ and\ \citenamefont
  {Girvan}}]{Srinivasan2022ParallelNetworks}%
  \BibitemOpen
  \bibfield  {author} {\bibinfo {author} {\bibfnamefont {K.}~\bibnamefont
  {Srinivasan}}, \bibinfo {author} {\bibfnamefont {N.}~\bibnamefont {Coble}},
  \bibinfo {author} {\bibfnamefont {J.}~\bibnamefont {Hamlin}}, \bibinfo
  {author} {\bibfnamefont {T.}~\bibnamefont {Antonsen}}, \bibinfo {author}
  {\bibfnamefont {E.}~\bibnamefont {Ott}},\ and\ \bibinfo {author}
  {\bibfnamefont {M.}~\bibnamefont {Girvan}},\ }\bibfield  {title} {\bibinfo
  {title} {{Parallel Machine Learning for Forecasting the Dynamics of Complex
  Networks}},\ }\bibfield  {journal} {\bibinfo  {journal} {Physical Review
  Letters}\ }\textbf {\bibinfo {volume} {128}},\ \href
  {https://doi.org/10.1103/PhysRevLett.128.164101}
  {10.1103/PhysRevLett.128.164101} (\bibinfo {year} {2022})\BibitemShut
  {NoStop}%
\bibitem [{\citenamefont {Pathak}\ \emph {et~al.}(2018)\citenamefont {Pathak},
  \citenamefont {Hunt}, \citenamefont {Girvan}, \citenamefont {Lu},\ and\
  \citenamefont {Ott}}]{pathak2018model}%
  \BibitemOpen
  \bibfield  {author} {\bibinfo {author} {\bibfnamefont {J.}~\bibnamefont
  {Pathak}}, \bibinfo {author} {\bibfnamefont {B.}~\bibnamefont {Hunt}},
  \bibinfo {author} {\bibfnamefont {M.}~\bibnamefont {Girvan}}, \bibinfo
  {author} {\bibfnamefont {Z.}~\bibnamefont {Lu}},\ and\ \bibinfo {author}
  {\bibfnamefont {E.}~\bibnamefont {Ott}},\ }\bibfield  {title} {\bibinfo
  {title} {Model-free prediction of large spatiotemporally chaotic systems from
  data: A reservoir computing approach},\ }\href@noop {} {\bibfield  {journal}
  {\bibinfo  {journal} {Physical review letters}\ }\textbf {\bibinfo {volume}
  {120}},\ \bibinfo {pages} {024102} (\bibinfo {year} {2018})}\BibitemShut
  {NoStop}%
\bibitem [{\citenamefont {Vapnik}\ and\ \citenamefont
  {Chervonenkis}(1991)}]{vapnik1991necessary}%
  \BibitemOpen
  \bibfield  {author} {\bibinfo {author} {\bibfnamefont {V.}~\bibnamefont
  {Vapnik}}\ and\ \bibinfo {author} {\bibfnamefont {A.}~\bibnamefont
  {Chervonenkis}},\ }\bibfield  {title} {\bibinfo {title} {The necessary and
  sufficient conditions for consistency in the empirical risk minimization
  method},\ }\href@noop {} {\bibfield  {journal} {\bibinfo  {journal} {Pattern
  Recognition and Image Analysis}\ }\textbf {\bibinfo {volume} {1}},\ \bibinfo
  {pages} {283} (\bibinfo {year} {1991})}\BibitemShut {NoStop}%
\bibitem [{\citenamefont {Bartlett}\ and\ \citenamefont
  {Mendelson}(2002)}]{bartlett2002rademacher}%
  \BibitemOpen
  \bibfield  {author} {\bibinfo {author} {\bibfnamefont {P.~L.}\ \bibnamefont
  {Bartlett}}\ and\ \bibinfo {author} {\bibfnamefont {S.}~\bibnamefont
  {Mendelson}},\ }\bibfield  {title} {\bibinfo {title} {Rademacher and gaussian
  complexities: Risk bounds and structural results},\ }\href@noop {} {\bibfield
   {journal} {\bibinfo  {journal} {Journal of Machine Learning Research}\
  }\textbf {\bibinfo {volume} {3}},\ \bibinfo {pages} {463} (\bibinfo {year}
  {2002})}\BibitemShut {NoStop}%
\bibitem [{\citenamefont {Tucker}(1959)}]{tucker1959generalization}%
  \BibitemOpen
  \bibfield  {author} {\bibinfo {author} {\bibfnamefont {H.~G.}\ \bibnamefont
  {Tucker}},\ }\bibfield  {title} {\bibinfo {title} {A generalization of the
  glivenko-cantelli theorem},\ }\href@noop {} {\bibfield  {journal} {\bibinfo
  {journal} {The Annals of Mathematical Statistics}\ }\textbf {\bibinfo
  {volume} {30}},\ \bibinfo {pages} {828} (\bibinfo {year} {1959})}\BibitemShut
  {NoStop}%
\bibitem [{\citenamefont {Gerbelot}\ \emph {et~al.}(2023)\citenamefont
  {Gerbelot}, \citenamefont {Karagulyan}, \citenamefont {Karp}, \citenamefont
  {Ravichandran}, \citenamefont {Stern},\ and\ \citenamefont
  {Srebro}}]{gerbelot2023applying}%
  \BibitemOpen
  \bibfield  {author} {\bibinfo {author} {\bibfnamefont {C.}~\bibnamefont
  {Gerbelot}}, \bibinfo {author} {\bibfnamefont {A.}~\bibnamefont
  {Karagulyan}}, \bibinfo {author} {\bibfnamefont {S.}~\bibnamefont {Karp}},
  \bibinfo {author} {\bibfnamefont {K.}~\bibnamefont {Ravichandran}}, \bibinfo
  {author} {\bibfnamefont {M.}~\bibnamefont {Stern}},\ and\ \bibinfo {author}
  {\bibfnamefont {N.}~\bibnamefont {Srebro}},\ }\bibfield  {title} {\bibinfo
  {title} {Applying statistical learning theory to deep learning},\ }\href@noop
  {} {\bibfield  {journal} {\bibinfo  {journal} {arXiv preprint
  arXiv:2311.15404}\ } (\bibinfo {year} {2023})}\BibitemShut {NoStop}%
\bibitem [{\citenamefont {Zhang}\ \emph {et~al.}(2019)\citenamefont {Zhang},
  \citenamefont {Bengio}, \citenamefont {Hardt}, \citenamefont {Mozer},\ and\
  \citenamefont {Singer}}]{zhang2019identity}%
  \BibitemOpen
  \bibfield  {author} {\bibinfo {author} {\bibfnamefont {C.}~\bibnamefont
  {Zhang}}, \bibinfo {author} {\bibfnamefont {S.}~\bibnamefont {Bengio}},
  \bibinfo {author} {\bibfnamefont {M.}~\bibnamefont {Hardt}}, \bibinfo
  {author} {\bibfnamefont {M.~C.}\ \bibnamefont {Mozer}},\ and\ \bibinfo
  {author} {\bibfnamefont {Y.}~\bibnamefont {Singer}},\ }\bibfield  {title}
  {\bibinfo {title} {Identity crisis: Memorization and generalization under
  extreme overparameterization},\ }\href@noop {} {\bibfield  {journal}
  {\bibinfo  {journal} {arXiv preprint arXiv:1902.04698}\ } (\bibinfo {year}
  {2019})}\BibitemShut {NoStop}%
\bibitem [{\citenamefont {Goyal}\ and\ \citenamefont
  {Bengio}(2022{\natexlab{a}})}]{Goyal2022InductiveCognition}%
  \BibitemOpen
  \bibfield  {author} {\bibinfo {author} {\bibfnamefont {A.}~\bibnamefont
  {Goyal}}\ and\ \bibinfo {author} {\bibfnamefont {Y.}~\bibnamefont {Bengio}},\
  }\bibfield  {title} {\bibinfo {title} {{Inductive biases for deep learning of
  higher-level cognition}},\ }\bibfield  {journal} {\bibinfo  {journal}
  {Proceedings of the Royal Society A: Mathematical, Physical and Engineering
  Sciences}\ }\textbf {\bibinfo {volume} {478}},\ \href
  {https://doi.org/10.1098/RSPA.2021.0068} {10.1098/RSPA.2021.0068} (\bibinfo
  {year} {2022}{\natexlab{a}})\BibitemShut {NoStop}%
\bibitem [{\citenamefont {Goyal}\ and\ \citenamefont
  {Bengio}(2022{\natexlab{b}})}]{goyal2022inductive}%
  \BibitemOpen
  \bibfield  {author} {\bibinfo {author} {\bibfnamefont {A.}~\bibnamefont
  {Goyal}}\ and\ \bibinfo {author} {\bibfnamefont {Y.}~\bibnamefont {Bengio}},\
  }\bibfield  {title} {\bibinfo {title} {Inductive biases for deep learning of
  higher-level cognition},\ }\href@noop {} {\bibfield  {journal} {\bibinfo
  {journal} {Proceedings of the Royal Society A}\ }\textbf {\bibinfo {volume}
  {478}},\ \bibinfo {pages} {20210068} (\bibinfo {year}
  {2022}{\natexlab{b}})}\BibitemShut {NoStop}%
\bibitem [{\citenamefont {Fotiadis}\ \emph {et~al.}(2023)\citenamefont
  {Fotiadis}, \citenamefont {Valencia}, \citenamefont {Hu}, \citenamefont
  {Garasto}, \citenamefont {Cantwell},\ and\ \citenamefont
  {Bharath}}]{fotiadis2023disentangled}%
  \BibitemOpen
  \bibfield  {author} {\bibinfo {author} {\bibfnamefont {S.}~\bibnamefont
  {Fotiadis}}, \bibinfo {author} {\bibfnamefont {M.~L.}\ \bibnamefont
  {Valencia}}, \bibinfo {author} {\bibfnamefont {S.}~\bibnamefont {Hu}},
  \bibinfo {author} {\bibfnamefont {S.}~\bibnamefont {Garasto}}, \bibinfo
  {author} {\bibfnamefont {C.~D.}\ \bibnamefont {Cantwell}},\ and\ \bibinfo
  {author} {\bibfnamefont {A.~A.}\ \bibnamefont {Bharath}},\ }\bibfield
  {title} {\bibinfo {title} {Disentangled generative models for robust
  prediction of system dynamics},\ }in\ \href@noop {} {\emph {\bibinfo
  {booktitle} {International Conference on Machine Learning}}}\ (\bibinfo
  {organization} {PMLR},\ \bibinfo {year} {2023})\ pp.\ \bibinfo {pages}
  {10222--10248}\BibitemShut {NoStop}%
\bibitem [{\citenamefont {Boccaletti}\ \emph {et~al.}(2006)\citenamefont
  {Boccaletti}, \citenamefont {Latora}, \citenamefont {Moreno}, \citenamefont
  {Chavez},\ and\ \citenamefont {Hwang}}]{boccaletti2006complex}%
  \BibitemOpen
  \bibfield  {author} {\bibinfo {author} {\bibfnamefont {S.}~\bibnamefont
  {Boccaletti}}, \bibinfo {author} {\bibfnamefont {V.}~\bibnamefont {Latora}},
  \bibinfo {author} {\bibfnamefont {Y.}~\bibnamefont {Moreno}}, \bibinfo
  {author} {\bibfnamefont {M.}~\bibnamefont {Chavez}},\ and\ \bibinfo {author}
  {\bibfnamefont {D.-U.}\ \bibnamefont {Hwang}},\ }\bibfield  {title} {\bibinfo
  {title} {Complex networks: Structure and dynamics},\ }\href@noop {}
  {\bibfield  {journal} {\bibinfo  {journal} {Physics reports}\ }\textbf
  {\bibinfo {volume} {424}},\ \bibinfo {pages} {175} (\bibinfo {year}
  {2006})}\BibitemShut {NoStop}%
\bibitem [{\citenamefont {Vasiliauskaite}\ and\ \citenamefont
  {Rosas}(2020)}]{vasiliauskaite2020understanding}%
  \BibitemOpen
  \bibfield  {author} {\bibinfo {author} {\bibfnamefont {V.}~\bibnamefont
  {Vasiliauskaite}}\ and\ \bibinfo {author} {\bibfnamefont {F.~E.}\
  \bibnamefont {Rosas}},\ }\bibfield  {title} {\bibinfo {title} {Understanding
  complexity via network theory: a gentle introduction},\ }\href
  {DOI:doi.org/10.47041/CVTD4629} {\bibfield  {journal} {\bibinfo  {journal}
  {arXiv preprint arXiv:2004.14845}\ } (\bibinfo {year} {2020})}\BibitemShut
  {NoStop}%
\bibitem [{\citenamefont {Sprot}(2008)}]{Sprot2008ChaoticNetworks}%
  \BibitemOpen
  \bibfield  {author} {\bibinfo {author} {\bibfnamefont {J.~C.}\ \bibnamefont
  {Sprot}},\ }\bibfield  {title} {\bibinfo {title} {{Chaotic dynamics on large
  networks}},\ }\href {https://doi.org/10.1063/1.2945229} {\bibfield  {journal}
  {\bibinfo  {journal} {Chaos: An Interdisciplinary Journal of Nonlinear
  Science}\ }\textbf {\bibinfo {volume} {18}},\ \bibinfo {pages} {023135}
  (\bibinfo {year} {2008})}\BibitemShut {NoStop}%
\bibitem [{\citenamefont {Delvenne}\ \emph {et~al.}(2015)\citenamefont
  {Delvenne}, \citenamefont {Lambiotte},\ and\ \citenamefont
  {Rocha}}]{delvenne2015diffusion}%
  \BibitemOpen
  \bibfield  {author} {\bibinfo {author} {\bibfnamefont {J.-C.}\ \bibnamefont
  {Delvenne}}, \bibinfo {author} {\bibfnamefont {R.}~\bibnamefont
  {Lambiotte}},\ and\ \bibinfo {author} {\bibfnamefont {L.~E.}\ \bibnamefont
  {Rocha}},\ }\bibfield  {title} {\bibinfo {title} {Diffusion on networked
  systems is a question of time or structure},\ }\href@noop {} {\bibfield
  {journal} {\bibinfo  {journal} {Nature communications}\ }\textbf {\bibinfo
  {volume} {6}},\ \bibinfo {pages} {7366} (\bibinfo {year} {2015})}\BibitemShut
  {NoStop}%
\bibitem [{\citenamefont {Rodrigues}\ \emph {et~al.}(2016)\citenamefont
  {Rodrigues}, \citenamefont {Peron}, \citenamefont {Ji},\ and\ \citenamefont
  {Kurths}}]{rodrigues2016kuramoto}%
  \BibitemOpen
  \bibfield  {author} {\bibinfo {author} {\bibfnamefont {F.~A.}\ \bibnamefont
  {Rodrigues}}, \bibinfo {author} {\bibfnamefont {T.~K.~D.}\ \bibnamefont
  {Peron}}, \bibinfo {author} {\bibfnamefont {P.}~\bibnamefont {Ji}},\ and\
  \bibinfo {author} {\bibfnamefont {J.}~\bibnamefont {Kurths}},\ }\bibfield
  {title} {\bibinfo {title} {The kuramoto model in complex networks},\
  }\href@noop {} {\bibfield  {journal} {\bibinfo  {journal} {Physics Reports}\
  }\textbf {\bibinfo {volume} {610}},\ \bibinfo {pages} {1} (\bibinfo {year}
  {2016})}\BibitemShut {NoStop}%
\bibitem [{\citenamefont {Rabinovich}\ \emph {et~al.}(2006)\citenamefont
  {Rabinovich}, \citenamefont {Varona}, \citenamefont {Selverston},\ and\
  \citenamefont {Abarbanel}}]{rabinovich2006dynamical}%
  \BibitemOpen
  \bibfield  {author} {\bibinfo {author} {\bibfnamefont {M.~I.}\ \bibnamefont
  {Rabinovich}}, \bibinfo {author} {\bibfnamefont {P.}~\bibnamefont {Varona}},
  \bibinfo {author} {\bibfnamefont {A.~I.}\ \bibnamefont {Selverston}},\ and\
  \bibinfo {author} {\bibfnamefont {H.~D.}\ \bibnamefont {Abarbanel}},\
  }\bibfield  {title} {\bibinfo {title} {Dynamical principles in
  neuroscience},\ }\href@noop {} {\bibfield  {journal} {\bibinfo  {journal}
  {Reviews of modern physics}\ }\textbf {\bibinfo {volume} {78}},\ \bibinfo
  {pages} {1213} (\bibinfo {year} {2006})}\BibitemShut {NoStop}%
\bibitem [{Note1()}]{Note1}%
  \BibitemOpen
  \bibinfo {note} {Such as the principle of relativity that leads to, e.g.\ the
  mass-energy equivalence.}\BibitemShut {Stop}%
\bibitem [{\citenamefont {Riascos}\ and\ \citenamefont
  {Mateos}(2021)}]{riascos2021random}%
  \BibitemOpen
  \bibfield  {author} {\bibinfo {author} {\bibfnamefont {A.~P.}\ \bibnamefont
  {Riascos}}\ and\ \bibinfo {author} {\bibfnamefont {J.~L.}\ \bibnamefont
  {Mateos}},\ }\bibfield  {title} {\bibinfo {title} {Random walks on weighted
  networks: a survey of local and non-local dynamics},\ }\href@noop {}
  {\bibfield  {journal} {\bibinfo  {journal} {Journal of Complex Networks}\
  }\textbf {\bibinfo {volume} {9}},\ \bibinfo {pages} {cnab032} (\bibinfo
  {year} {2021})}\BibitemShut {NoStop}%
\bibitem [{\citenamefont {Gleeson}(2013)}]{gleeson2013binary}%
  \BibitemOpen
  \bibfield  {author} {\bibinfo {author} {\bibfnamefont {J.~P.}\ \bibnamefont
  {Gleeson}},\ }\bibfield  {title} {\bibinfo {title} {Binary-state dynamics on
  complex networks: Pair approximation and beyond},\ }\href@noop {} {\bibfield
  {journal} {\bibinfo  {journal} {Physical Review X}\ }\textbf {\bibinfo
  {volume} {3}},\ \bibinfo {pages} {021004} (\bibinfo {year}
  {2013})}\BibitemShut {NoStop}%
\bibitem [{\citenamefont {Scarselli}\ \emph {et~al.}(2009)\citenamefont
  {Scarselli}, \citenamefont {Gori}, \citenamefont {Tsoi}, \citenamefont
  {Hagenbuchner},\ and\ \citenamefont {Monfardini}}]{Scarselli2009TheModel}%
  \BibitemOpen
  \bibfield  {author} {\bibinfo {author} {\bibfnamefont {F.}~\bibnamefont
  {Scarselli}}, \bibinfo {author} {\bibfnamefont {M.}~\bibnamefont {Gori}},
  \bibinfo {author} {\bibfnamefont {A.~C.}\ \bibnamefont {Tsoi}}, \bibinfo
  {author} {\bibfnamefont {M.}~\bibnamefont {Hagenbuchner}},\ and\ \bibinfo
  {author} {\bibfnamefont {G.}~\bibnamefont {Monfardini}},\ }\bibfield  {title}
  {\bibinfo {title} {{The graph neural network model}},\ }\href
  {https://doi.org/10.1109/TNN.2008.2005605} {\bibfield  {journal} {\bibinfo
  {journal} {IEEE Transactions on Neural Networks}\ }\textbf {\bibinfo {volume}
  {20}},\ \bibinfo {pages} {61} (\bibinfo {year} {2009})}\BibitemShut {NoStop}%
\bibitem [{\citenamefont {Xu}\ \emph {et~al.}(2018{\natexlab{b}})\citenamefont
  {Xu}, \citenamefont {Jegelka}, \citenamefont {Hu},\ and\ \citenamefont
  {Leskovec}}]{Xu2018HowNetworks}%
  \BibitemOpen
  \bibfield  {author} {\bibinfo {author} {\bibfnamefont {K.}~\bibnamefont
  {Xu}}, \bibinfo {author} {\bibfnamefont {S.}~\bibnamefont {Jegelka}},
  \bibinfo {author} {\bibfnamefont {W.}~\bibnamefont {Hu}},\ and\ \bibinfo
  {author} {\bibfnamefont {J.}~\bibnamefont {Leskovec}},\ }\bibfield  {title}
  {\bibinfo {title} {{How Powerful are Graph Neural Networks?}},\ }\bibfield
  {journal} {\bibinfo  {journal} {7th International Conference on Learning
  Representations, ICLR 2019}\ }\href
  {https://doi.org/10.48550/arxiv.1810.00826} {10.48550/arxiv.1810.00826}
  (\bibinfo {year} {2018}{\natexlab{b}})\BibitemShut {NoStop}%
\bibitem [{Note2()}]{Note2}%
  \BibitemOpen
  \bibinfo {note} {In place of $\protect \mathbf {A}$, one may consider, e.g.,
  a single-layer graph convolution~\cite {kipf2016semi}, degree-scaled
  adjacency matrix, or a laplacian matrix.}\BibitemShut {Stop}%
\bibitem [{\citenamefont {Morris}\ \emph {et~al.}(2019)\citenamefont {Morris},
  \citenamefont {Ritzert}, \citenamefont {Fey}, \citenamefont {Hamilton},
  \citenamefont {Lenssen}, \citenamefont {Rattan},\ and\ \citenamefont
  {Grohe}}]{GraphConv}%
  \BibitemOpen
  \bibfield  {author} {\bibinfo {author} {\bibfnamefont {C.}~\bibnamefont
  {Morris}}, \bibinfo {author} {\bibfnamefont {M.}~\bibnamefont {Ritzert}},
  \bibinfo {author} {\bibfnamefont {M.}~\bibnamefont {Fey}}, \bibinfo {author}
  {\bibfnamefont {W.~L.}\ \bibnamefont {Hamilton}}, \bibinfo {author}
  {\bibfnamefont {J.~E.}\ \bibnamefont {Lenssen}}, \bibinfo {author}
  {\bibfnamefont {G.}~\bibnamefont {Rattan}},\ and\ \bibinfo {author}
  {\bibfnamefont {M.}~\bibnamefont {Grohe}},\ }\bibfield  {title} {\bibinfo
  {title} {Weisfeiler and leman go neural: Higher-order graph neural
  networks},\ }in\ \href@noop {} {\emph {\bibinfo {booktitle} {Proceedings of
  the AAAI conference on artificial intelligence}}},\ Vol.~\bibinfo {volume}
  {33}\ (\bibinfo {year} {2019})\ pp.\ \bibinfo {pages}
  {4602--4609}\BibitemShut {NoStop}%
\bibitem [{\citenamefont {Bresson}\ and\ \citenamefont
  {Laurent}(2017{\natexlab{a}})}]{ResGatedGraphConv}%
  \BibitemOpen
  \bibfield  {author} {\bibinfo {author} {\bibfnamefont {X.}~\bibnamefont
  {Bresson}}\ and\ \bibinfo {author} {\bibfnamefont {T.}~\bibnamefont
  {Laurent}},\ }\bibfield  {title} {\bibinfo {title} {Residual gated graph
  convnets},\ }\href@noop {} {\bibfield  {journal} {\bibinfo  {journal} {arXiv
  preprint arXiv:1711.07553}\ } (\bibinfo {year}
  {2017}{\natexlab{a}})}\BibitemShut {NoStop}%
\bibitem [{\citenamefont {Hamilton}\ \emph {et~al.}(2017)\citenamefont
  {Hamilton}, \citenamefont {Ying},\ and\ \citenamefont {Leskovec}}]{SAGEConv}%
  \BibitemOpen
  \bibfield  {author} {\bibinfo {author} {\bibfnamefont {W.}~\bibnamefont
  {Hamilton}}, \bibinfo {author} {\bibfnamefont {Z.}~\bibnamefont {Ying}},\
  and\ \bibinfo {author} {\bibfnamefont {J.}~\bibnamefont {Leskovec}},\
  }\bibfield  {title} {\bibinfo {title} {Inductive representation learning on
  large graphs},\ }\href@noop {} {\bibfield  {journal} {\bibinfo  {journal}
  {Advances in neural information processing systems}\ }\textbf {\bibinfo
  {volume} {30}} (\bibinfo {year} {2017})}\BibitemShut {NoStop}%
\bibitem [{\citenamefont {Defferrard}\ \emph {et~al.}(2016)\citenamefont
  {Defferrard}, \citenamefont {Bresson},\ and\ \citenamefont
  {Vandergheynst}}]{ChebConv}%
  \BibitemOpen
  \bibfield  {author} {\bibinfo {author} {\bibfnamefont {M.}~\bibnamefont
  {Defferrard}}, \bibinfo {author} {\bibfnamefont {X.}~\bibnamefont
  {Bresson}},\ and\ \bibinfo {author} {\bibfnamefont {P.}~\bibnamefont
  {Vandergheynst}},\ }\bibfield  {title} {\bibinfo {title} {Convolutional
  neural networks on graphs with fast localized spectral filtering},\
  }\href@noop {} {\bibfield  {journal} {\bibinfo  {journal} {Advances in neural
  information processing systems}\ }\textbf {\bibinfo {volume} {29}} (\bibinfo
  {year} {2016})}\BibitemShut {NoStop}%
\bibitem [{\citenamefont {Veli{\v{c}}kovi{\'c}}\ \emph
  {et~al.}(2017)\citenamefont {Veli{\v{c}}kovi{\'c}}, \citenamefont {Cucurull},
  \citenamefont {Casanova}, \citenamefont {Romero}, \citenamefont {Lio},\ and\
  \citenamefont {Bengio}}]{GATConv}%
  \BibitemOpen
  \bibfield  {author} {\bibinfo {author} {\bibfnamefont {P.}~\bibnamefont
  {Veli{\v{c}}kovi{\'c}}}, \bibinfo {author} {\bibfnamefont {G.}~\bibnamefont
  {Cucurull}}, \bibinfo {author} {\bibfnamefont {A.}~\bibnamefont {Casanova}},
  \bibinfo {author} {\bibfnamefont {A.}~\bibnamefont {Romero}}, \bibinfo
  {author} {\bibfnamefont {P.}~\bibnamefont {Lio}},\ and\ \bibinfo {author}
  {\bibfnamefont {Y.}~\bibnamefont {Bengio}},\ }\bibfield  {title} {\bibinfo
  {title} {Graph attention networks},\ }\href@noop {} {\bibfield  {journal}
  {\bibinfo  {journal} {arXiv preprint arXiv:1710.10903}\ } (\bibinfo {year}
  {2017})}\BibitemShut {NoStop}%
\bibitem [{\citenamefont {Fey}\ and\ \citenamefont {Lenssen}(2019)}]{pyG}%
  \BibitemOpen
  \bibfield  {author} {\bibinfo {author} {\bibfnamefont {M.}~\bibnamefont
  {Fey}}\ and\ \bibinfo {author} {\bibfnamefont {J.~E.}\ \bibnamefont
  {Lenssen}},\ }\bibfield  {title} {\bibinfo {title} {Fast graph representation
  learning with pytorch geometric},\ }\href@noop {} {\bibfield  {journal}
  {\bibinfo  {journal} {arXiv preprint arXiv:1903.02428}\ } (\bibinfo {year}
  {2019})}\BibitemShut {NoStop}%
\bibitem [{\citenamefont {Bresson}\ and\ \citenamefont
  {Laurent}(2017{\natexlab{b}})}]{bresson2017residual}%
  \BibitemOpen
  \bibfield  {author} {\bibinfo {author} {\bibfnamefont {X.}~\bibnamefont
  {Bresson}}\ and\ \bibinfo {author} {\bibfnamefont {T.}~\bibnamefont
  {Laurent}},\ }\bibfield  {title} {\bibinfo {title} {Residual gated graph
  convnets},\ }\href@noop {} {\bibfield  {journal} {\bibinfo  {journal} {arXiv
  preprint arXiv:1711.07553}\ } (\bibinfo {year}
  {2017}{\natexlab{b}})}\BibitemShut {NoStop}%
\bibitem [{\citenamefont {Aggarwal}\ \emph {et~al.}(2001)\citenamefont
  {Aggarwal}, \citenamefont {Hinneburg},\ and\ \citenamefont
  {Keim}}]{aggarwal2001surprising}%
  \BibitemOpen
  \bibfield  {author} {\bibinfo {author} {\bibfnamefont {C.~C.}\ \bibnamefont
  {Aggarwal}}, \bibinfo {author} {\bibfnamefont {A.}~\bibnamefont
  {Hinneburg}},\ and\ \bibinfo {author} {\bibfnamefont {D.~A.}\ \bibnamefont
  {Keim}},\ }\bibfield  {title} {\bibinfo {title} {On the surprising behavior
  of distance metrics in high dimensional space},\ }in\ \href@noop {} {\emph
  {\bibinfo {booktitle} {Database Theory—ICDT 2001: 8th International
  Conference London, UK, January 4--6, 2001 Proceedings 8}}}\ (\bibinfo
  {organization} {Springer},\ \bibinfo {year} {2001})\ pp.\ \bibinfo {pages}
  {420--434}\BibitemShut {NoStop}%
\bibitem [{\citenamefont {Belkin}\ \emph {et~al.}(2019)\citenamefont {Belkin},
  \citenamefont {Hsu}, \citenamefont {Ma},\ and\ \citenamefont
  {Mandal}}]{belkin2019reconciling}%
  \BibitemOpen
  \bibfield  {author} {\bibinfo {author} {\bibfnamefont {M.}~\bibnamefont
  {Belkin}}, \bibinfo {author} {\bibfnamefont {D.}~\bibnamefont {Hsu}},
  \bibinfo {author} {\bibfnamefont {S.}~\bibnamefont {Ma}},\ and\ \bibinfo
  {author} {\bibfnamefont {S.}~\bibnamefont {Mandal}},\ }\bibfield  {title}
  {\bibinfo {title} {Reconciling modern machine-learning practice and the
  classical bias--variance trade-off},\ }\href@noop {} {\bibfield  {journal}
  {\bibinfo  {journal} {Proceedings of the National Academy of Sciences}\
  }\textbf {\bibinfo {volume} {116}},\ \bibinfo {pages} {15849} (\bibinfo
  {year} {2019})}\BibitemShut {NoStop}%
\bibitem [{\citenamefont {Zhang}\ \emph {et~al.}(2021)\citenamefont {Zhang},
  \citenamefont {Bengio}, \citenamefont {Hardt}, \citenamefont {Recht},\ and\
  \citenamefont {Vinyals}}]{zhang2021understanding}%
  \BibitemOpen
  \bibfield  {author} {\bibinfo {author} {\bibfnamefont {C.}~\bibnamefont
  {Zhang}}, \bibinfo {author} {\bibfnamefont {S.}~\bibnamefont {Bengio}},
  \bibinfo {author} {\bibfnamefont {M.}~\bibnamefont {Hardt}}, \bibinfo
  {author} {\bibfnamefont {B.}~\bibnamefont {Recht}},\ and\ \bibinfo {author}
  {\bibfnamefont {O.}~\bibnamefont {Vinyals}},\ }\bibfield  {title} {\bibinfo
  {title} {Understanding deep learning (still) requires rethinking
  generalization},\ }\href@noop {} {\bibfield  {journal} {\bibinfo  {journal}
  {Communications of the ACM}\ }\textbf {\bibinfo {volume} {64}},\ \bibinfo
  {pages} {107} (\bibinfo {year} {2021})}\BibitemShut {NoStop}%
\bibitem [{\citenamefont {Jakubovitz}\ \emph {et~al.}(2019)\citenamefont
  {Jakubovitz}, \citenamefont {Giryes},\ and\ \citenamefont
  {Rodrigues}}]{jakubovitz2019generalization}%
  \BibitemOpen
  \bibfield  {author} {\bibinfo {author} {\bibfnamefont {D.}~\bibnamefont
  {Jakubovitz}}, \bibinfo {author} {\bibfnamefont {R.}~\bibnamefont {Giryes}},\
  and\ \bibinfo {author} {\bibfnamefont {M.~R.}\ \bibnamefont {Rodrigues}},\
  }\bibfield  {title} {\bibinfo {title} {Generalization error in deep
  learning},\ }in\ \href@noop {} {\emph {\bibinfo {booktitle} {Compressed
  Sensing and Its Applications: Third International MATHEON Conference 2017}}}\
  (\bibinfo {organization} {Springer},\ \bibinfo {year} {2019})\ pp.\ \bibinfo
  {pages} {153--193}\BibitemShut {NoStop}%
\bibitem [{Note3()}]{Note3}%
  \BibitemOpen
  \bibinfo {note} {Novel SLT frameworks, including algorithm stability~\cite
  {hardt2016train, bousquet2002stability}, algorithm robustness~\cite
  {xu2012robustness}, PAC-Bayes theory~\cite
  {bartlett2017spectrally,mcallester1999pac}, compression and sampling~\cite
  {arora2018stronger,giryes2022function} are active fields of research and
  could possibly shed light on generalization in this class of
  models.}\BibitemShut {Stop}%
\bibitem [{Note4()}]{Note4}%
  \BibitemOpen
  \bibinfo {note} {Some UAT results cover density in non-compact domains,
  e.g.~\cite {kidger2020universal}. Nonetheless, the authors proceeded with the
  assumption that a target function maps to zero outside of a given
  support.}\BibitemShut {Stop}%
\bibitem [{\citenamefont {Anysz}\ \emph {et~al.}(2016)\citenamefont {Anysz},
  \citenamefont {Zbiciak},\ and\ \citenamefont {Ibadov}}]{anysz2016influence}%
  \BibitemOpen
  \bibfield  {author} {\bibinfo {author} {\bibfnamefont {H.}~\bibnamefont
  {Anysz}}, \bibinfo {author} {\bibfnamefont {A.}~\bibnamefont {Zbiciak}},\
  and\ \bibinfo {author} {\bibfnamefont {N.}~\bibnamefont {Ibadov}},\
  }\bibfield  {title} {\bibinfo {title} {The influence of input data
  standardization method on prediction accuracy of artificial neural
  networks},\ }\href@noop {} {\bibfield  {journal} {\bibinfo  {journal}
  {Procedia Engineering}\ }\textbf {\bibinfo {volume} {153}},\ \bibinfo {pages}
  {66} (\bibinfo {year} {2016})}\BibitemShut {NoStop}%
\bibitem [{\citenamefont {Bianconi}(2009)}]{bianconi2009entropy}%
  \BibitemOpen
  \bibfield  {author} {\bibinfo {author} {\bibfnamefont {G.}~\bibnamefont
  {Bianconi}},\ }\bibfield  {title} {\bibinfo {title} {Entropy of network
  ensembles},\ }\href@noop {} {\bibfield  {journal} {\bibinfo  {journal}
  {Physical Review E}\ }\textbf {\bibinfo {volume} {79}},\ \bibinfo {pages}
  {036114} (\bibinfo {year} {2009})}\BibitemShut {NoStop}%
\bibitem [{Note5()}]{Note5}%
  \BibitemOpen
  \bibinfo {note} {Alternatively, one may consider a micro-canonical ensemble
  by employing, e.g.\ a configuration model~\cite {park2004statistical},
  thereby imposing harder constraints of fixed number of edges.}\BibitemShut
  {Stop}%
\bibitem [{\citenamefont {Hastie}\ \emph {et~al.}(2009)\citenamefont {Hastie},
  \citenamefont {Tibshirani}, \citenamefont {Friedman},\ and\ \citenamefont
  {Friedman}}]{hastie2009elements}%
  \BibitemOpen
  \bibfield  {author} {\bibinfo {author} {\bibfnamefont {T.}~\bibnamefont
  {Hastie}}, \bibinfo {author} {\bibfnamefont {R.}~\bibnamefont {Tibshirani}},
  \bibinfo {author} {\bibfnamefont {J.~H.}\ \bibnamefont {Friedman}},\ and\
  \bibinfo {author} {\bibfnamefont {J.~H.}\ \bibnamefont {Friedman}},\
  }\href@noop {} {\emph {\bibinfo {title} {The elements of statistical
  learning: data mining, inference, and prediction}}},\ Vol.~\bibinfo {volume}
  {2}\ (\bibinfo  {publisher} {Springer},\ \bibinfo {year} {2009})\BibitemShut
  {NoStop}%
\bibitem [{\citenamefont {Newman}(2018)}]{newman2018networks}%
  \BibitemOpen
  \bibfield  {author} {\bibinfo {author} {\bibfnamefont {M.}~\bibnamefont
  {Newman}},\ }\href@noop {} {\emph {\bibinfo {title} {Networks}}}\ (\bibinfo
  {publisher} {Oxford university press},\ \bibinfo {year} {2018})\BibitemShut
  {NoStop}%
\bibitem [{\citenamefont {Erd{\H{o}}s}\ \emph {et~al.}(1960)\citenamefont
  {Erd{\H{o}}s}, \citenamefont {R{\'e}nyi} \emph
  {et~al.}}]{erdHos1960evolution}%
  \BibitemOpen
  \bibfield  {author} {\bibinfo {author} {\bibfnamefont {P.}~\bibnamefont
  {Erd{\H{o}}s}}, \bibinfo {author} {\bibfnamefont {A.}~\bibnamefont
  {R{\'e}nyi}}, \emph {et~al.},\ }\bibfield  {title} {\bibinfo {title} {On the
  evolution of random graphs},\ }\href@noop {} {\bibfield  {journal} {\bibinfo
  {journal} {Publ. math. inst. hung. acad. sci}\ }\textbf {\bibinfo {volume}
  {5}},\ \bibinfo {pages} {17} (\bibinfo {year} {1960})}\BibitemShut {NoStop}%
\bibitem [{\citenamefont {White}\ \emph {et~al.}(1986)\citenamefont {White},
  \citenamefont {Southgate}, \citenamefont {Thomson}, \citenamefont {Brenner}
  \emph {et~al.}}]{white1986structure}%
  \BibitemOpen
  \bibfield  {author} {\bibinfo {author} {\bibfnamefont {J.~G.}\ \bibnamefont
  {White}}, \bibinfo {author} {\bibfnamefont {E.}~\bibnamefont {Southgate}},
  \bibinfo {author} {\bibfnamefont {J.~N.}\ \bibnamefont {Thomson}}, \bibinfo
  {author} {\bibfnamefont {S.}~\bibnamefont {Brenner}}, \emph {et~al.},\
  }\bibfield  {title} {\bibinfo {title} {The structure of the nervous system of
  the nematode caenorhabditis elegans},\ }\href@noop {} {\bibfield  {journal}
  {\bibinfo  {journal} {Philos Trans R Soc Lond B Biol Sci}\ }\textbf {\bibinfo
  {volume} {314}},\ \bibinfo {pages} {1} (\bibinfo {year} {1986})}\BibitemShut
  {NoStop}%
\bibitem [{Note6()}]{Note6}%
  \BibitemOpen
  \bibinfo {note} {Retrieved from \protect \url
  {https://networks.skewed.de/net/celegansneural}}\BibitemShut {NoStop}%
\bibitem [{\citenamefont {Gao}\ and\ \citenamefont
  {Yan}(2022)}]{Gao2022AutonomousData}%
  \BibitemOpen
  \bibfield  {author} {\bibinfo {author} {\bibfnamefont {T.~T.}\ \bibnamefont
  {Gao}}\ and\ \bibinfo {author} {\bibfnamefont {G.}~\bibnamefont {Yan}},\
  }\bibfield  {title} {\bibinfo {title} {{Autonomous inference of complex
  network dynamics from incomplete and noisy data}},\ }\href
  {https://doi.org/10.1038/s43588-022-00217-0} {\bibfield  {journal} {\bibinfo
  {journal} {Nature Computational Science 2022 2:3}\ }\textbf {\bibinfo
  {volume} {2}},\ \bibinfo {pages} {160} (\bibinfo {year} {2022})}\BibitemShut
  {NoStop}%
\bibitem [{\citenamefont {Rocsoreanu}\ \emph {et~al.}(2012)\citenamefont
  {Rocsoreanu}, \citenamefont {Georgescu},\ and\ \citenamefont
  {Giurgiteanu}}]{rocsoreanu2012fitzhugh}%
  \BibitemOpen
  \bibfield  {author} {\bibinfo {author} {\bibfnamefont {C.}~\bibnamefont
  {Rocsoreanu}}, \bibinfo {author} {\bibfnamefont {A.}~\bibnamefont
  {Georgescu}},\ and\ \bibinfo {author} {\bibfnamefont {N.}~\bibnamefont
  {Giurgiteanu}},\ }\href@noop {} {\emph {\bibinfo {title} {The FitzHugh-Nagumo
  model: bifurcation and dynamics}}},\ Vol.~\bibinfo {volume} {10}\ (\bibinfo
  {publisher} {Springer Science \& Business Media},\ \bibinfo {year}
  {2012})\BibitemShut {NoStop}%
\bibitem [{\citenamefont {Rahaman}\ \emph {et~al.}(2019)\citenamefont
  {Rahaman}, \citenamefont {Baratin}, \citenamefont {Arpit}, \citenamefont
  {Draxler}, \citenamefont {Lin}, \citenamefont {Hamprecht}, \citenamefont
  {Bengio},\ and\ \citenamefont {Courville}}]{rahaman2019spectral}%
  \BibitemOpen
  \bibfield  {author} {\bibinfo {author} {\bibfnamefont {N.}~\bibnamefont
  {Rahaman}}, \bibinfo {author} {\bibfnamefont {A.}~\bibnamefont {Baratin}},
  \bibinfo {author} {\bibfnamefont {D.}~\bibnamefont {Arpit}}, \bibinfo
  {author} {\bibfnamefont {F.}~\bibnamefont {Draxler}}, \bibinfo {author}
  {\bibfnamefont {M.}~\bibnamefont {Lin}}, \bibinfo {author} {\bibfnamefont
  {F.}~\bibnamefont {Hamprecht}}, \bibinfo {author} {\bibfnamefont
  {Y.}~\bibnamefont {Bengio}},\ and\ \bibinfo {author} {\bibfnamefont
  {A.}~\bibnamefont {Courville}},\ }\bibfield  {title} {\bibinfo {title} {On
  the spectral bias of neural networks},\ }in\ \href@noop {} {\emph {\bibinfo
  {booktitle} {International Conference on Machine Learning}}}\ (\bibinfo
  {organization} {PMLR},\ \bibinfo {year} {2019})\ pp.\ \bibinfo {pages}
  {5301--5310}\BibitemShut {NoStop}%
\bibitem [{\citenamefont {Ronen}\ \emph {et~al.}(2019)\citenamefont {Ronen},
  \citenamefont {Jacobs}, \citenamefont {Kasten},\ and\ \citenamefont
  {Kritchman}}]{ronen2019convergence}%
  \BibitemOpen
  \bibfield  {author} {\bibinfo {author} {\bibfnamefont {B.}~\bibnamefont
  {Ronen}}, \bibinfo {author} {\bibfnamefont {D.}~\bibnamefont {Jacobs}},
  \bibinfo {author} {\bibfnamefont {Y.}~\bibnamefont {Kasten}},\ and\ \bibinfo
  {author} {\bibfnamefont {S.}~\bibnamefont {Kritchman}},\ }\bibfield  {title}
  {\bibinfo {title} {The convergence rate of neural networks for learned
  functions of different frequencies},\ }\href@noop {} {\bibfield  {journal}
  {\bibinfo  {journal} {Advances in Neural Information Processing Systems}\
  }\textbf {\bibinfo {volume} {32}} (\bibinfo {year} {2019})}\BibitemShut
  {NoStop}%
\bibitem [{\citenamefont {Xu}\ \emph {et~al.}(2019)\citenamefont {Xu},
  \citenamefont {Zhang},\ and\ \citenamefont {Xiao}}]{xu2019training}%
  \BibitemOpen
  \bibfield  {author} {\bibinfo {author} {\bibfnamefont {Z.-Q.~J.}\
  \bibnamefont {Xu}}, \bibinfo {author} {\bibfnamefont {Y.}~\bibnamefont
  {Zhang}},\ and\ \bibinfo {author} {\bibfnamefont {Y.}~\bibnamefont {Xiao}},\
  }\bibfield  {title} {\bibinfo {title} {Training behavior of deep neural
  network in frequency domain},\ }in\ \href@noop {} {\emph {\bibinfo
  {booktitle} {Neural Information Processing: 26th International Conference,
  ICONIP 2019, Sydney, NSW, Australia, December 12--15, 2019, Proceedings, Part
  I 26}}}\ (\bibinfo {organization} {Springer},\ \bibinfo {year} {2019})\ pp.\
  \bibinfo {pages} {264--274}\BibitemShut {NoStop}%
\bibitem [{Note7()}]{Note7}%
  \BibitemOpen
  \bibinfo {note} {For efficiency, an adjoint sensitivity method could also be
  used~\cite {chen2018neural}}\BibitemShut {NoStop}%
\bibitem [{\citenamefont {Chen}\ \emph
  {et~al.}(2018{\natexlab{a}})\citenamefont {Chen}, \citenamefont
  {Pennington},\ and\ \citenamefont {Schoenholz}}]{chen2018dynamical}%
  \BibitemOpen
  \bibfield  {author} {\bibinfo {author} {\bibfnamefont {M.}~\bibnamefont
  {Chen}}, \bibinfo {author} {\bibfnamefont {J.}~\bibnamefont {Pennington}},\
  and\ \bibinfo {author} {\bibfnamefont {S.}~\bibnamefont {Schoenholz}},\
  }\bibfield  {title} {\bibinfo {title} {Dynamical isometry and a mean field
  theory of rnns: Gating enables signal propagation in recurrent neural
  networks},\ }in\ \href@noop {} {\emph {\bibinfo {booktitle} {International
  Conference on Machine Learning}}}\ (\bibinfo {organization} {PMLR},\ \bibinfo
  {year} {2018})\ pp.\ \bibinfo {pages} {873--882}\BibitemShut {NoStop}%
\bibitem [{\citenamefont {Pennington}\ \emph {et~al.}(2017)\citenamefont
  {Pennington}, \citenamefont {Schoenholz},\ and\ \citenamefont
  {Ganguli}}]{pennington2017resurrecting}%
  \BibitemOpen
  \bibfield  {author} {\bibinfo {author} {\bibfnamefont {J.}~\bibnamefont
  {Pennington}}, \bibinfo {author} {\bibfnamefont {S.}~\bibnamefont
  {Schoenholz}},\ and\ \bibinfo {author} {\bibfnamefont {S.}~\bibnamefont
  {Ganguli}},\ }\bibfield  {title} {\bibinfo {title} {Resurrecting the sigmoid
  in deep learning through dynamical isometry: theory and practice},\
  }\href@noop {} {\bibfield  {journal} {\bibinfo  {journal} {Advances in neural
  information processing systems}\ }\textbf {\bibinfo {volume} {30}} (\bibinfo
  {year} {2017})}\BibitemShut {NoStop}%
\bibitem [{\citenamefont {Gouk}\ \emph {et~al.}(2021)\citenamefont {Gouk},
  \citenamefont {Frank}, \citenamefont {Pfahringer},\ and\ \citenamefont
  {Cree}}]{Gouk2021RegularisationContinuity}%
  \BibitemOpen
  \bibfield  {author} {\bibinfo {author} {\bibfnamefont {H.}~\bibnamefont
  {Gouk}}, \bibinfo {author} {\bibfnamefont {E.}~\bibnamefont {Frank}},
  \bibinfo {author} {\bibfnamefont {B.}~\bibnamefont {Pfahringer}},\ and\
  \bibinfo {author} {\bibfnamefont {M.~J.}\ \bibnamefont {Cree}},\ }\bibfield
  {title} {\bibinfo {title} {{Regularisation of neural networks by enforcing
  Lipschitz continuity}},\ }\href
  {https://doi.org/10.1007/S10994-020-05929-W/FIGURES/4} {\bibfield  {journal}
  {\bibinfo  {journal} {Machine Learning}\ }\textbf {\bibinfo {volume} {110}},\
  \bibinfo {pages} {393} (\bibinfo {year} {2021})}\BibitemShut {NoStop}%
\bibitem [{\citenamefont {Kipf}\ and\ \citenamefont
  {Welling}(2016)}]{kipf2016semi}%
  \BibitemOpen
  \bibfield  {author} {\bibinfo {author} {\bibfnamefont {T.~N.}\ \bibnamefont
  {Kipf}}\ and\ \bibinfo {author} {\bibfnamefont {M.}~\bibnamefont {Welling}},\
  }\bibfield  {title} {\bibinfo {title} {Semi-supervised classification with
  graph convolutional networks},\ }\href@noop {} {\bibfield  {journal}
  {\bibinfo  {journal} {arXiv preprint arXiv:1609.02907}\ } (\bibinfo {year}
  {2016})}\BibitemShut {NoStop}%
\bibitem [{\citenamefont {Hardt}\ \emph {et~al.}(2016)\citenamefont {Hardt},
  \citenamefont {Recht},\ and\ \citenamefont {Singer}}]{hardt2016train}%
  \BibitemOpen
  \bibfield  {author} {\bibinfo {author} {\bibfnamefont {M.}~\bibnamefont
  {Hardt}}, \bibinfo {author} {\bibfnamefont {B.}~\bibnamefont {Recht}},\ and\
  \bibinfo {author} {\bibfnamefont {Y.}~\bibnamefont {Singer}},\ }\bibfield
  {title} {\bibinfo {title} {Train faster, generalize better: Stability of
  stochastic gradient descent},\ }in\ \href@noop {} {\emph {\bibinfo
  {booktitle} {International conference on machine learning}}}\ (\bibinfo
  {organization} {PMLR},\ \bibinfo {year} {2016})\ pp.\ \bibinfo {pages}
  {1225--1234}\BibitemShut {NoStop}%
\bibitem [{\citenamefont {Bousquet}\ and\ \citenamefont
  {Elisseeff}(2002)}]{bousquet2002stability}%
  \BibitemOpen
  \bibfield  {author} {\bibinfo {author} {\bibfnamefont {O.}~\bibnamefont
  {Bousquet}}\ and\ \bibinfo {author} {\bibfnamefont {A.}~\bibnamefont
  {Elisseeff}},\ }\bibfield  {title} {\bibinfo {title} {Stability and
  generalization},\ }\href@noop {} {\bibfield  {journal} {\bibinfo  {journal}
  {The Journal of Machine Learning Research}\ }\textbf {\bibinfo {volume}
  {2}},\ \bibinfo {pages} {499} (\bibinfo {year} {2002})}\BibitemShut {NoStop}%
\bibitem [{\citenamefont {Xu}\ and\ \citenamefont
  {Mannor}(2012)}]{xu2012robustness}%
  \BibitemOpen
  \bibfield  {author} {\bibinfo {author} {\bibfnamefont {H.}~\bibnamefont
  {Xu}}\ and\ \bibinfo {author} {\bibfnamefont {S.}~\bibnamefont {Mannor}},\
  }\bibfield  {title} {\bibinfo {title} {Robustness and generalization},\
  }\href@noop {} {\bibfield  {journal} {\bibinfo  {journal} {Machine learning}\
  }\textbf {\bibinfo {volume} {86}},\ \bibinfo {pages} {391} (\bibinfo {year}
  {2012})}\BibitemShut {NoStop}%
\bibitem [{\citenamefont {Bartlett}\ \emph {et~al.}(2017)\citenamefont
  {Bartlett}, \citenamefont {Foster},\ and\ \citenamefont
  {Telgarsky}}]{bartlett2017spectrally}%
  \BibitemOpen
  \bibfield  {author} {\bibinfo {author} {\bibfnamefont {P.~L.}\ \bibnamefont
  {Bartlett}}, \bibinfo {author} {\bibfnamefont {D.~J.}\ \bibnamefont
  {Foster}},\ and\ \bibinfo {author} {\bibfnamefont {M.~J.}\ \bibnamefont
  {Telgarsky}},\ }\bibfield  {title} {\bibinfo {title} {Spectrally-normalized
  margin bounds for neural networks},\ }\href@noop {} {\bibfield  {journal}
  {\bibinfo  {journal} {Advances in neural information processing systems}\
  }\textbf {\bibinfo {volume} {30}} (\bibinfo {year} {2017})}\BibitemShut
  {NoStop}%
\bibitem [{\citenamefont {McAllester}(1999)}]{mcallester1999pac}%
  \BibitemOpen
  \bibfield  {author} {\bibinfo {author} {\bibfnamefont {D.~A.}\ \bibnamefont
  {McAllester}},\ }\bibfield  {title} {\bibinfo {title} {Pac-bayesian model
  averaging},\ }in\ \href@noop {} {\emph {\bibinfo {booktitle} {Proceedings of
  the twelfth annual conference on Computational learning theory}}}\ (\bibinfo
  {year} {1999})\ pp.\ \bibinfo {pages} {164--170}\BibitemShut {NoStop}%
\bibitem [{\citenamefont {Arora}\ \emph {et~al.}(2018)\citenamefont {Arora},
  \citenamefont {Ge}, \citenamefont {Neyshabur},\ and\ \citenamefont
  {Zhang}}]{arora2018stronger}%
  \BibitemOpen
  \bibfield  {author} {\bibinfo {author} {\bibfnamefont {S.}~\bibnamefont
  {Arora}}, \bibinfo {author} {\bibfnamefont {R.}~\bibnamefont {Ge}}, \bibinfo
  {author} {\bibfnamefont {B.}~\bibnamefont {Neyshabur}},\ and\ \bibinfo
  {author} {\bibfnamefont {Y.}~\bibnamefont {Zhang}},\ }\bibfield  {title}
  {\bibinfo {title} {Stronger generalization bounds for deep nets via a
  compression approach},\ }in\ \href@noop {} {\emph {\bibinfo {booktitle}
  {International Conference on Machine Learning}}}\ (\bibinfo {organization}
  {PMLR},\ \bibinfo {year} {2018})\ pp.\ \bibinfo {pages}
  {254--263}\BibitemShut {NoStop}%
\bibitem [{\citenamefont {Giryes}(2022)}]{giryes2022function}%
  \BibitemOpen
  \bibfield  {author} {\bibinfo {author} {\bibfnamefont {R.}~\bibnamefont
  {Giryes}},\ }\bibfield  {title} {\bibinfo {title} {A function space analysis
  of finite neural networks with insights from sampling theory},\ }\href@noop
  {} {\bibfield  {journal} {\bibinfo  {journal} {IEEE Transactions on Pattern
  Analysis and Machine Intelligence}\ }\textbf {\bibinfo {volume} {45}},\
  \bibinfo {pages} {27} (\bibinfo {year} {2022})}\BibitemShut {NoStop}%
\bibitem [{\citenamefont {Park}\ and\ \citenamefont
  {Newman}(2004)}]{park2004statistical}%
  \BibitemOpen
  \bibfield  {author} {\bibinfo {author} {\bibfnamefont {J.}~\bibnamefont
  {Park}}\ and\ \bibinfo {author} {\bibfnamefont {M.~E.}\ \bibnamefont
  {Newman}},\ }\bibfield  {title} {\bibinfo {title} {Statistical mechanics of
  networks},\ }\href@noop {} {\bibfield  {journal} {\bibinfo  {journal}
  {Physical Review E}\ }\textbf {\bibinfo {volume} {70}},\ \bibinfo {pages}
  {066117} (\bibinfo {year} {2004})}\BibitemShut {NoStop}%
\bibitem [{\citenamefont {Chen}\ \emph
  {et~al.}(2018{\natexlab{b}})\citenamefont {Chen}, \citenamefont {Rubanova},
  \citenamefont {Bettencourt},\ and\ \citenamefont
  {Duvenaud}}]{chen2018neural}%
  \BibitemOpen
  \bibfield  {author} {\bibinfo {author} {\bibfnamefont {R.~T.}\ \bibnamefont
  {Chen}}, \bibinfo {author} {\bibfnamefont {Y.}~\bibnamefont {Rubanova}},
  \bibinfo {author} {\bibfnamefont {J.}~\bibnamefont {Bettencourt}},\ and\
  \bibinfo {author} {\bibfnamefont {D.~K.}\ \bibnamefont {Duvenaud}},\
  }\bibfield  {title} {\bibinfo {title} {Neural ordinary differential
  equations},\ }\href@noop {} {\bibfield  {journal} {\bibinfo  {journal}
  {Advances in neural information processing systems}\ }\textbf {\bibinfo
  {volume} {31}} (\bibinfo {year} {2018}{\natexlab{b}})}\BibitemShut {NoStop}%
\end{thebibliography}
%

\end{document}


\preprint{APS/123-QED}

\title{Supplemental Information:\\ Stretched and measured neural predictions of complex network dynamics}

\author{Vaiva Vasiliauskaite}
\affiliation{Computational Social Science, ETH Z\"urich, 8092 Z\"urich, Switzerland}
\author{Nino Antulov-Fantulin}\email{anino@ethz.ch}
\affiliation{%
Computational Social Science, ETH Z\"urich \& Aisot Technologies AG, Z\"urich, Switzerland}
\date{\today}

\maketitle

\section{Neural network mappings} Now we detail each neural network's effects on an input vector $\vvec{x}\in\mathbb{R}^{n\times k}$, layer by layer.
Before propagation, we tensorize $\vvec{x}\in\mathbb{R}^{n\times k}\rightarrow \vvec{h}^1\in\mathbb{R}^{n\times1\times k}$. All dimensions are counted from $1$.

\paragraph{Self-interactions} The function $\pmb{\psi}^{\text{self}}$ performs a mapping $\reals^{n\times1\times k}\rightarrow \mathbb{R}^{n\times 1\times k}$. It consists of input, hidden, and output layers.
\textbf{The input layer} performs a mapping $\reals^{n\times 1\times k}\times_3\reals^{k\times h_1}\in \reals^{n\times 1\times h_1}$ realized through a 3-mode product of a tensor with a matrix. Overall, the layer is defined as
\begin{equation}\label{3_mode_product}
   \vvec{h}^{2} = \sigma_1(\vvec{h}^{1} \times_3 \vvec{W}_1 + \vvec{b}_1),
\end{equation}
where $\vvec{W}_1 \in \reals^{k\times h_1}$,
$\vvec{b}_1 \in \reals^{1\times h_1}$. 
 Element-wise, \eqnref{3_mode_product} is defined as
  \begin{equation*}
[\vvec{h}^{2} ]_{i,j,k}
= \sigma_1 \left( \sum_{m} {h}^{1}_{i,j,m} (W_1)_{m,k} + (b_1)_{j,k} \right).
 \end{equation*}
Note that all layers in a neural networks that constitute Eq.\ 2 of the main text compute \eqnref{3_mode_product}, with the difference in the dimensions of the input vector $\textbf{h}$, the matrix $\textbf{W}$ and the vector $\textbf{b}$ that determine the dimensions of the mapping, as well as non-linearity $\sigma$. 

Subsequent \textbf{hidden layer} has $\vvec{W}_2 \in \reals^{h_1\times h_1}$ and $\vvec{b}_2 \in \reals^{1\times h_1}$. After applying this layer, we have a vector in $\reals^{n\times 1\times h_1}$. In the current analysis, we found one hidden layer to be sufficient, but their number can be increased arbitrarily. Lastly, the \textbf{output layer} consists of $\vvec{W}_3 \in \reals^{h_1\times k}$, $\vvec{b}_3 \in \reals^{1\times k}$ and maps the vector back to dimension $\reals^{n\times 1\times k}$. In all layers, the non-linearity $\sigma=\tanh$.
     
\paragraph{Neighbor interaction} The function $\pmb{\psi}^{\text{nbr}}$ consists of three feed forward neural networks. Firstly, we apply $ \pmb{\psi}^{q_1}, \pmb{\psi}^{q_2}$ that both map the input to an embedded dimension: $\mathbb{R}^{n\times 1\times k}\rightarrow \mathbb{R}^{n\times 1\times h_2}$. These neural networks consist of two layers. The \textbf{input layer} that is parameterized with $\vvec{W}_1\in \reals^{k\times h_2}$ and $\vvec{b}_1 \in \reals^{1 \times h_2}$, $\sigma_1=\tanh$ and the \textbf{hidden layer}, parameterized with $\vvec{W}_2\in \reals^{h_2\times h_2}$, $\vvec{b}_2 \in \reals^{1 \times h_2}$ and identity function as $\sigma_2$. 

To combine the functions $\pmb{\psi}^{q_1} $ and $ \pmb{\psi}^{q_2}$, we first transpose the outputs of these functions. The first transpose is defined as $ \pmb{\psi}^{q_1}(\vvec{x})\in  \reals^{n\times 1\times h_2}\rightarrow \mathbb{R}^{h_2\times n\times 1}$ and is realized through two transpose operations: $ \reals^{n\times 1\times h_2} \rightarrow_{1,3}  \reals^{h_2 \times 1 \times n} \rightarrow_{2,3} \reals^{h_2\times n \times 1}$ denoted collectively as ``$\top_1$''. The output of $ \pmb{\psi}^{q_2}$ is transposed $ \reals^{n\times 1\times h_2} \rightarrow_{1,3} \reals^{h_2\times 1 \times n}$, realized through an operation denoted as ``${\top_{2}}$''. We then perform a ``batched" matrix-matrix product of the transposed outputs, i.e.\ $\mathbb{R}^{h_2\times n\times 1}\times_b \mathbb{R}^{h_2\times 1\times n}\in \reals^{h_2\times n \times n}$. 
This operation element-wise is defined as 
\begin{eqnarray*}
    \left[\pmb{\psi}^{q_{1,2}}(\vvec{x})
    \right]_{i,j,k}= 
    \sum_{m}{\pmb{\psi}^{q_1}(\vvec{x})^{\top_{1}} }_{i,j,m}{\pmb{\psi}^{q_2}(\vvec{x})^{\top_{2}}}_{i,m,k}.
\end{eqnarray*}
The result of this product is an approximation of interactions of all pairs of nodes in the network. To filter out the non-neighbor interactions, we compute 
$\pmb{\Phi} \odot  \pmb{\psi}^{q_{1,2}}$ that maps $\mathbb{R}^{h_2\times n\times n} \rightarrow \reals^{h_2\times n \times n}$. Here the operator $\odot$ denotes a standard ``broadcasted" element-wise multiplication. 
Element-wise this multiplication is defined
\begin{equation*}
    \big[\pmb{\Phi}\odot\pmb{\psi}^{q_{1,2}}(\vvec{x})\big]_{i,j,k} = \Phi_{j,k}\left[ \pmb{\psi}^{q_{1,2}}(\vvec{x}).\right]_{i,j,k}
\end{equation*}

Lastly, to return to the original dimension, we apply an invariant pooling layer $\pmb{\phi}^{\mysum}$:
\begin{equation*}\small
    \pmb{\psi}^{\mysum}(\pmb{\Phi}\odot\pmb{\psi}^{q_{1,2}}(\vvec{x}))=\left[\sum_{j}\big[\pmb{\Phi}\odot\pmb{\psi}^{q_{1,2}}(\vvec{x})\big]_{i,j,k}\right]^{\top_3}\times_3 \mathbf{W}_3+\mathbf{b}_3.
\end{equation*}
Here $\top_3$ denotes a transpose $\reals^{h_2\times 1 \times n} \rightarrow_{1,3} \reals^{n \times 1 \times h_2}$, $\mathbf{W}_3\in \reals^{h_2\times k},\mathbf{b}_3\in \reals^{1\times k} $.

Note that we assumed that the function $\textbf{Q}$ can be approximated by a product of two neural network functions. It is a reasonable assumption for factorizable $\textbf{Q}$, i.e.\ $\textbf{Q}(\vvec{x}_1,\vvec{x}_2) = \textbf{q}(\vvec{x}_1) \textbf{q}(\vvec{x}_2)$. Most of the dynamics we will discuss are indeed factorizable. One exception is heat diffusion on graphs, which is factorizable only for homogeneous degree distributions. Nevertheless, our simulations show the described neural network model can also approximate Heat dynamics. More generally, one may consider a neural network of the following form
\begin{eqnarray}
\dot{\vvec{x}}_i
&=& \pmb{\psi}^{\text{self}}(\vvec{x}_i)+ \pmb{\psi}^{\mysum}\left( \sum_{j : A_{ij}\neq 0} \pmb{\psi}^{Q}(\vvec{x}_i, \vvec{x}_j) \right), 
\end{eqnarray}
where $ \pmb{\psi}^{Q}$ acts on edges, and the $2^\text{nd}$ term overall is a universal approximation of functions on edge sets~\cite{wagstaff2022universal,Xu2018HowNetworks}, while the $1^\text{st}$ term is as before.

A GNN can be further constrained to achieve physical realism of a model. For example, if the system is closed, it does not exchange energy or mass with the environment, therefore $\sum_i \dot{x}_i(t) = C \quad \forall t$ (assuming $k=1$). A neural network can be weakly or strongly constrained to adhere to such conservation of a derivative. 

If Lipschitz continuity of $\Fbcal$ is known, it can be used to regularize and potentially improve the neural approximation $\pmb{\Psi}$~\cite{Gouk2021RegularisationContinuity}. 
 In order to guarantee the local existence and uniqueness of the solution to the initial value problem $\pmb{\psi}^{\text{self}}(\vvec{x})+ \pmb{\psi}^{\text{nbr}}(\vvec{x})$ for different $(\vvec{x}_0)$, the learned vector field $\Fbcal$ has to be Lipschitz continuous (see Picard–Lindelöf theorem~\cite{coddington1956theory}).
To enforce Lipschitz continuity of $\pmb{\Psi}$, we are using 1-Lipschitz activation function \textsc{tanh}~\cite{Gouk2021RegularisationContinuity}.

\section{Two dimensional vector field}\label{sec:small}

\begin{figure}[!ht]
    \centering
    \includegraphics[width=1\linewidth]{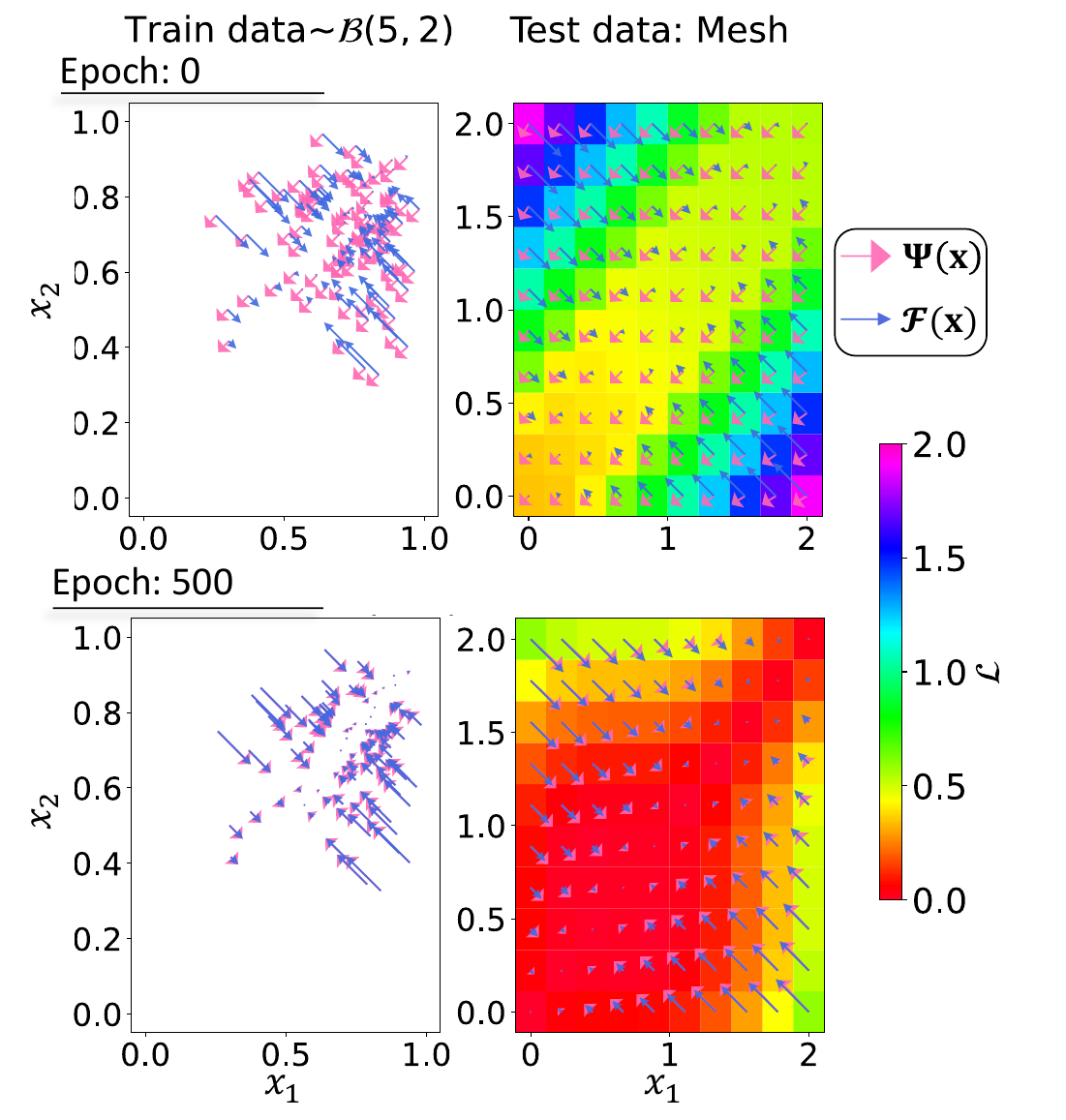}
    \caption{Learnt (pink arrows) and true (purple arrows) vector field of diffusion dynamics on a connected $n=2$ node graph. The training sample are taken from a beta distribution $\mathcal{B}(5,2)$ and tested on a uniform lattice defined within range $[0,2]$. Color shows normalized $\ell^1$ loss at a given square, calculated for an arrow that originates within this square. }
    \label{fig_diffusion}
\end{figure}

To test our intuition about the neural network model, we first consider dynamics in the simplest possible network: heat diffusion on a connected network of $n=2$ nodes. Both datasets include $100$ samples, the training set is sampled from a beta distribution $\psi_X\sim\mathcal{B}(5,2)$, whereas the test set is composed of a uniformly spaced set of points of a two-dimensional lattice in the region $[0,2]$. As \figref{fig_diffusion} shows, after $500$ training epochs, the neural network approximates the state space well within the region of support of the beta distribution ($[0,1]$) even when $\psi_X\not\equiv \omega_X$. Furthermore, we observe a monotonic increase in loss as a function of distance from the training support.

\section{Simulation details}\label{sec:hierarchy}

\textbf{Dynamics} The following parameters were used to simulate the dynamics. 
\begin{enumerate}
    \item \textbf{Heat}: $B = 0.5$.
    \item \textbf{MAK}: $B = 0.1, R = 1, F = 0.5, b= 3$. 
    \item \textbf{PD}:  $B=2, R=0.3, a=1.5, b= 3$. 
    \item \textbf{MM}: $B=4, R=0.5, h=3$. 
    \item \textbf{SIS}: $B=4, R = 0.5$. 
    \item \textbf{FHN}: $\rho=1, a = 0.28, b=0.5, c= -0.04$.
\end{enumerate}

\textbf{Loss} In all cases, during training we optimize an $\ell_1$ loss function, as defined in Eq.\ 4 of the main text. In all cases, unless otherwise stated, $\lambda=1$.

\textbf{Integration} For numerical integration we used an explicit Runge-Kutta 5(4) method~\cite{10.2307/2008219}. Implemented in \textsc{torchdiffeq} python package~\cite{torchdiffeq}.

\paragraph*{Figure 1:} Neural networks (\textbf{a} our model and \textbf{b} SAGEConv~\cite{hamilton2017inductive}) were trained using $200$ training samples with $x_1,x_2$ sampled from a beta distribution $\mathcal{B}(5,2)$. Both models were trained in $1000$ epochs at a learning rate of $0.005$ with no weight decay. The strength of the regularizer in the loss function was set to $\lambda=0.1$. The architectures are described in \secref{sec:gcn_comp}.

\paragraph*{Figure 2:} 10 overparameterized feedforward neural networks were trained with 251 parameters ($h=10$) each with 4 fully connected layers and Tanh activations. Each network was trained in 3000 epochs, using 40 bootstrapped samples from 50 total number of samples, using the learning rate of $0.003$ and no weight decay. Samples were taken from a uniform distribution $\mathcal{U}[-2,2]$. The target function is $\Fcal(x)=\cos(2x)+\mathcal{N}(0, \sigma=0.01)$. 

\paragraph*{Table 1, Figures 3,4:} Each neural network was trained for $2000$ epochs at a learning rate of $0.001$ and the weight decay of $0.001$. We used a total of $5000$ training samples, where $\vvec{x}$ was sampled from a uniform distribution $\mathcal{U}(0,1)$. The batch size was set to $100$. The hidden layers have dimensions $h_1  = h_2  = 30$. Both $\pmb{\psi}^{\text{self}}$ and $\pmb{\psi}^{\text{nbr}}$ contain one hidden layer. After $1000$ training epochs, we also used \href{https://pytorch.org/docs/stable/generated/torch.optim.lr_scheduler.ReduceLROnPlateau.html}{\textsc{ReduceOnPlateau} scheduler} with the patience parameter set to $50$ and the cooldown parameter set to $10$.

The reported values are an average and a standard deviation across $10^3$ samples. For the two right-most columns, we sampled a new ER graph with $n=100,p=0.1$ and $n=100,p=0.6$. In Fig.\ 3, the results for each parameter value of the beta distribution ($a$ and $b$) are also averaged over $10^3$ independent samples. The heatmap shows the range $[0.1,12]$. In Fig.\ 4a) and c), the reported results are an average and standard deviation across $10$ graphs at each $p$ (in a)) or $n$ (in c)), where for each graph we computed the mean loss over $100$ input samples from $\mathcal{U}(0,1)$. In Fig.\ 4b), the results are an average and standard deviation across $1000$ samples from $\mathcal{U}(0,1)+ \Delta$.

\paragraph*{Figure 5:} For each dynamics, a total of $50$ neural networks were trained, each with $90\%$ of data. The training data were generated by solving five initial value problems (IVPs). In each problem, the initial state was sampled from a uniform distribution $\mathcal{U}(0,1)$. Each initial value was integrated over the time interval $t\in [0,2]$ with a step size $\text{d}t = 0.0101$, totalling to $995$ samples in a training dataset. To generate labels, $\Fbcal$ was evaluated analytically for each $\vvec{x}(t)$. The models were trained for $2000$ epochs at a learning rate of $0.005$ and the weight decay of $0.001$. The batch size was set to $50$. The hidden layers have dimensions $h_1  = h_2  = 30$. $\pmb{\psi}^{\text{self}}$, $\pmb{\psi}^{{q_1}}$ and $\pmb{\psi}^{{q_2}}$ all contain one hidden layer. After $1000$ training epochs, we also used \href{https://pytorch.org/docs/stable/generated/torch.optim.lr_scheduler.ReduceLROnPlateau.html}{\textsc{ReduceOnPlateau} scheduler} with the patience parameter set to $50$ and the cooldown parameter set to $10$.

\textbf{Fig.\ 5a)} The test graph is equivalent to the training graph. For a given $\Delta$ value, $10$ independent time series trajectories were computed, starting from an initial value sampled from a shifted uniform distribution $\mathcal{U}(0,1)+ \Delta$. For each initial value, a neural network prediction was computed by taking a random neural network from an ensemble and using it for integration. $100$ samples $\vvec{x}(t)$ from the predicted trajectory were then used to compute the distribution of the $d$-statistic. The $d$-statistic for the full sample (the null distribution) was evaluated by taking $100$ samples from the training data. $m=20$ subsample of $M=50$ neural networks were taken for each sample. We used a statistical significance level of $\alpha=0.05$ to compute the number of accepted datapoints by comparing the null distribution to the test distribution following Eq.\ 4 of the main text. The orange circles show the average and the standard deviation for accepted datapoints across the samples. The purple circles report the loss, computed for each trajectory, averaged over trajectories. 

\textbf{Fig.\ 5b)} We computed the $d$-statistic of the full sample and of the test sample in the same way, however now using $1000$ draws for estimating the statistics. The figures show predicted trajectories computed using a random neural network from the ensemble. The title reports the fraction of datapoints that passed the significance test. Here the test graph is not equivalent to the training graph, but is sampled from and Erd\"os-R\'enyi ensemble with $n=15, p=0.3$.

\paragraph*{Figure 6:} A total of $20$ neural networks were trained, each with $90\%$. To increase the expressivity of the model, $\pmb{\psi}^{\text{self}}$ has $2$ hidden layers with $h_1=30$, whereas $\pmb{\psi}^{q_1},\pmb{\psi}^{q_2}$ have $3$ hidden layers each with $h_2=30$. Note that FHN dynamics is not fully described by Eq.1 of the main text, since the interaction term is proportional to node's degree. Therefore we use a degree-scaled undirected adjacency matrix $\tilde{\mathbf{A}}=\frac{\mathbf{A}}{\textbf{k}}$ as an input to a neural network. Here $\textbf{k}$ is a vector that contains nodes' degrees. 

The training data were generated by solving one initial value problem. The initial state was sampled from $\mathcal{U}(0,1)$ and integrated over the time interval $t\in[0,50]$ with a step size $\text{d}t= 0.01$, totalling to $5000$ training samples. To generate labels, $\Fbcal$ was evaluated numerically via Newton's difference quotient: $\vvec{y}(t)=\frac{\vvec{x}(t+\text{d}t)-\vvec{x}(t)}{\text{d}t}$. The models were trained for $1000$ epochs at a learning rate of $0.005$ and weight decay of $0$. The batch size was set to $200$ samples.

For testing results presented in the figure, we sampled a new initial value from a uniform distribution and compared the ground truth time series trajectories to trajectories predicted by one of the trained models. The initial values were integrated over the time interval $t\in[0,100]$ with a step size $\text{d}t= 0.01$. 

We then zoomed in to predicted dynamics for two nodes, namely node $i=0$, for which the prediction was highly accurate, and node $i=137$, for which the prediction during testing was inaccurate. To compute the null distribution of the $d$-statistic for each of these nodes, we took $1000$ samples $\textbf{x}_i(t)$ from the training data computed $d=\text{Var}(\pmb{\Psi}_m(\textbf{x}(t))_i)$ for each dimension $k$ across $m=5$ randomly selected models for each sample. To evaluate the prediction accuracy, we similarly computed the distribution of the $d$-statistic using the test data. Note that since $k=2$ for the Fitzhugh-Nagumo dynamics, the $d$-statistic was computed for each dimension $k$ independently.   

\paragraph*{Figure 7 b):} A neural network was trained for $2000$ epochs at a learning rate of $0.0001$ and a weight decay $0.001$. The batch size was set to $50$. The hidden layers have dimensions $h_1=h_2=30$. $\pmb{\psi}^{\text{self}}$, $\pmb{\psi}^{{q_1}}$ and $\pmb{\psi}^{{q_2}}$ all contain one hidden layer. 

The training data were generated by solving one initial value problem of Heat Diffusion dynamics with $B=1.5$ on an Erd\"os-R\'enyi graph with $n=10$ and $p=0.3$. The initial state was sampled from $\mathcal{U}(0,1)$ and integrated over the time interval $t\in[0,1]$. To generate a time-varying step size, we subsampled $100$ time stamps from a list $(0,\Delta t, 2\Delta t,...,1)$, where $\Delta t = 0.0001$, therefore $\langle t_r\rangle\approx 0.01$. The time series were then obtained by integrating $\Fbcal$ using the time-varying timesteps $t_r$. We then added 
normally distributed noise with $\sigma=0.01$ to the time series to obtain $\vvec{z}(t_r)$.

To test the effect of noise, we trained another neural network using the same data without observational noise. We then considered a test loss discussed in the main text, which we evaluated by drawing $1000$ samples $\vvec{x}$ from a uniform distribution. Lastly, we also compared the observed losses for these two cases to the loss, observed in one untrained neural network.

\section{Comparison with other graph neural networks}\label{sec:gcn_comp}
\begin{table*}[!ht]
\centering
\begin{threeparttable}\footnotesize
  \caption{This table presents the performance of various GNN models with multiple graph convolution layers under different testing settings. The $|\pmb{\Psi}|$ column indicates the number of trainable parameters. The accuracy is defined as a percent relative error, defined in Eq.\ 7 of the main text. }
  \label{gcn_compare_multi}
  \begin{tabular}{l c c c c}
    \toprule
    \textbf{Model} & 
    \begin{tabular}{@{}c@{}}$|\pmb{\Psi}|$\end{tabular} 
                
    & \begin{tabular}{@{}c@{}}Train data \\ $\vvec{x}\sim\mathcal{U}(0,1)$ \\ 
                $\mathcal{G}\equiv \mathcal{G}_{\text{train}}$
                \end{tabular} 
                & \begin{tabular}{@{}c@{}}Test data \\ $\vvec{x}\sim\mathcal{U}(0,1)+0.5$ \\ 
                $\mathcal{G}\equiv \mathcal{G}_{\text{train}}$\end{tabular} 
                & \begin{tabular}{@{}c@{}} Test data \\$\vvec{x}\sim\mathcal{U}(0,1)$ \\ $\mathcal{G}\sim \mathcal{P}_{\text{test}}(\mathfrak{G})$\end{tabular} 
                \\
    \midrule
    \midrule
    \multicolumn{5}{c}{\textbf{Heat}} \\
    \midrule
GNVF  &  242 & $ 3.69 \pm 0.56 $ & $ 6.75 \pm 0.46 $ & $ 3.45 \pm 0.44 $\\
SAGEConv  &  3781 & $ 22.87 \pm 1.2 $ & $ 43.34 \pm 1.6 $ & $ 59.39 \pm 0.48 $\\
GraphConv  &  3781 & $ 7.48 \pm 0.82 $ & $ 54.76 \pm 5.18 $ & $ 60.51 \pm 12.45 $\\
ResGatedGraphConv  &  7711 & $ 2.99 \pm 0.52 $ & $ 16.85 \pm 2.19 $ & $ 24.28 \pm 1.65 $\\
GATConv  &  2131 & $ 91.43 \pm 5.55 $ & $ 100.34 \pm 0.7 $ & $ 99.68 \pm 2.57 $\\
ChebConv  &  18421 & $ 3.16 \pm 0.54 $ & $ 12.88 \pm 1.68 $ & $ 56.84 \pm 0.23 $\\
 \hline
    \midrule
    \multicolumn{5}{c}{\textbf{MAK}} \\
    \midrule
GNVF  &  242 & $ 3.85 \pm 0.53 $ & $ 11.17 \pm 0.99 $ & $ 3.11 \pm 0.37 $\\
SAGEConv  &  3781 & $ 27.68 \pm 1.79 $ & $ 59.0 \pm 2.53 $ & $ 64.08 \pm 0.73 $\\
GraphConv  &  3781 & $ 8.36 \pm 1.17 $ & $ 33.24 \pm 2.48 $ & $ 49.62 \pm 2.47 $\\
ResGatedGraphConv  &  7711 & $ 2.0 \pm 0.32 $ & $ 46.06 \pm 3.05 $ & $ 34.31 \pm 1.21 $\\
GATConv  &  2131 & $ 58.04 \pm 5.15 $ & $ 61.39 \pm 2.4 $ & $ 76.21 \pm 4.26 $\\
ChebConv  &  18421 & $ 3.71 \pm 0.82 $ & $ 55.33 \pm 2.73 $ & $ 64.42 \pm 0.56 $\\ \midrule
    \midrule
    \multicolumn{5}{c}{\textbf{MM}} \\
    \midrule
GNVF  &  242 & $ 1.75 \pm 0.31 $ & $ 15.7 \pm 1.89 $ & $ 5.3 \pm 1.01 $\\
SAGEConv  &  3781 & $ 12.86 \pm 1.09 $ & $ 34.54 \pm 1.91 $ & $ 109.46 \pm 11.29 $\\
GraphConv  &  3781 & $ 3.84 \pm 0.73 $ & $ 27.63 \pm 1.24 $ & $ 57.98 \pm 9.13 $\\
ResGatedGraphConv  &  7711 & $ 1.52 \pm 0.17 $ & $ 17.74 \pm 1.33 $ & $ 68.91 \pm 6.46 $\\
GATConv  &  2131 & $ 64.19 \pm 5.34 $ & $ 60.78 \pm 3.04 $ & $ 158.75 \pm 20.38 $\\
ChebConv  &  18421 & $ 2.43 \pm 0.39 $ & $ 13.15 \pm 1.18 $ & $ 103.73 \pm 10.1 $\\
\midrule
    \midrule
    \multicolumn{5}{c}{\textbf{PD}} \\
    \midrule
GNVF  &  242 & $ 5.5 \pm 0.44 $ & $ 28.77 \pm 2.39 $ & $ 9.46 \pm 0.55 $\\
SAGEConv  &  3781 & $ 30.99 \pm 1.89 $ & $ 61.62 \pm 3.11 $ & $ 71.56 \pm 0.85 $\\
GraphConv  &  3781 & $ 6.27 \pm 0.88 $ & $ 63.42 \pm 4.32 $ & $ 32.27 \pm 2.77 $\\
ResGatedGraphConv  &  7711 & $ 3.19 \pm 0.64 $ & $ 54.28 \pm 4.12 $ & $ 60.28 \pm 1.46 $\\
GATConv  &  2131 & $ 48.49 \pm 3.73 $ & $ 103.24 \pm 1.43 $ & $ 70.56 \pm 2.05 $\\
ChebConv  &  18421 & $ 2.9 \pm 0.79 $ & $ 53.78 \pm 3.48 $ & $ 72.75 \pm 0.91 $\\
\midrule
    \midrule
    \multicolumn{5}{c}{\textbf{SIS}} \\
    \midrule
GNVF  &  242 & $ 0.65 \pm 0.07 $ & $ 15.21 \pm 1.74 $ & $ 0.88 \pm 0.1 $\\
SAGEConv  &  3781 & $ 25.59 \pm 1.49 $ & $ 45.91 \pm 2.73 $ & $ 61.68 \pm 0.9 $\\
GraphConv  &  3781 & $ 6.55 \pm 0.72 $ & $ 70.91 \pm 4.7 $ & $ 50.21 \pm 1.71 $\\
ResGatedGraphConv  &  7711 & $ 2.36 \pm 0.51 $ & $ 53.89 \pm 4.99 $ & $ 27.01 \pm 2.38 $\\
GATConv  &  2131 & $ 73.08 \pm 4.52 $ & $ 130.39 \pm 5.78 $ & $ 74.54 \pm 1.82 $\\
ChebConv  &  18421 & $ 2.46 \pm 0.63 $ & $ 86.15 \pm 8.32 $ & $ 62.52 \pm 0.21 $\\
\bottomrule
  \end{tabular}
\end{threeparttable}
\end{table*}

In this section, we conduct a comparative analysis of several graph neural networks (GNNs) and their effectiveness in approximating complex network dynamics, discussed throughout the paper. Specifically, we examined five well-known GNNs: \textsc{SAGEConv}~\cite{hamilton2017inductive}, \textsc{ChebConv}~\cite{defferrard2016convolutional}, \textsc{GraphConv}~\cite{morris2019weisfeiler}, \textsc{ResGatedGraphConv}~\cite{bresson2017residual}, and \textsc{GATConv}~\cite{velivckovic2017graph}, where the naming conventions are from PyTorch Geometric, which was employed for their implementation. We train all neural networks, including ours (denoted \textsc{GNVF} -- ``Graph Neural Vector Field''), using the same training setting. The graph used for training is sampled from an Erd\"os-R\'enyi ensemble with parameters $p=0.5$ and $n=10$. Training was done in $1000$ epochs using Adam optimizer with a learning rate of $0.005$ and no weight decay. The training data consists of $200$ samples from a uniform distribution $\mathcal{U}(0,1)$, whereas test data contains $500$ samples taken from the same distribution, or a shifted uniform distribution. 

Each graph neural network has 3 graph convolution layers: the first graph layer maps from original input's dimension $k$ to a hidden dimension $h=30$, the next two layers map from the hidden dimension to the same hidden dimension. Lastly, we apply a linear decoding layer to revert back to the original input's dimension. We also reduced the size of the GNVF, by removing the hidden layers in $\pmb{\psi}^{\text{self}}$, $\pmb{\psi}^{{q_1}}$ and $\pmb{\psi}^{{q_2}}$.

In \tabref{gcn_compare_multi}, we present the in-sample error and out-of-sample error as a relative percent error defined in Eq.\ 7 of the main text under two distinct conditions: firstly, when the test data is sampled from a partially unobserved range (specifically, $\vvec{x}\sim\mathcal{U}(0,1) + 0.5$), and secondly, when the training graph is replaced with a new graph sampled from a different ensemble, namely an Erd\"os-R\'enyi graph with parameters $p=0.2$ and $n=120$. Comparing our model to other architectures, we observe a substantially smaller error both in-sample and out-of-sample. Furthermore, our proposed model shows significantly better performance with novel test graphs compared to other GNNs, suggesting that these models may impose other types of biases to the system thereby causing poorer generalization. Amongst GNN architectures we observe that those which distinguish between self- and neighbor- interactions --- specifically, \textsc{GraphConv}, \textsc{ResGatedGraphConv}, and our model detailed in the main text --- demonstrate superior performance compared to those that do not, such as \textsc{SAGEConv}, \textsc{ChebConv}, and \textsc{GATConv}.

If we removed all but one graph convolution layer and enforced the spatio-temporal locality constraint that we discussed in the main text, we observe in \tabref{gcn_compare_single} that, somewhat counter-intuitively, the out-of-sample loss is improved in almost all cases. 

\begin{table*}[!ht]
\centering
\begin{threeparttable}\footnotesize
  \caption{This table presents the performance of various GNN models with a single graph convolution layer under different testing settings. The $|\pmb{\Psi}|$ column indicates the number of trainable parameters. The accuracy is defined as a percent relative error, defined in Eq.\ 7 of the main text. }
  \label{gcn_compare_single}
  \begin{tabular}{l c c c c c}
    \toprule
    \textbf{Model} & 
    \begin{tabular}{@{}c@{}}$|\pmb{\Psi}|$\end{tabular} 
                
    & \begin{tabular}{@{}c@{}}Train data \\ $\vvec{x}\sim\mathcal{U}(0,1)$ \\ 
                $\mathcal{G}\equiv \mathcal{H}$\end{tabular} 
                & \begin{tabular}{@{}c@{}}Test data \\ $\vvec{x}\sim\mathcal{U}(0,1)+0.5$ \\ 
                $\mathcal{G}\equiv \mathcal{H}$\end{tabular} 
                & \begin{tabular}{@{}c@{}} Test data \\$\vvec{x}\sim\mathcal{U}(0,1)$ \\ $\mathcal{H}\sim \mathcal{P}(\mathfrak{H})$\end{tabular} 
                \\
    \midrule
    \midrule
    \multicolumn{5}{c}{\textbf{Heat}} \\
    \midrule
SAGEConv  &  121 & $ 23.59 \pm 1.04 $ & $ 25.56 \pm 1.06 $ & $ 61.64 \pm 0.37 $\\
GraphConv  &  121 & $ 24.24 \pm 2.4 $ & $ 186.26 \pm 18.61 $ & $ 59.49 \pm 3.19 $\\
ResGatedGraphConv  &  271 & $ 1.73 \pm 0.16 $ & $ 18.83 \pm 1.57 $ & $ 31.89 \pm 1.09 $\\
GATConv  &  151 & $ 73.89 \pm 6.95 $ & $ 126.18 \pm 4.44 $ & $ 79.01 \pm 2.1 $\\
ChebConv  &  361 & $ 13.39 \pm 4.47 $ & $ 251.08 \pm 26.02 $ & $ 59.18 \pm 3.75 $\\
 \hline
    \midrule
    \multicolumn{5}{c}{\textbf{MAK}} \\
    \midrule
SAGEConv  &  121 & $ 28.36 \pm 1.69 $ & $ 36.23 \pm 1.69 $ & $ 63.81 \pm 0.54 $\\
GraphConv  &  121 & $ 0.87 \pm 0.22 $ & $ 19.34 \pm 2.32 $ & $ 17.41 \pm 2.89 $\\
ResGatedGraphConv  &  271 & $ 0.91 \pm 0.19 $ & $ 26.11 \pm 2.82 $ & $ 17.52 \pm 2.43 $\\
GATConv  &  151 & $ 59.71 \pm 5.35 $ & $ 42.46 \pm 1.64 $ & $ 74.68 \pm 1.69 $\\
ChebConv  &  361 & $ 8.69 \pm 1.96 $ & $ 20.77 \pm 1.35 $ & $ 65.01 \pm 2.31 $\\
 \midrule
    \midrule
    \multicolumn{5}{c}{\textbf{MM}} \\
    \midrule
SAGEConv  &  121 & $ 13.82 \pm 1.0 $ & $ 32.67 \pm 1.64 $ & $ 119.63 \pm 11.92 $\\
GraphConv  &  121 & $ 5.76 \pm 0.56 $ & $ 22.62 \pm 1.8 $ & $ 27.07 \pm 5.79 $\\
ResGatedGraphConv  &  271 & $ 0.82 \pm 0.14 $ & $ 8.98 \pm 0.86 $ & $ 7.52 \pm 1.33 $\\
GATConv  &  151 & $ 71.6 \pm 5.02 $ & $ 67.36 \pm 2.9 $ & $ 132.3 \pm 7.86 $\\
ChebConv  &  361 & $ 5.52 \pm 0.68 $ & $ 25.84 \pm 1.22 $ & $ 117.72 \pm 10.48 $\\
\midrule
    \midrule
    \multicolumn{5}{c}{\textbf{PD}} \\
    \midrule
SAGEConv  &  121 & $ 32.97 \pm 1.67 $ & $ 53.46 \pm 2.13 $ & $ 73.81 \pm 1.1 $\\
GraphConv  &  121 & $ 6.31 \pm 0.64 $ & $ 29.93 \pm 1.79 $ & $ 8.62 \pm 1.41 $\\
ResGatedGraphConv  &  271 & $ 1.86 \pm 0.25 $ & $ 40.33 \pm 3.08 $ & $ 10.65 \pm 1.08 $\\
GATConv  &  151 & $ 66.95 \pm 4.02 $ & $ 97.08 \pm 4.47 $ & $ 68.92 \pm 1.27 $\\
ChebConv  &  361 & $ 10.65 \pm 1.48 $ & $ 46.97 \pm 3.23 $ & $ 70.85 \pm 1.39 $\\
\midrule
    \midrule
    \multicolumn{5}{c}{\textbf{SIS}} \\
    \midrule
SAGEConv  &  121 & $ 27.47 \pm 1.43 $ & $ 68.36 \pm 5.59 $ & $ 64.33 \pm 0.72 $\\
GraphConv  &  121 & $ 1.02 \pm 0.36 $ & $ 46.54 \pm 5.59 $ & $ 23.51 \pm 3.54 $\\
ResGatedGraphConv  &  271 & $ 1.27 \pm 0.39 $ & $ 43.01 \pm 4.4 $ & $ 21.15 \pm 2.64 $\\
GATConv  &  151 & $ 72.61 \pm 4.66 $ & $ 138.6 \pm 6.46 $ & $ 74.49 \pm 1.79 $\\
ChebConv  &  361 & $ 7.28 \pm 1.92 $ & $ 134.55 \pm 9.6 $ & $ 61.58 \pm 1.91 $\\
    \bottomrule
  \end{tabular}
\end{threeparttable}
\end{table*}

\section{Derivation of the Computational Graph and Gradient Computation in a Neural ODE model}\label{sec:realdata} 

In the example discussed in Sec.\ VI of the main text, we use a neural ODE~\cite{chen2018neural} model realized through a forward computational graph with an automatic differentiation~\cite{baydin2018automatic}. Here we derive the computational graph and its gradient computation for the custom loss function Eq.\ 8 of the main text.

For simplicity, let us assume a scalar variable, i.e.\ $d=1$ and that its ground truth trajectory is $\{ x_{r} \}$, where index $r$ denotes specific times $\{t_0, t_1,...,t_R\}$ at some irregular time samplings. We also assume that observations are corrupted with additive noise: $z[x_r]=x_r+\epsilon_r$, where $\epsilon_r\sim \mathcal{N}(0,\sigma)$. A neural network represents a scalar function ${\Psi}:\mathbb{R}^1 \rightarrow \mathbb{R}^1 $ and is parameterized with weights ${w}$. The goal of learning is finding parameters ${w}$ that minimize the following objective:
\begin{eqnarray}
    \mathcal{L} &=& \sum_{r=0}^{R} \mathcal{L}_r\left( z[x_{r}] , \hat{x}_r \right) \\ \nonumber
 &=& \sum_{r=0}^{R} \mathcal{L}_r\left( x_{r} +\epsilon_r, x_{r-1} +\epsilon_{r-1}+ \int_{t_{r-1}}^{t_{r}} {\Psi}(\hat{x}(\tau)) d\tau \right).
\end{eqnarray}
As in regular non-convex optimization, the weights are updated according to the following rule:
\begin{eqnarray}\label{eq:wupdate}
{w}_k^{l+1} &=&  {w}_k^{l} - \eta 
\frac{\partial \mathcal{{L}}(\hat{{x}}_{r}, z[{x}_{r}]; {w})}{\partial {w}_{k}}, \\ \nonumber
&=&  {w}_k^{l} - \eta 
 \sum_{r=0}^R\frac{\partial \mathcal{L}_r(\hat{{x}}_{r}, z[{x}_{r}]; {w})}{\partial {w}_{k}},
\end{eqnarray}
where $\eta$ is a learning rate and $l$ is a training step.

Let us first find an expression of the derivative for $\mathcal{L}$:
\begin{equation}\label{eq:dl_by_dw}
\frac{\partial \mathcal{L}_r(\hat{{x}}_{r}, z[{x}_{r}]; {w})}{\partial {w}_{k}} =
\frac{\partial \mathcal{L}_r(\hat{{x}}_{r}, z[{x}_{r}]; {w})}{\partial {x}_{r}} 
\frac{\partial \hat{{x}}_{r}}{\partial {w}_{k}} .
\end{equation}
To compute \eqnref{eq:dl_by_dw}, let us subdivide an interval $[t_{r-1},t_r]$ into $N$ intervals, each of granularity $\Delta t=(t_r-t_{r-1})/N$ as follows: $\{t_{r,j}\}_{j=0}^N$, where $t_{r,j}=t_{r-1}+j\Delta t$. The dynamics of $\hat{{x}}_r$ is
\begin{eqnarray*}
\hat{{x}}_r &=& \hat{{x}}_{r,N} = \hat{{x}}_{r,N-1} + \Delta t\,{\Psi}_{r,N-1}\\ \nonumber
{\Psi}_{r,N-1} &=& {\Psi}(\hat{{x}}_{r,N-1}), \nonumber
\end{eqnarray*}
which means that $\hat{{x}}_{r,N}$ is a function of two variables, $\hat{{x}}_{r,N}=f(\hat{{x}}_{r,N-1}, {\Psi}_{r,N-1})$. The derivative of $\hat{x}_r$ therefore is
\begin{eqnarray*}
    \frac{\partial \hat{{x}}_{r,N}}{\partial {w}_{k}} =
\frac{\partial \hat{{x}}_{r,N}}{ \partial \hat{{x}}_{r,n-1}}
\frac{\partial \hat{{x}}_{r,N-1}}{\partial {w}_{k}}+
\frac{\partial \hat{{x}}_{r,N}}{\partial {\Psi}_{r,N-1}} 
\frac{\partial {\Psi}_{r,N-1}}{\partial {w}_{k}}.
\end{eqnarray*}
Since $\frac{\partial \hat{{x}}_{r,N}}{ \partial \hat{{x}}_{r,N-1}}=1$ and $\frac{\partial \hat{{x}}_{r,N}}{\partial {\Psi}_{r,N-1}}=\Delta t$, we get
\begin{equation*}
\frac{\partial \hat{{x}}_{r,N}}{\partial {w}_{k}} =
\frac{\partial \hat{{x}}_{r,N-1}}{\partial {w}_{k}}+
\Delta t
\frac{\partial {\Psi}_{r,N-1}}{\partial {w}_{k}},
\end{equation*}
which is a recurrence relation
\begin{equation}\label{eq:dhatx_dw}
\frac{\partial \hat{{x}}_{r,N}}{\partial {w}_{k}} =
\sum_{j=0}^{N-1}
\Delta t
\frac{\partial {\Psi}_{r,j}}{\partial {w}_{k}}.
\end{equation}
Note that $\frac{\partial \hat{{x}}_{r,0}}{\partial {w}_{k}}=\frac{\partial \hat{{x}}_{r-1}}{\partial {w}_{k}}=0$, since $\hat{{x}}_{r,0}$ is by definition ``a temporary initial condition" for a segment $[t_{r-1},t_r]$. 

By substituting \eqnref{eq:dhatx_dw} back to the \eqnref{eq:dl_by_dw}, we get
\begin{equation}\label{eq:dl_dw_derivative_final}
\frac{\partial \mathcal{L}_r(\hat{{x}}_{r}, z[{x}_{r}]; {w})}{\partial {w}_{k}} =
\frac{\partial \mathcal{L}_r(\hat{{x}}_{r}, z[{x}_{r}]; {w})}{\partial \hat{{x}}_{r}} 
\sum_{j=0}^{N-1}
\Delta t
\frac{\partial {\Psi}_{r,j}}{\partial {w}_{k}},
\end{equation}
where $\frac{\partial {\Psi}_{r,j}}{\partial {w}_{k}}$ is just the regular gradient of differentiable neural network ${\Psi}_{r,j} = {\Psi}(\hat{{x}}_{r,j};{w})$. Putting \eqnref{eq:dl_dw_derivative_final} back to the \eqnref{eq:wupdate}, we get that weights are updated as follows:
\begin{equation}\label{eq:w_update_final}
{w}_k^{l+1} = {w}_k^l - \eta   \sum_{r=0}^{R} 
\frac{\partial \mathcal{L}_r(\hat{{x}}_{r}, z[{x}_{r}]; {w})}{\partial \hat{{x}}_{r}} 
\sum_{j=0}^{N-1}
\Delta t
\frac{\partial {\Psi}_{r,j}}{\partial {w}_{k}}.
\end{equation}
From \eqnref{eq:w_update_final}, one can see how observational noise affects the computation of the gradient: it affects only the derivative of the loss w.r.t.\ $\hat{x}_r$. E.g.\ in case $\mathcal{L}_r$ is computed using $\ell^2$ norm: $\mathcal{L}_r(\hat{{x}}_{r}, z[{x}_{r}]; {w})=(\hat{{x}}_{r}-z[{x}_{r}])^2$, we get $\frac{\partial \mathcal{L}_r(\hat{{x}}_{r}, z[{x}_{r}]; {w})}{\partial \hat{{x}}_{r}} =2(\hat{{x}}_{r}-z[{x}_{r}])$. For $\ell^1$ norm, $\mathcal{L}_r(\hat{{x}}_{r}, z[{x}_{r}]; {w})=|\hat{{x}}_{r}-z[{x}_{r}]|$, we get $\frac{\partial \mathcal{L}_r(\hat{{x}}_{r}, z[{x}_{r}]; {w})}{\partial \hat{{x}}_{r}} = \textbf{sgn}[\hat{{x}}_{r}-z[{x}_{r}]]$ for $\hat{{x}}_{r} \neq z[{x}_{r}]$.

\FloatBarrier